\documentclass[apj]{emulateapj}
\usepackage{apjfonts}
\usepackage{times}

\usepackage{amsmath}
\usepackage{graphicx}
\usepackage{natbib}

\usepackage{color}

 \font\sevenrm=cmr7 scaled 1000

\slugcomment{Draft Version, }
\shorttitle{}
\shortauthors{Woo et al.}

\newcommand{\Hb}{\rm H{$\beta$}}

\newcommand{\FeII}{\ion{Fe}{2}}
\newcommand{\HeII}{\ion{He}{2}}
\newcommand{\MgII}{\ion{Mg}{2}}
\newcommand{\CIV}{\ion{C}{4}}

\newcommand{\OIII}{\ion{O}{3}}
\newcommand{\OII}{\ion{O}{2}}

\newcommand{\mbh}{M$_{\rm BH}$}

\newcommand{\kms}{km~s$^{\rm -1}$}

\begin{document}

\title{Calibration and LIMITATIONS of the \ion{Mg}{2} line-based BLACK HOLE MASSES}

\author{Jong-Hak Woo$^{1}$}
\author{Huynh Anh N. Le$^{1}$}
\author{Marios Karouzos$^{1}$}
\author{Dawoo Park$^{1}$}
\author{Daeseong Park$^{2}$}
\author{Matthew A. Malkan$^{3}$}
\author{Tommaso Treu$^{3}$}
\author{Vardha N. Bennert$^{4}$}

\affil{$^{1}$Astronomy Program, Department of Physics and Astronomy, Seoul National University, Seoul, 08826\\
$^{2}$Korea Astronomy and Space science Institute, Daejeon, Republic of Korea, 34055\\
$^{3}$Department of Physics and Astronomy, University of California, Los Angeles, CA 90095, USA\\
$^{4}$Physics Department, California Polytechnic State University, San Luis Obispo, CA 93407, USA\\
}

\begin{abstract}
 
We present the single-epoch black hole mass (\mbh) calibrations based on the rest-frame UV and optical measurements of \ion{Mg}{2} 2798\AA\ and \Hb\ 4861\AA\ lines and AGN continuum, using a sample of 52 moderate-luminosity AGNs at z$\sim$0.4 and z$\sim$0.6 with high-quality Keck spectra.
We combine this sample with a large number of luminous AGNs from the Sloan Digital Sky Survey to increase the dynamic range for a better comparison of UV and optical velocity and luminosity measurements. With respect to the reference \mbh\ based on the line dispersion of \Hb\ and continuum luminosity at 5100\AA, we calibrate the UV and optical mass estimators,
by determining the best-fit values of the coefficients in the mass equation.
By investigating whether the UV estimators show systematic trend with Eddington ratio, FWHM of \Hb, the \FeII\ strength, and the UV/optical slope, we find no significant bias except for the slope. By fitting the systematic difference of \MgII-based and \Hb-based masses with the L$_{3000}$/L$_{5100}$ ratio, we provide a correction term as a function of the spectral index as $\Delta$C = 0.24 (1+$\alpha_{\lambda}$) + 0.17, which can be added to the \MgII-based mass estimators if the spectral slope can be well determined.
The derived UV mass estimators typically show $>$$\sim$0.2 dex intrinsic scatter with respect to \Hb-based \mbh, suggesting that the UV-based mass has an additional uncertainty of $\sim$0.2 dex, even if high quality rest-frame UV spectra are available.

\end{abstract}
\keywords{galaxies: active -- galaxies: nuclei -- galaxies: Seyfert}

\section{INTRODUCTION} \label{section:intro}

One of the most fundamental properties of active galactic nuclei (AGNs) is black hole mass (\mbh), which sets the upper limit of AGN energetics via the Eddington limit. \mbh\ also represents the integration of mass accretion over the life time of a given black hole (BH), connecting the growth histories of galaxies and BHs as implied by the observed correlation between \mbh\ and host galaxy properties \citep[e.g.,][]{Kormendy&Ho13, Woo+13}. 

Estimating \mbh\ became a routine process for type 1 AGNs, which are characterized by the presence of broad emission lines, since various single-epoch mass estimators were developed based on the empirical results from reverberation-mapping studies. While the size of the broad-line region (BLR) is measured from the time lag between the light curves of AGN continuum and broad emission line flux in the reverberation-mapping studies \citep{BM82, Peterson93}, which requires a long-term monitoring campaign \citep[e.g.,][]{Wandel1999, Kaspi+00, Peterson+04, Bentz+09, Barth2011,Grier2013, Barth2015,Fausnaugh+17, Park+17}, the single-epoch estimators utilize the empirical size-luminosity relation obtained from the reverberation studies. As a proxy for the BLR size, the monochromatic luminosity at 5100\AA\ can be used to indirectly infer the BLR size based on the empirical relation \citep[e.g.,][]{Kaspi+00, Kaspi+05, Bentz+06, Bentz+13}.

Since the size-luminosity relation provides a powerful simple method for determining \mbh, which only requires single spectroscopic observation, it has been applied to a large sample of type 1 AGNs \citep[e.g.,][]{Woo&Urry02, MD04, Shen+11}. While the \Hb\ line was the main tool to measure the time lag in the reverberation-mapping studies, hence, the size-luminosity relation is best-calibrated with the \Hb\ line, the single epoch method with various recipes became applicable to type 1 AGNs at higher redshift. In this case, the rest-frame UV continuum and emission lines obtained in large optical spectroscopic surveys, e.g., Sloan Digital Sky Survey (SDSS) are typically used to estimate \mbh. For example, a combination of the \ion{Mg}{2} 2798\AA\ line velocity and near-UV continuum luminosity has been used for AGNs at 0.4 $<$ z $<$ 2 \citep[e.g.,][]{MD04, VP06, McGill+08, Wang+09, Shen+11}, while the pair of the velocity of \ion{C}{4} 1549\AA\ line and AGN contiuum luminosity in the Far-UV is often adopted for AGNs at z$>$$\sim$2 \citep[e.g.,][]{Shen+11, Karouzos+15}. 

By combining the virial assumption, i.e., the BLR gas is mainly governed by the gravitational potential of the central BH, and the size-luminosity relation between BLR size (R$_{BLR}$) and continuum luminosity (L) as R$_{BLR}$ $\propto$ L$^{\sim 0.5}$, $M_{\rm BH}$ can be expressed as,
\begin{equation}
	\log M_{\rm BH}= \alpha + \beta \log V + \gamma \log L
	\label{eq:mbh}
\end{equation}
where $V$ is the velocity measured from the width of a broad emission line. 
For \Hb-based mass estimators, $\beta$ is fixed as 2, based on the virial theorem. However the value of $\beta$ can differ from 2 as the comparison between \Hb\ line width and the line width of other broad emission lines often shows a non-linear relationship \citep[e.g.,][]{Wang+09, Marziani+13}.
In the case of $\gamma$, the most updated \Hb\ size luminosity relation study reported $\gamma$=0.533$^{0.035}_{-0.033}$ \citep{Bentz+13}, which is consistent with a naive photoionization assumption \citep{Wandel1999}, while for other luminosity measures, i.e., UV continuum or line luminosities, the value can also vary.

The alternative UV mass estimators are calibrated compared to the \Hb-based mass estimators. 
Since the directly measured time-lag (i.e., BLR size) based on the variability of \MgII\ has been limited to only a small number of objects \citep{Reichert+94, Metzroth+06, Shen+16}, due to the observational difficulties and/or the lack of consistent variability between line and continuum \citep{Woo+08, Cackett+15}, there is no available \MgII-based size-luminosity relation. Thus, UV mass estimators need to be calibrated with \Hb-based measurements, i.e., time lags and mass estimates.  
Using the reverberation-mapped AGN sample, for example, \citet{MJ02} compared the \Hb-based BLR size with the UV luminosity at 3000 \AA\ (L$_{3000}$), and provided 
a recipe of \mbh\ determination based on L$_{3000}$ and the FWHM of \MgII, \citep[see also][]{MD04, VP06}. Note that while the AGNs with the measured reverberation lags and the line width measurements based on the rms spectra have been used to calibrate \mbh\ estimators, this sample is limited to relatively low-z objects with low-to-moderate luminosity \citep{Bentz+06}. 
Alternatively, the calibration has been performed by determining the best coefficients in Eq. 1, which provides the most consistent masses compared to \Hb-based single-epoch masses \citep[e.g.,][]{McGill+08,Shen+11}

In this work, we present the UV mass estimators by updating the result of \citet{McGill+08}, who reported a new calibration of \MgII-based mass estimators using a small sample of 19 AGNs with very high quality Keck spectroscopic data, and investigated the systematic uncertainties due to the variability, line width measurements, and Eddington ratio, which may affect the calibration of UV mass estimators. The improvements of the current work are as follows: (i) the enlarged sample size from 19 AGNs to $\sim$ 50 objects; (ii) the enlarged dynamic range by a factor of $\sim$2; (iii) the improved and updated spectral decomposition, particularly with a better UV \FeII\ template; (iv) the updated virial factor and normalization, reflecting the progress of the calibration of \Hb-based \mbh\ studies. 
The high quality spectra from the Keck telescope enable us to reliably remove non-broad line components from the observed spectra to accurately measure the continuum and emission line properties. In \S \ref{section:obs}, we describe the sample, observation, and data reduction. \S \ref{section:meas} describes the fitting procedure and analysis. \S \ref{section:scaling} presents the calibration. 
Finally, we provide discussion and summary in \S 5 and \S 6, respectively. The following cosmological parameters are used throughout the paper: $H_0 = 70$~km~s$^{-1}$~Mpc$^{-1}$, $\Omega_{\rm m} = 0.30$, and $\Omega_{\Lambda} = 0.70$.
\begin{turnpage}
\centering
\begin{deluxetable*}{lccccccccccccccc}
\tablecolumns{16}
\tablewidth{\textwidth}
\tabletypesize{\scriptsize}
\tablecaption{Target Properties}
\tablehead{
\colhead{Name}&
\colhead{z}&
\colhead{R.A.}&
\colhead{Dec.}&
\colhead{{\it i'}}&
\colhead{t(s)}&
\colhead{Run}&
\colhead{FWHM$_{\rm H\beta}$} &
\colhead{FWHM$_{\rm Mg \sevenrm II}$} &
\colhead{$\sigma_{\rm H\beta}$} &
\colhead{$\sigma_{\rm Mg \sevenrm II}$} &
\colhead{L$_{3000}$} &
\colhead{L$_{5100}$} &
\colhead{L$_{\rm H\beta}$} &
\colhead{L$_{\rm Mg \sevenrm II}$} 
\\
\colhead{(1)}&
\colhead{(2)}&
\colhead{(3)}&
\colhead{(4)}&
\colhead{(5)}&
\colhead{(6)}&
\colhead{(7)}&
\colhead{(8)} &
\colhead{(9)} &
\colhead{(10)} &
\colhead{(11)} &
\colhead{(12)} &
\colhead{(13)} &
\colhead{(14)} &
\colhead{(15)} 
}

\startdata
S01	&	$0.3593$	&	$15\,	39\,	16.24$	&	$+03\,	23\,	22.07$	&	$18.89$	&	$10400$	&	$1,4$	&	4662	$\pm$	26	&	3324	$\pm$	66	&	2194	$\pm$	21	&	1856	$\pm$	35	&	2.38	$\pm$	0.01	&	1.37	$\pm$	0.02	&	2.22	$\pm$	0.01	&	5.25	$\pm$	0.05	\\
S02	&	$0.3545$	&	$16\,	11\,	11.66$	&	$+51\,	31\,	31.16$	&	$19.00$	&	$3000$	&	$1$	&	4841	$\pm$	35	&	3332	$\pm$	48	&	2274	$\pm$	25	&	2088	$\pm$	35	&	1.61	$\pm$	0.01	&	1.25	$\pm$	0.02	&	2.95	$\pm$	0.01	&	6.43	$\pm$	0.06	\\
S03	&	$0.3582$	&	$17\,	32\,	03.08$	&	$+61\,	17\,	51.89$	&	$18.30$	&	$5500$	&	$1,8$	&	3018	$\pm$	17	&	2221	$\pm$	42	&	1716	$\pm$	9	&	1249	$\pm$	26	&	4.11	$\pm$	0.01	&	2.11	$\pm$	0.04	&	3.80	$\pm$	0.01	&	4.14	$\pm$	0.04	\\
S04	&	$0.3579$	&	$21\,	02\,	11.50$	&	$-06\,	46\,	45.01$	&	$18.57$	&	$2400$	&	$1$	&	2821	$\pm$	46	&	3079	$\pm$	60	&	1749	$\pm$	46	&	1708	$\pm$	45	&	2.05	$\pm$	0.03	&	1.19	$\pm$	0.05	&	1.07	$\pm$	0.02	&	3.99	$\pm$	0.06	\\
S05	&	$0.3530$	&	$21\,	04\,	51.83$	&	$-07\,	12\,	09.41$	&	$18.54$	&	$12600$	&	$1,4$	&	4908	$\pm$	26	&	4013	$\pm$	114	&	3333	$\pm$	17	&	2637	$\pm$	90	&	3.14	$\pm$	0.09	&	2.23	$\pm$	0.03	&	4.52	$\pm$	0.05	&	6.58	$\pm$	0.14	\\
S06	&	$0.3684$	&	$21\,	20\,	34.18$	&	$-06\,	41\,	22.24$	&	$18.84$	&	$3300$	&	$1$	&	4527	$\pm$	65	&	3056	$\pm$	153	&	1413	$\pm$	106	&	1423	$\pm$	123	&	2.44	$\pm$	0.02	&	1.10	$\pm$	0.09	&	1.30	$\pm$	0.04	&	2.74	$\pm$	0.08	\\
S07	&	$0.3517$	&	$23\,	09\,	46.14$	&	$+00\,	00\,	48.87$	&	$18.18$	&	$7200$	&	$1,4$	&	4635	$\pm$	28	&	3429	$\pm$	125	&	2547	$\pm$	15	&	2107	$\pm$	55	&	3.66	$\pm$	0.03	&	1.81	$\pm$	0.08	&	3.89	$\pm$	0.01	&	3.50	$\pm$	0.05	\\
S08	&	$0.3585$	&	$23\,	59\,	53.44$	&	$-09\,	36\,	55.63$	&	$18.49$	&	$2400$	&	$1$	&	2909	$\pm$	63	&	2212	$\pm$	72	&	1217	$\pm$	33	&	1167	$\pm$	75	&	2.89	$\pm$	0.02	&	1.59	$\pm$	0.05	&	0.89	$\pm$	0.03	&	1.70	$\pm$	0.05	\\
S09	&	$0.3542$	&	$00\,	59\,	16.10$	&	$+15\,	38\,	16.10$	&	$18.38$	&	$1800$	&	$1$	&	2655	$\pm$	28	&	2946	$\pm$	52	&	1748	$\pm$	21	&	1652	$\pm$	44	&	2.68	$\pm$	0.03	&	1.76	$\pm$	0.06	&	2.80	$\pm$	0.03	&	6.74	$\pm$	0.08	\\
S10	&	$0.3505$	&	$01\,	01\,	12.06$	&	$-09\,	45\,	00.81$	&	$17.97$	&	$3300$	&	$1,7$	&	4850	$\pm$	20	&	3388	$\pm$	97	&	2597	$\pm$	12	&	2034	$\pm$	96	&	6.65	$\pm$	0.06	&	2.77	$\pm$	0.04	&	6.12	$\pm$	0.02	&	6.84	$\pm$	0.11	\\
S11	&	$0.3558$	&	$01\,	07\,	15.97$	&	$-08\,	34\,	29.37$	&	$18.47$	&	$10200$	&	$1,4$	&	2595	$\pm$	14	&	2650	$\pm$	50	&	1354	$\pm$	8	&	1410	$\pm$	39	&	3.33	$\pm$	0.05	&	1.57	$\pm$	0.03	&	2.55	$\pm$	0.01	&	2.74	$\pm$	0.04	\\
S12	&	$0.3574$	&	$02\,	13\,	40.59$	&	$+13\,	47\,	56.05$	&	$18.37$	&	$1800$	&	$1$	&	8800	$\pm$	333	&	7014	$\pm$	172	&	4256	$\pm$	56	&	3376	$\pm$	107	&	3.58	$\pm$	0.05	&	1.82	$\pm$	0.12	&	4.67	$\pm$	0.03	&	8.30	$\pm$	0.13	\\
S16	&	$0.3702$	&	$11\,	19\,	37.59$	&	$+00\,	56\,	20.36$	&	$19.10$	&	$600$	&	$9$	&	3749	$\pm$	784	&	7008	$\pm$	704	&	1867	$\pm$	445	&	3518	$\pm$	579	&	0.21	$\pm$	0.02	&	0.69	$\pm$	0.11	&	0.76	$\pm$	0.08	&	0.41	$\pm$	0.03	\\
S21	&	$0.3532$	&	$11\,	05\,	56.18$	&	$+03\,	12\,	43.15$	&	$17.31$	&	$1500$	&	$2$	&	8296	$\pm$	145	&	4311	$\pm$	211	&	3897	$\pm$	105	&	2037	$\pm$	131	&	1.92	$\pm$	0.01	&	5.33	$\pm$	0.09	&	8.10	$\pm$	0.03	&	2.36	$\pm$	0.05	\\
S23	&	$0.3511$	&	$14\,	00\,	16.65$	&	$-01\,	08\,	22.16$	&	$18.16$	&	$1800$	&	$2,4$	&	9629	$\pm$	146	&	5482	$\pm$	151	&	4251	$\pm$	168	&	2604	$\pm$	55	&	2.20	$\pm$	0.01	&	1.78	$\pm$	0.03	&	2.70	$\pm$	0.02	&	4.63	$\pm$	0.05	\\
S24	&	$0.3616$	&	$14\,	00\,	34.70$	&	$+00\,	47\,	33.43$	&	$18.29$	&	$9600$	&	$2,4$	&	7061	$\pm$	49	&	4466	$\pm$	72	&	2635	$\pm$	17	&	2288	$\pm$	30	&	1.81	$\pm$	0.03	&	1.49	$\pm$	0.02	&	2.39	$\pm$	0.01	&	5.00	$\pm$	0.03	\\
S26	&	$0.3691$	&	$15\,	29\,	22.26$	&	$+59\,	28\,	54.54$	&	$18.92$	&	$3600$	&	$2$	&	5386	$\pm$	22	&	4642	$\pm$	149	&	1914	$\pm$	10	&	2305	$\pm$	130	&	1.59	$\pm$	0.01	&	0.83	$\pm$	0.02	&	2.36	$\pm$	0.02	&	2.43	$\pm$	0.06	\\
S27	&	$0.3667$	&	$15\,	36\,	51.27$	&	$+54	\,14\,	42.63$	&	$18.86$	&	$7200$	&	$2$	&	2508	$\pm$	28	&	2682	$\pm$	65	&	1409	$\pm$	17	&	1234	$\pm$	47	&	1.52	$\pm$	0.04	&	1.26	$\pm$	0.05	&	1.72	$\pm$	0.01	&	1.88	$\pm$	0.03	\\
S28	&	$0.3679$	&	$16\,	11\,	56.29$	&	$+45\,	16\,	10.91$	&	$18.63$	&	$5760$	&	$3,4$	&	4600	$\pm$	51	&	4974	$\pm$	87	&	2532	$\pm$	36	&	2690	$\pm$	47	&	2.06	$\pm$	0.02	&	0.97	$\pm$	0.03	&	1.79	$\pm$	0.02	&	4.43	$\pm$	0.04	\\
S29	&	$0.3575$	&	$21\,	58\,	41.92$	&	$-01\,	15\,	00.32$	&	$18.95$	&	$3600$	&	$3$	&	3533	$\pm$	44	&	3036	$\pm$	72	&	1847	$\pm$	28	&	1780	$\pm$	57	&	1.14	$\pm$	0.02	&	1.20	$\pm$	0.04	&	1.48	$\pm$	0.01	&	1.59	$\pm$	0.02	\\
S31	&	$0.3505$	&	$10\,	15\,	27.26$	&	$+62\,	59\,	11.52$	&	$18.15$	&	$9000$	&	$9,10$	&	4012	$\pm$	27	&	3099	$\pm$	123	&	2117	$\pm$	20	&	1887	$\pm$	74	&	1.75	$\pm$	0.02	&	0.93	$\pm$	0.03	&	2.24	$\pm$	0.01	&	1.93	$\pm$	0.04	\\
SS1	&	$0.3566$	&	$08\,	04\,	27.98$	&	$+52\,	23\,	06.21$	&	$18.55$	&	$9000$	&	$5$	&	2620	$\pm$	49	&	2458	$\pm$	71	&	1501	$\pm$	32	&	1255	$\pm$	66	&	1.59	$\pm$	0.02	&	1.04	$\pm$	0.08	&	1.88	$\pm$	0.03	&	1.62	$\pm$	0.03	\\
SS2	&	$0.3672$	&	$09\,	34\,	55.60$	&	$+05\,	14\,	09.15$	&	$18.82$	&	$7200$	&	$5$	&	2815	$\pm$	61	&	2777	$\pm$	69	&	1316	$\pm$	41	&	1296	$\pm$	46	&	0.90	$\pm$	0.01	&	0.83	$\pm$	0.04	&	0.81	$\pm$	0.02	&	0.97	$\pm$	0.02	\\
SS4	&	$0.3630$	&	$09\,	58\,	50.15$	&	$+40\,	03\,	42.33$	&	$18.74$	&	$5400$	&	$9$	&	2213	$\pm$	35	&	2302	$\pm$	29	&	1378	$\pm$	15	&	1198	$\pm$	21	&	2.28	$\pm$	0.01	&	1.35	$\pm$	0.05	&	1.93	$\pm$	0.01	&	3.83	$\pm$	0.03	\\
SS5	&	$0.3733$	&	$10\,	07\,	06.25$	&	$+08\,	42\,	28.41$	&	$18.69$	&	$3600$	&	$9$	&	2790	$\pm$	63	&	1954	$\pm$	83	&	1612	$\pm$	40	&	1092	$\pm$	65	&	1.39	$\pm$	0.01	&	1.40	$\pm$	0.04	&	0.90	$\pm$	0.01	&	0.49	$\pm$	0.02	\\
SS6	&	$0.3584$	&	$10\,	21\,	03.57$	&	$+30\,	47\,	55.87$	&	$18.92$	&	$5400$	&	$9$	&	1947	$\pm$	21	&	2069	$\pm$	86	&	1031	$\pm$	13	&	868	$\pm$	57	&	1.33	$\pm$	0.01	&	0.69	$\pm$	0.03	&	1.12	$\pm$	0.01	&	0.85	$\pm$	0.02	\\
SS7	&	$0.3618$	&	$10\,	43\,	31.50$	&	$-01\,	07\,	32.88$	&	$18.82$	&	$5400$	&	$9$	&	2959	$\pm$	56	&	2020	$\pm$	79	&	1371	$\pm$	27	&	1163	$\pm$	47	&	2.06	$\pm$	0.01	&	0.98	$\pm$	0.04	&	1.19	$\pm$	0.01	&	1.44	$\pm$	0.03	\\
SS8	&	$0.3656$	&	$10\,	46\,	10.60$	&	$+03\,	50\,	31.26$	&	$18.45$	&	$9900$	&	$9,10,11$	&	2733	$\pm$	43	&	2446	$\pm$	48	&	1532	$\pm$	11	&	1298	$\pm$	28	&	3.20	$\pm$	0.01	&	1.54	$\pm$	0.02	&	2.27	$\pm$	0.01	&	3.03	$\pm$	0.03	\\
SS9	&	$0.3701$	&	$12\,	58\,	38.71$	&	$+45\,	55\,	15.55$	&	$18.56$	&	$5400$	&	$9$	&	2787	$\pm$	27	&	3014	$\pm$	50	&	1569	$\pm$	14	&	1501	$\pm$	25	&	2.87	$\pm$	0.04	&	1.25	$\pm$	0.02	&	2.08	$\pm$	0.02	&	4.25	$\pm$	0.04	\\
SS10	&	$0.3658$	&	$13\,	34\,	14.84$	&	$+11\,	42\,	21.52$	&	$17.83$	&	$3600$	&	$10$	&	2232	$\pm$	36	&	1429	$\pm$	58	&	1431	$\pm$	20	&	820	$\pm$	42	&	4.97	$\pm$	0.13	&	4.09	$\pm$	0.13	&	6.19	$\pm$	0.02	&	4.87	$\pm$	0.12	\\
SS11	&	$0.3732$	&	$13\,	52\,	26.90$	&	$+39\,	24\,	26.84$	&	$18.39$	&	$2400$	&	$10$	&	3505	$\pm$	78	&	2661	$\pm$	117	&	1466	$\pm$	70	&	1630	$\pm$	96	&	2.67	$\pm$	0.02	&	2.07	$\pm$	0.02	&	1.01	$\pm$	0.04	&	1.84	$\pm$	0.06	\\
SS12	&	$0.3625$	&	$15\,	01\,	16.82$	&	$+53\,	31\,	02.13$	&	$17.80$	&	$5500$	&	$7$	&	2101	$\pm$	10	&	1865	$\pm$	94	&	1371	$\pm$	6	&	930	$\pm$	59	&	3.34	$\pm$	0.04	&	4.34	$\pm$	0.02	&	6.87	$\pm$	0.01	&	1.87	$\pm$	0.05	\\
SS13	&	$0.3745$	&	$15\,	05\,	41.78$	&	$+49\,	35\,	19.99$	&	$18.73$	&	$11100$	&	$8,9$	&	2169	$\pm$	12	&	2182	$\pm$	35	&	1143	$\pm$	9	&	1123	$\pm$	28	&	2.16	$\pm$	0.01	&	1.49	$\pm$	0.02	&	1.10	$\pm$	0.01	&	1.76	$\pm$	0.02	\\
SS14	&	$0.3706$	&	$21\,	15\,	31.68$	&	$-07\,	26\,	27.50$	&	$19.24$	&	$9000$	&	$7$	&	2143	$\pm$	27	&	2114	$\pm$	50	&	1212	$\pm$	17	&	1061	$\pm$	36	&	1.21	$\pm$	0.01	&	0.65	$\pm$	0.03	&	1.05	$\pm$	0.01	&	1.46	$\pm$	0.02	\\
SS15	&	$0.3595$	&	$01\,	44\,	12.77$	&	$-00\,	06\,	10.54$	&	$19.46$	&	$8700$	&	$7$	&	1604	$\pm$	28	&	1870	$\pm$	93	&	1000	$\pm$	15	&	1008	$\pm$	80	&	0.46	$\pm$	0.00	&	0.64	$\pm$	0.02	&	0.65	$\pm$	0.01	&	0.19	$\pm$	0.01	\\
SS17	&	$0.3554$	&	$21\,	44\,	10.62$	&	$-01\,	01\,	13.42$	&	$18.47$	&	$5400$	&	$7$	&	1631	$\pm$	82	&	1730	$\pm$	34	&	1029	$\pm$	51	&	819	$\pm$	28	&	3.64	$\pm$	0.02	&	1.90	$\pm$	0.10	&	1.01	$\pm$	0.01	&	3.10	$\pm$	0.04	\\
SS18	&	$0.3582$	&	$23\,	40\,	50.52$	&	$+01\,	06\,	35.47$	&	$18.50$	&	$7200$	&	$7$	&	1890	$\pm$	179	&	1484	$\pm$	36	&	957	$\pm$	50	&	756	$\pm$	38	&	2.39	$\pm$	0.02	&	1.36	$\pm$	0.03	&	1.91	$\pm$	0.02	&	1.59	$\pm$	0.04	\\
W01	&	$0.5736$	&	$08\,	36\,	54.98$	&	$+07\,	57\,	12.46$	&	$18.59$	&	$10800$	&	$6$	&	7378	$\pm$	43	&	5904	$\pm$	53	&	3152	$\pm$	9	&	3004	$\pm$	24	&	5.28	$\pm$	0.09	&	4.71	$\pm$	0.00	&	7.06	$\pm$	0.03	&	7.52	$\pm$	0.04	\\
W02	&	$0.5720$	&	$11\,	06\,	41.86$	&	$+61\,	41\,	46.57$	&	$18.96$	&	$12600$	&	$6$	&	12647	$\pm$	88	&	4573	$\pm$	340	&	4811	$\pm$	34	&	3137	$\pm$	50	&	3.75	$\pm$	0.01	&	3.03	$\pm$	0.08	&	6.79	$\pm$	0.02	&	7.57	$\pm$	0.20	\\
W03	&	$0.5760$	&	$00\,	20\,	05.69$	&	$-00\,	50\,	16.25$	&	$19.38$	&	$10800$	&	$7$	&	7461	$\pm$	61	&	4400	$\pm$	50	&	3508	$\pm$	27	&	2299	$\pm$	26	&	1.78	$\pm$	0.01	&	1.47	$\pm$	0.04	&	3.13	$\pm$	0.01	&	3.52	$\pm$	0.02	\\
W04	&	$0.5766$	&	$09\,	32\,	10.96$	&	$+43\,	38\,	13.03$	&	$18.96$	&	$16200$	&	$5,6$	&	3490	$\pm$	30	&	3103	$\pm$	184	&	1728	$\pm$	18	&	1607	$\pm$	83	&	4.34	$\pm$	0.01	&	3.68	$\pm$	0.04	&	5.04	$\pm$	0.06	&	6.32	$\pm$	0.15	\\
W05	&	$0.5767$	&	$09\,	48\,	52.73$	&	$+36\,	31\,	20.55$	&	$18.59$	&	$10800$	&	$5$	&	2722	$\pm$	13	&	1840	$\pm$	51	&	1738	$\pm$	8	&	977	$\pm$	28	&	4.66	$\pm$	0.01	&	4.94	$\pm$	0.03	&	8.30	$\pm$	0.03	&	4.19	$\pm$	0.06	\\
W08	&	$0.5712$	&	$16\,	32\,	52.42$	&	$+26\,	37\,	49.11$	&	$18.70$	&	$6800$	&	$8$	&	7340	$\pm$	50	&	5582	$\pm$	158	&	2977	$\pm$	23	&	3390	$\pm$	112	&	5.08	$\pm$	0.02	&	4.17	$\pm$	0.06	&	5.86	$\pm$	0.04	&	7.46	$\pm$	0.08	\\
W09	&	$0.5654$	&	$15\,	52\,	27.81$	&	$+56\,	22\,	36.46$	&	$19.04$	&	$9200$	&	$6$	&	5273	$\pm$	86	&	2673	$\pm$	85	&	2747	$\pm$	47	&	1594	$\pm$	97	&	1.34	$\pm$	0.01	&	2.64	$\pm$	0.05	&	3.79	$\pm$	0.03	&	1.08	$\pm$	0.03	\\
W10	&	$0.5711$	&	$11\,	14\,	15.83$	&	$-00\,	59\,	20.41$	&	$19.60$	&	$7200$	&	$5$	&	3636	$\pm$	83	&	2775	$\pm$	184	&	1477	$\pm$	58	&	1488	$\pm$	203	&	1.40	$\pm$	0.01	&	2.92	$\pm$	0.04	&	2.12	$\pm$	0.05	&	0.65	$\pm$	0.04	\\
W11	&	$0.5650$	&	$01\,	55\,	16.18$	&	$-09\,	45\,	55.94$	&	$20.09$	&	$10800$	&	$5$	&	3812	$\pm$	89	&	3593	$\pm$	115	&	2026	$\pm$	72	&	1693	$\pm$	114	&	1.76	$\pm$	0.03	&	0.78	$\pm$	0.06	&	1.39	$\pm$	0.04	&	2.24	$\pm$	0.05	\\
W12	&	$0.5623$	&	$14\,	39\,	55.10$	&	$+35\,	53\,	05.31$	&	$19.20$	&	$9000$	&	$6$	&	7698	$\pm$	221	&	2370	$\pm$	82	&	3859	$\pm$	27	&	1769	$\pm$	46	&	3.96	$\pm$	0.01	&	3.62	$\pm$	0.09	&	9.01	$\pm$	0.02	&	7.83	$\pm$	0.11	\\
W14	&	$0.5617$	&	$12\,	56\,	31.89$	&	$-02\,	31\,	30.60$	&	$18.77$	&	$3000$	&	$5$	&	5001	$\pm$	15	&	3042	$\pm$	102	&	2616	$\pm$	16	&	1747	$\pm$	65	&	5.04	$\pm$	0.06	&	5.56	$\pm$	0.03	&	11.29	$\pm$	0.04	&	4.16	$\pm$	0.07	\\
W16	&	$0.5780$	&	$15\,	26\,	54.93$	&	$-00\,	32\,	43.27$	&	$19.99$	&	$7500$	&	$8$	&	2392	$\pm$	19	&	2331	$\pm$	44	&	1564	$\pm$	17	&	1121	$\pm$	34	&	2.30	$\pm$	0.01	&	1.05	$\pm$	0.05	&	2.16	$\pm$	0.01	&	2.34	$\pm$	0.03	\\
W17	&	$0.5617$	&	$10\,	07\,	28.38$	&	$+39\,	26\,	51.81$	&	$19.75$	&	$12800$	&	$5,6$	&	5556	$\pm$	94	&	3807	$\pm$	97	&	2483	$\pm$	59	&	2153	$\pm$	36	&	0.92	$\pm$	0.01	&	0.86	$\pm$	0.03	&	1.52	$\pm$	0.02	&	3.06	$\pm$	0.03	\\
W20	&	$0.5761$	&	$15\,	00\,	14.81$	&	$+32\,	29\,	40.38$	&	$19.60$	&	$5400$	&	$7$	&	10861	$\pm$	360	&	3846	$\pm$	256	&	3806	$\pm$	77	&	2438	$\pm$	51	&	1.18	$\pm$	0.01	&	1.33	$\pm$	0.20	&	1.28	$\pm$	0.03	&	3.51	$\pm$	0.07	\\
W22	&	$0.5652$	&	$03\,	42\,	29.70$	&	$-05\,	23\,	19.44$	&	$18.70$	&	$9000$	&	$5$	&	5835	$\pm$	80	&	3344	$\pm$	54	&	2654	$\pm$	20	&	2044	$\pm$	35	&	6.03	$\pm$	0.01	&	4.65	$\pm$	0.05	&	5.98	$\pm$	0.04	&	4.89	$\pm$	0.04	\\\enddata
\label{table:targets}
\tablecomments{
Col. (1) : Name.
Col. (2) : redshift from SDSS DR7.
Col. (3) : Right ascension (J2000.0).
Col. (4) : Declination (J2000.0).
Col. (5) : Extinction-corrected {\it i'} AB magnitude from SDSS DR7 photometry.
Col. (6) : Exposure time (s)
Col. (7) : Observation date (Table \ref{table:observation}).
Col. (8): FWHM of \Hb\  (km s$^{-1}$).
Col. (9): FWHM of \ion{Mg}{2} (km s$^{-1}$).
Col. (10): Line dispersion of \Hb\ (km s$^{-1}$).
Col. (11): Line dispersion of \ion{Mg}{2} (km s$^{-1}$). 
Col. (12): Luminosity at 3000 \AA\  after \ion{Fe}{2} subtraction (10$^{44}$ erg s$^{-1}$).
Col. (13): Luminosity  at 5100 \AA\ after \ion{Fe}{2} and stellar model subtraction (10$^{44}$ erg s$^{-1}$).
Col. (14): Luminosity of \Hb\ (10$^{42}$ erg s$^{-1}$).
Col. (15): Luminosity of \ion{Mg}{2} (10$^{42}$ erg s$^{-1}$).
}
\end{deluxetable*}
\end{turnpage}

\begin{figure*}
\figurenum{1}
\centering
	\includegraphics[width = 0.49\textwidth, height = 13 cm]{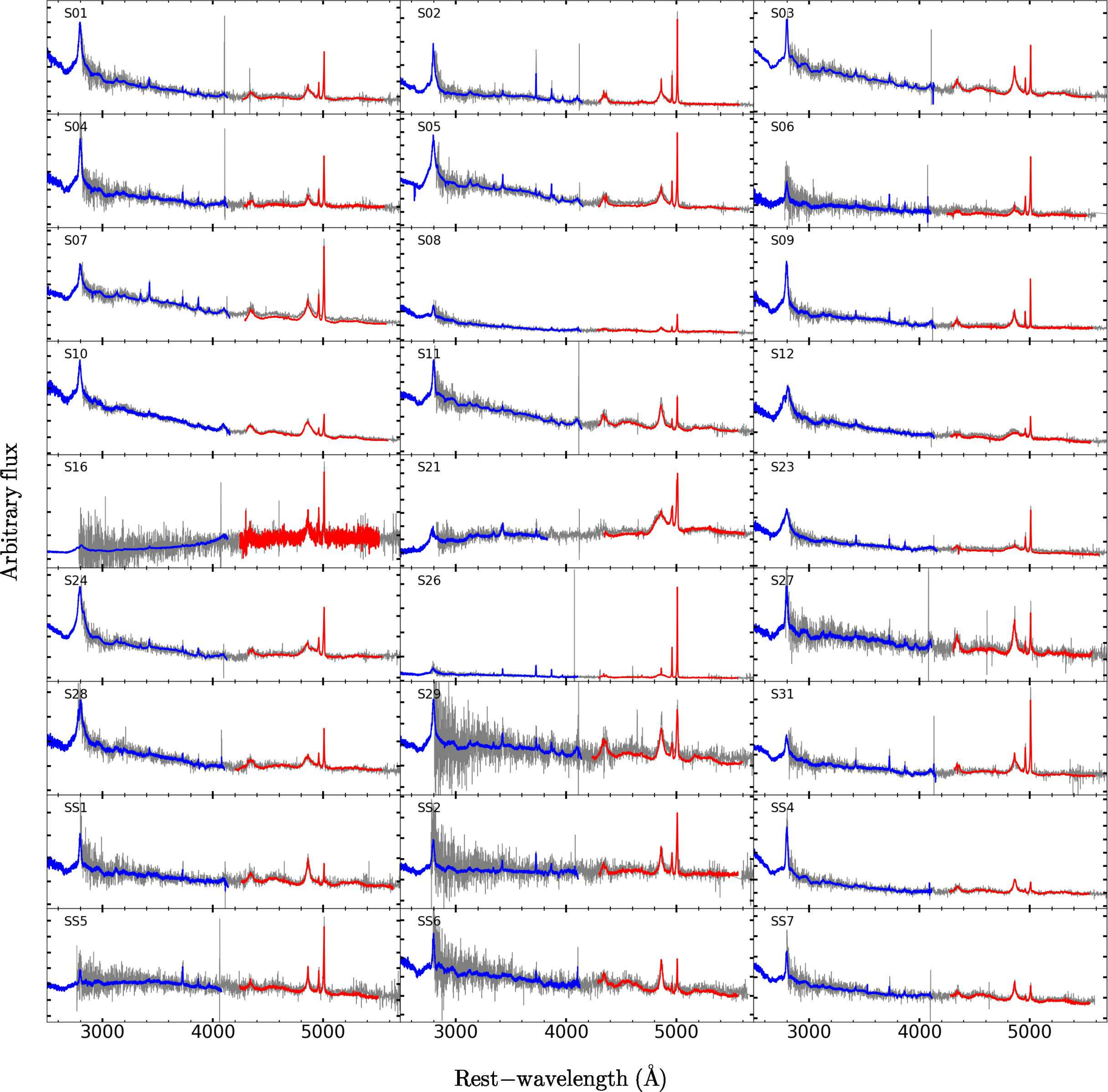}
		\includegraphics[width = 0.49\textwidth, height = 13 cm ]{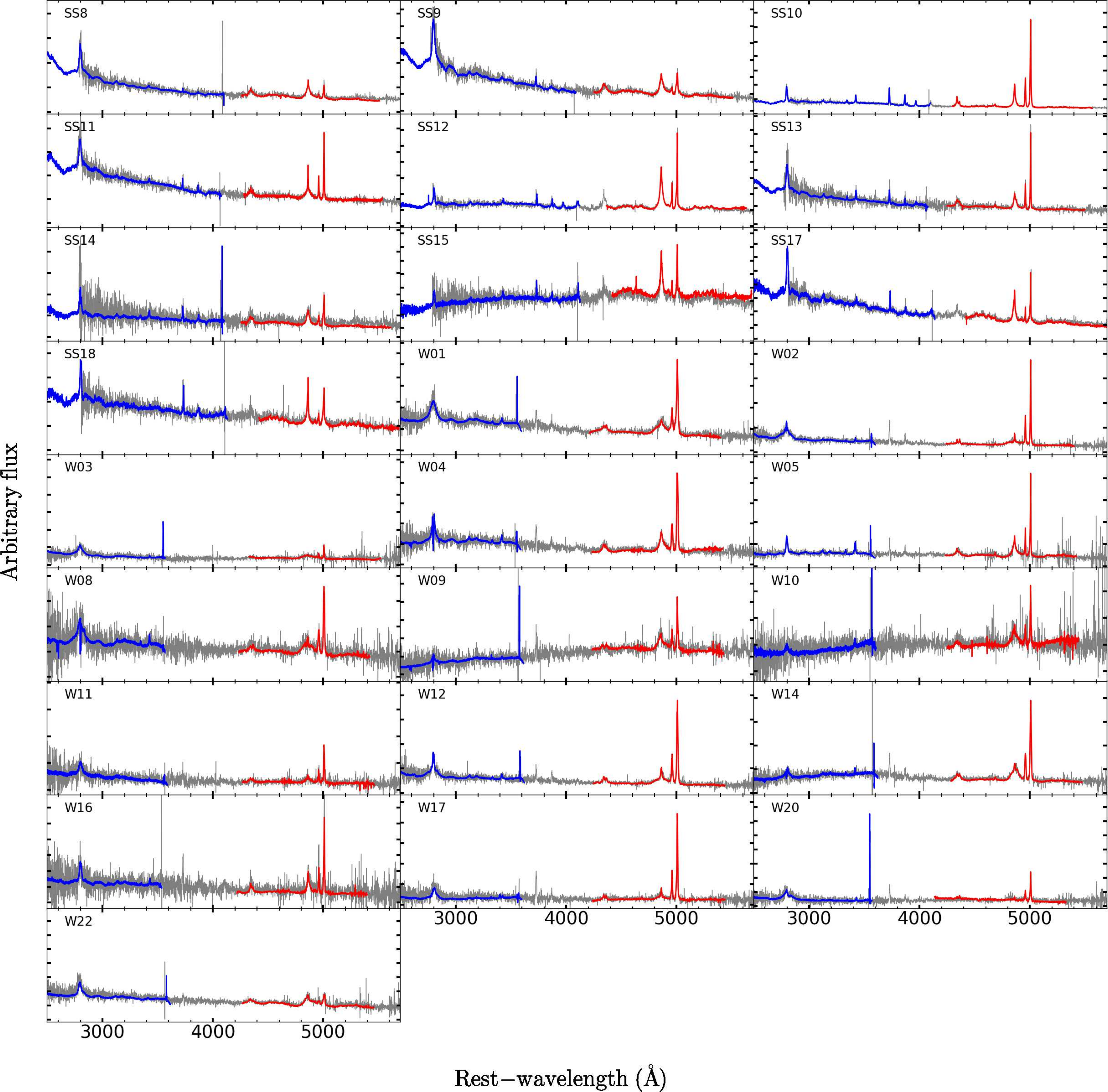}
	\caption{Rest-frame spectra, covering the \MgII\ (blue) and the \Hb\ regions (red) obtained with Keck LRIS. The SDSS spectra after Galactic extinction correction are shown in gray. 
	\label{fig:allspec1}}
\end{figure*}

\section{Observations \& Data Reduction}
\label{section:obs}

The sample was initially selected for measuring stellar velocity dispersions of AGN host galaxies to study the evolution of the
$M_{\rm BH} - \sigma_{*}$ relation \citep{Treu04, Woo06, Woo+08, Bennert+10}. We selected moderate luminosity AGNs from SDSS, at particular redshift ranges, at $\sim 0.36$ and $\sim 0.52$ in order to observe the broad \Hb\ emission line and stellar absorption lines in the rest-frame 5100-5500 \AA. 

We observed the sample at the Keck telescope between 2003 September and 2009 April as summarized in Table~\ref{table:targets} using the Low-Resolution Imaging Spectrometer \citep{Oke+95}, which provided two spectral ranges, containing \ion{Mg}{2} (2798{\AA}) and \Hb \ (4861{\AA}) broad emission lines at blue and red CCDs, respectively. The data reduction and calibration for the red and blue CCD data were reported by \citet{Woo06} and \citet{McGill+08}, respectively. Here, we briefly summarize the observations and reduction for the blue CCD (see Table 2). We used the 600 lines mm$^{-1}$ grism with a pixel scale of 0.63{\AA} and a resolution of ~145 km s$^{-1}$ in line dispersion.

Standard spectroscopic data reduction processes, including bias subtraction, flat fielding, flux calibration, and wavelength calibration were performed using IRAF\footnote{IRAF is distributed by the National Optical Astronomy Observatories, which are operated by the Association of Universities for Research in Astronomy, Inc., under cooperative agreement with the National Science Foundation (NSF).}.
We extracted one dimensional spectra with a 10 pixel (i.e., 1.35 \arcsec) wide aperture. Wavelength calibration was applied using Hg, Ne, Cd arc lamp images. After the flux calibration based on the observation of spectroscopic standard stars (i.e., Feige 34), we rescaled the flux level of our targets to that of SDSS spectrophotometry to compensate the uncertainties of flux calibration due to slit loss, seeing effect, etc. Finally, Galactic extinction was corrected based on the method given by \citet{schlegel98} (see Figure 1).

\begin{deluxetable*}{lccccc}
\tablecolumns{5}
\tablewidth{\textwidth}
\tabletypesize{\scriptsize}
\tablecaption{Observation Log}
\tablehead{
\colhead{Run}&
\colhead{Date}&
\colhead{Slit}&
\colhead{Seeing}&
\colhead{Weather}
\\
\colhead{}&
\colhead{}&
\colhead{(arcsec)}&
\colhead{(arcsec)}&
\colhead{}
\\
\colhead{(1)}&
\colhead{(2)}&
\colhead{(3)}&
\colhead{(4)}&
\colhead{(5)}
}
  
\startdata
1	&	2003	Sep	 3	&	$1.5$	&	$\sim	1$	&	Cirrus\\
2	&	2004	May 14	&	$1$	&	$\sim	1$	&	Cirrus\\
3	&	2004	May 22	&	$1$	&	$\sim	0.8$	&	Clear\\
4	&	2005	Jul	7,8	&	$1$	&	$\sim	0.7-0.9$	&	Clear\\
5	&	2007	Jan	23,24,25		&	$1$	&	$\sim	0.6-1.2$	&	Clear\\
6	&	2007	Apr	19,20,21		&	$1$	&	$\sim	0.6-0.8$	&	Clear\\
7	&	2007	Aug	18,19	&		$1$	&	$\sim	1-1.7$	&	Clear\\
8	&	2008	Aug	2,3	&		$1$	&	$\sim	0.8$	&	Clear\\
9	&	2009	Jan	21,22	&	$1$	&	$\sim	1.1-1.5$	&	Clear\\
10	&	2009	Apr	2	&		$1$	&	$\sim	1.2$	&	Cirrus\\
11	&	2009	Apr	16	&		$1$	&	$\sim	0.8$	&	Clear
\enddata
\tablecomments{
Col. (1) : Observing run.
Col. (2) : Observing date.
Col. (3) : Slit width.
Col. (4) : Seeing.
Col. (5) : weather condition
}
\label{table:observation}
\end{deluxetable*}

\section{measurements}\label{section:meas}

We measured the line width of \Hb\ and \MgII\ and the luminosity of the AGN continuum at 5100\AA\ and 3000\AA\ as well as \Hb\ and \MgII\ lines, based on the multi-component spectral analysis (see Table~\ref{table:targets}). Here, we describe the fitting process for \Hb\ and \MgII, respectively.


\subsection{\Hb}

To measure the properties of the broad \Hb\ line, we performed a multi-component decomposition analysis, following the procedure described by \citet{Woo06} \citep[see also][]{Woo+10, Park12, Woo+13, Woo15}. All measurements were reported by \citet{Park15} and here we briefly summarize the fitting procedure for completeness. First, we modeled the AGN continuum with a power-law. The \ion{Fe}{2} models were constructed by broadening the I Zw I template from \citet{BG92} with a series of Gaussian velocities, while the stellar component was fitted with a simple stellar population synthesis model of \citet{BC03} with solar metallicity and age of 11~Gyr. The stellar model improves the \Hb\ line fitting since the \Hb\ absorption line attributed from stellar component is blended with a peak of the \Hb\ emission line originated from AGN. In particular, the FWHM of the line profile is sensitive to the shape of the peak while the line dispersion is not significantly affected. 

The fitting process was carried out using the non-linear Levenberg-Marquardt least-squares fitting routine \texttt{mpfit} \citep{markwardt09} in IDL, using two spectral fitting regions: 4430{\AA} - 4730{\AA} and 5100{\AA} - 5400{\AA}, where the \FeII\ blends are strong. The blue end of the fitting regions were slightly adjusted to avoid the H{$\gamma$}
and \Hb\ contamination if necessary.

For the broad component of \Hb, we used a sixth order Gauss-Hermite series model. If the broad component of \Hb\ is blended with the \HeII\ $\lambda$4686{\AA} line, we fitted the \HeII\ line, using two single Gaussian models, respectively for the broad and narrow components of \HeII, simultaneously with the \Hb\ model. We separetly fitted the narrow component of \Hb\ using the best-fit model of [\OIII] 5007{\AA}.  
Based on the best model of the broad \Hb\ component, we measured FWHM, line dispersion $(\sigma_{\rm H \beta})$, and line luminosity. We also measured the monochromatic luminosity at 5100\AA ($L_{5100}$) by averaging the continuum flux in the 50\AA\ window, using the power-law model representing the AGN continuum.
The measurement errors of the line width and continuum luminosity were determined based on the Monte Carlo simulations by generating 100 mock spectra by randomly fluctuating fluxes using the flux errors, and performing the decomposition analysis for each spectrum. We used the 1-$\sigma$ dispersion of the distribution as the measurement uncertainty.

\subsection{\ion{\rm Mg}{2}} \label{section:meas:MgII}

For measuring the width and luminosity of the \ion{Mg}{2} line, we followed the procedure outlined by \citet{McGill+08}. The multi-component fitting procedure is similar to that applied to the \Hb\ region, and here we briefly describe the fitting process. First, we fitted AGN power-law continuum and \ion{Fe}{2} blends, using the two windows: 2600{\AA} - 2750{\AA} and 2850{\AA} - 3090{\AA}. Second, we fitted the \ion{Mg}{2} line with a sixth order Gauss-Hermite series. The purpose of the line fitting is to measure the flux-weighted width of the line profle, for representing the velocity distribution of the gas. Thus, while we do not attempt to interpret the meaning of each coefficient in the Gauss-Hermite series, we determine the best model to reproduce the line profile. Since the Gauss-Hermite series can have negative values at the wing of the line profile, we empirically limit or adjust the fitting range,  in order to prevent negative fluxes in the best-fit model. 

We decided not to use a separate model to ft a narrow component of \ion{Mg}{2} since we do not see a clear signature that suggests the presence of the narrow component in the \MgII\ line profile (see \S~5.1 for more details on the narrow component subtraction). Note that the FWHM measurements of the broad \MgII\ can be underestimated if the existing narrow component in \MgII\ is not subtracted, while the effect on the line dispersion measurements will be insignificant. Several objects show strong absorption features in the \MgII\ line profile (e.g., W04, W08, W09, W14), for which we applied Gaussian models and simultaneously fitted them with the \MgII\ line profile. When the absorption features are close to the line center, increasing the uncertainty of the line width measurements, we checked whether the uncertainties of these objects introduces any systematic trend and found no significant effect (see Section 4.1). Third, the measurement errors of the \MgII\ line width and the continuum luminosity at 3000\AA\ were determined based on the Monte Carlo simulations. By randomly fluctuating fluxes using the flux errors, we generated 100 mock spectra and performed the decomposition analysis for each spectrum. Then, the 1-$\sigma$ dispersion of the distribution was taken as the measurement uncertainty.

As investigated by \cite{Wang+09}, a careful treatment is required for fitting \ion{Fe}{2} blends in the \MgII\ region. We used the \ion{Fe}{2} template from \citet{VW01} in our previous study \citep{McGill+08}. However, this template contains no information of \ion{Fe}{2} underneath the \ion{Mg}{2} line because it was constructed from the observed spectrum of the narrow line Seyfert 1 galaxy, I Zwicky 1 after masking the \MgII\ line. Instead, \citet{Tsuzuki06} suggested a new template calculated based on the one dimensional photoionization model combined with the observed spectrum of I Zw 1. The merit of this template is that the \FeII\ emission at the location of \MgII\ is available. Thus, we investigated the difference of the line fitting results using the two different templates.

\begin{figure}
\figurenum{2}
\center
	\includegraphics[width = 0.35\textwidth]{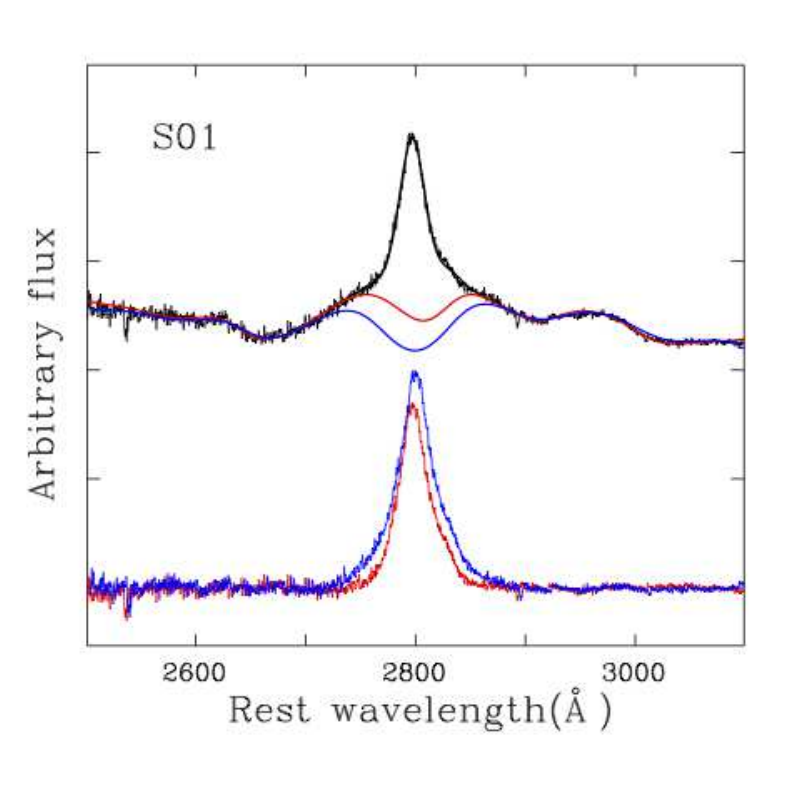}
	\caption{
	Top: The raw spectrum of S01 is shown ({\it black}) along with the best-fit \FeII\ models, respectively, using the \FeII\ templates from \citealt{VW01} ({\it blue}) and \citealt{Tsuzuki06} ({\it red}).
       Bottom: \FeII-subtracted \ion{Mg}{2} line profiles are presented, using the two different \FeII\ templates.	
       \label{fig:mgtemp}}	
\endcenter
\end{figure}

In Figure~\ref{fig:mgtemp} we present an example of the best-fit results based on the \ion{Fe}{2} templates of \citet{VW01} (blue) and \citet{Tsuzuki06} (red), respectively, for one AGN from our sample.  There is a clear difference between the two fitting results: the best-fit  \ion{Mg}{2} line profile is broader and stronger when the template of \citet{VW01} was used. This is due to the fact that there is no \FeII\ flux underneath of \MgII\ in \citet{VW01} template, hence, the \MgII\ line model takes more flux from the blended region (i.e., close to the wing of \MgII) into the \MgII\ line flux. 
Thus, in the following analysis, we will use the results based on the \FeII\ template of \citet{Tsuzuki06} for UV mass estimators.

Note that the template mismatch in subtracting the \ion{Fe}{2} blends may cause additional systematic errors on the line width measurements. \citet{Wang+09} reported that the revised \FeII\ template by \citet{Tsuzuki06} can provide reliable measurements within a 20\% uncertainty in the case of \MgII\ FWHM with the SDSS quality data. Thus, we expect that the uncertainty due to the template mismatch would be even smaller than 20\% for given high quality of our Keck data. 
Considering the possibility that the errors based on the Monte Carlo simulation underestimate the true uncertainty of the line width measurements, we assume an average error of 5\%, 10\%, and 20\%, respectively, in comparing the \MgII\ line width with that of \Hb. 
We find that regardless of the adopted errors, the best-fit slopes are consistent among each other, indicating that the fitting results are independent of the width measurement errors, unless the uncertainty is significantly larger than 20\%. Also, we investigate how the larger errors of the line width measurements affect the calibration of the best mass estimators. We obtain consistent results regardless of the adopted errors while the intrinsic scatter between mass estimates decreases with the increasing measurement errors as expected (see Section 4.3.2). 

\begin{figure*}
\figurenum{3}
 \centering
	\includegraphics[width = 0.98\textwidth]{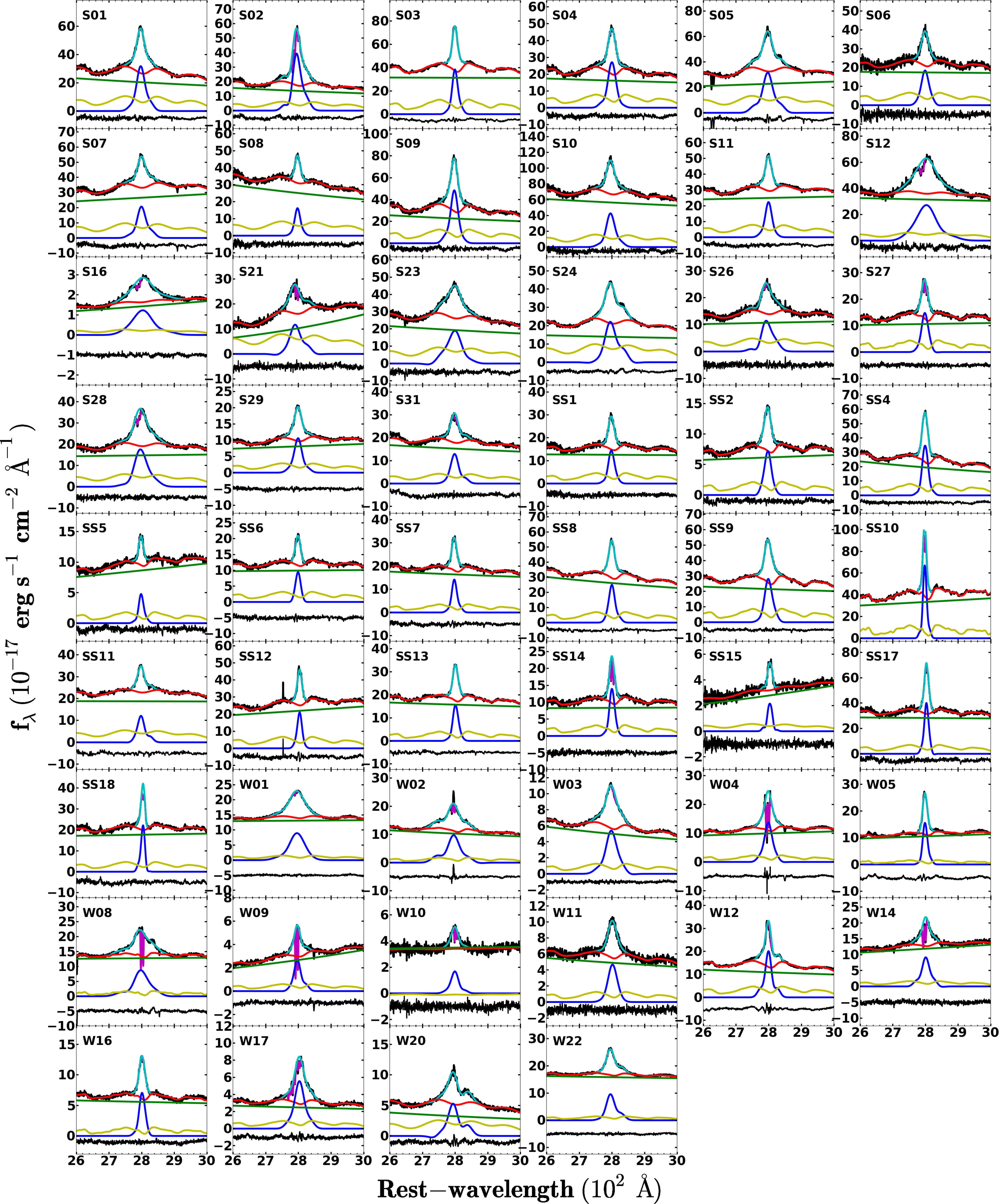}
	\caption{Multi-component fitting results for the \ion{Mg}{2} emission line region, using the \FeII\ template of \citet{Tsuzuki06}. In each panel, the rest-frame Keck spectrurm (thick black), power-law continuum + \ion{Fe}{2} model (red), total model including \ion{Mg}{2} (cyan), and models for the absorption line features in \MgII\ (magenta) are presented in the top, while the best-fit pseudo-continuum, i.e., AGN power-law continuum (green) and \ion{Fe}{2} model (yellow), and the \MgII\ line model (blue) are presented in the middle. In the bottom, the residual (black) between the observed spectrum and the combined models is shown, after shifting it down arbitrarily for clarity. 
	\label{fig:fit1}}
\end{figure*}

 The best-fit results for the \MgII\ line region based on the \FeII\ template by \citet{Tsuzuki06} are presented in Figure~\ref{fig:fit1}, while the multicomponent fitting results for the H$\beta$ region were presented by \citet{Park15}. There are several objects with relatively strong internal extinction,
namely, S16, S21, SS15, W09, W10, for which the spectral slope is very different compared to that of other AGNs in Figure 1. Thus, we will exclude these five AGNs from the \mbh\  estimator calibration since the luminosity and line width measurements are uncertain without a proper extinction correction. There are a couple of other AGNs with a hint of internal extinction in the \MgII\ region (e.g., SS5, SS12, W14), however, we decided to exclude only the five objects based on the spectral shape in the total UV to optical range (see more discussion in \S\ 4.2).

\subsection{Comparison of line profiles}

\begin{figure}
\figurenum{4}
 \centering
	\includegraphics[width = 0.40\textwidth]{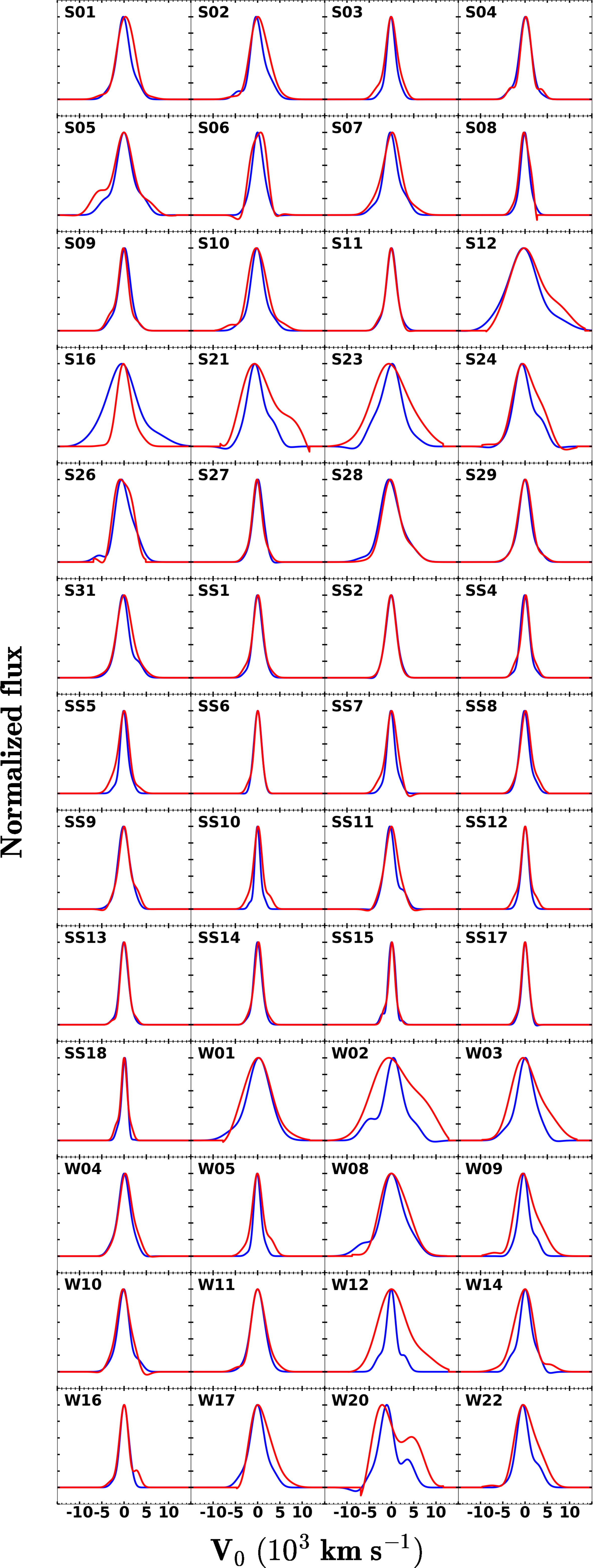} 
	\caption{Comparison of the best-fit model profiles of \MgII\ (blue) and \Hb\ (red) line profiles. The flux is normalized for comparison.}
	\label{fig:model1}
\end{figure}

We directly compare the best-fit model of the \Hb\ and \MgII\ line profiles after normalizing the flux with the peak value in Figure \ref{fig:model1}. While the majority of objects shows a similar line profiles between \Hb\ and \MgII, the \Hb\ width is somewhat broader than that of \MgII, as previously reported. For example, \cite{Marziani+13} showed that on average \Hb\ is 20\% broader than \MgII. However, there are cases with a much larger difference in their line profiles. In the case of S21, S24, S23, W02, W03, W09, W12, W20, the \Hb\ line profile is clearly different from that of \MgII\ and the line width of \Hb\ is much broader than that of \MgII\ by more than a factor of $\sim$two. Note that this discrepancy is mainly observed in AGNs with a very broad \Hb\ line \citep[see the discussion on population A in][]{Marziani+13}. In contrast, we found that one object, S16 shows a much broader line in \MgII\ than in \Hb\ by a factor of $\sim$1.8. It is not clear why the line profiles are very different between \Hb\ and \MgII. For the purpose of this study, we will compare the line width of \Hb\ and \MgII\ to provide UV mass estimators. However, it is clear that the \mbh\ based on the \MgII\ line will be systematically different for these objects. If we assume that the \Hb-based mass represents the true \mbh, then \MgII-based mass will suffer from systematic uncertainties due to the intrinsic difference of the line profiles between \Hb\ and \MgII. Thus, we will investigate the effect of these AGNs (six objects after excluding S16 and S21 due to heavy extinction) in our mass calibration (see \S~4.3).

To understand the characteristics of the line profiles, we compare line dispersion and FWHM of \MgII, using the measurements  based on the template of \cite{Tsuzuki06} (Figure 5 left). The average ratio between FWHM and line dispersion of \MgII\ is < log FWHM/$\sigma$> $= 0.27\pm0.05$, corresponding to 1.86 in linear scale, which is smaller than 2.36, the ratio of a Gaussian profile. 
The linear regression between FWHM and line dispersion ($\sigma$) of \ion{Mg}{2} shows a slope of 0.90 $\pm$ 0.04, indicating that FWHM and line dispersion shows almost linear relationship. In other words, the shape of the line profile of \MgII\ does not significantly change as a function of the line width although there is a slight hint that broader \MgII\ lines tend to have broader wings and narrower core than narrower \MgII\ lines.

In the case of \Hb, the FWHM-to-line dispersion ratio is $\sim$2 with a scatter larger than a factor of 2.
Also, the ratio increases with increasing line width, suggesting that there may be a systematic difference in the line profile between AGNs with a very broad line and AGNs with a relatively narrow line. In contrast, the FWHM/$\sigma$ ratio of \MgII\ is similar to that of \Hb, but with a much smaller scatter, indicating the \MgII\ line may not suffer a strong systematic trend as a function of the line width.

\begin{figure*}
\figurenum{5}
\center
	\includegraphics[width = 0.3\textwidth]{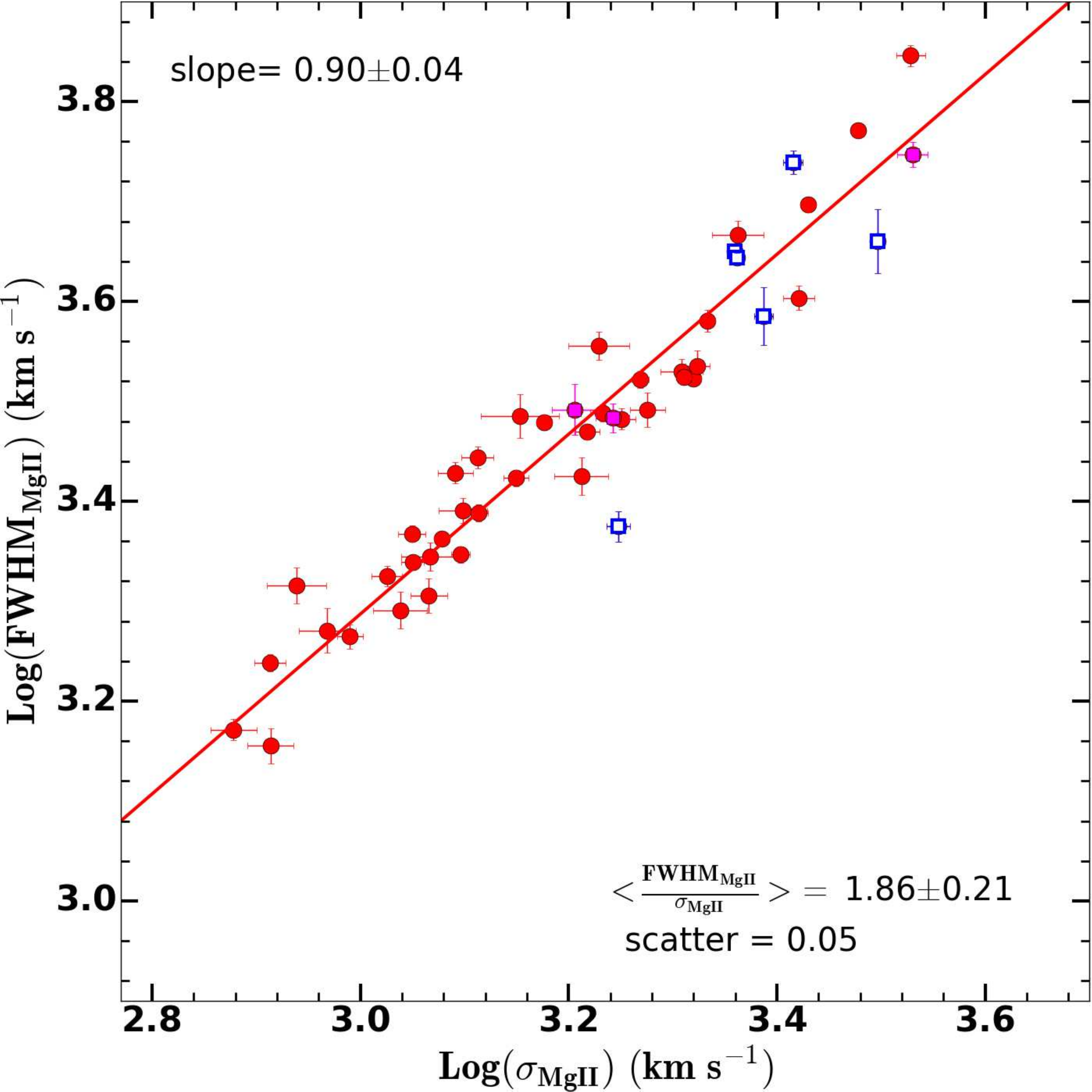}
		\includegraphics[width = 0.3\textwidth]{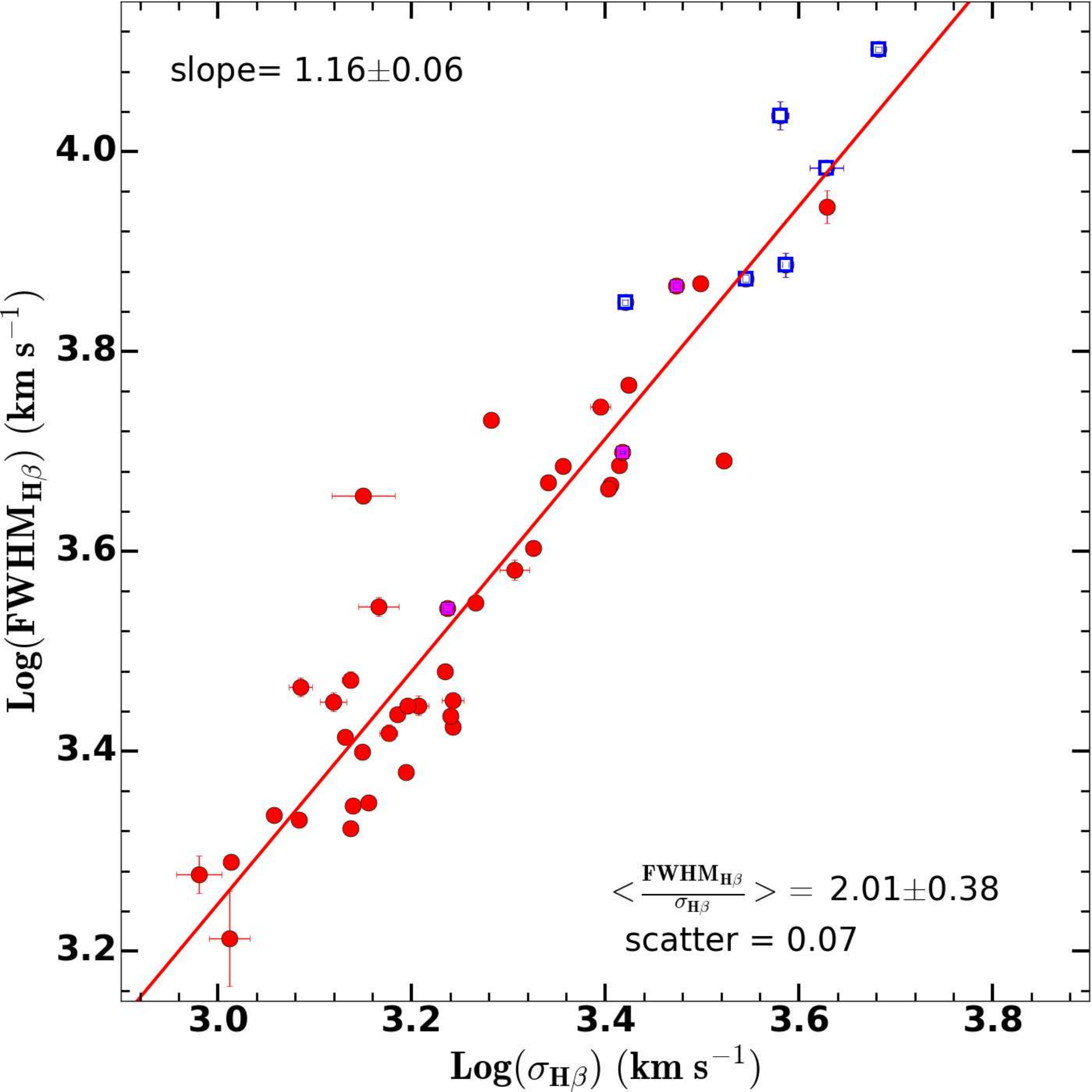}
			\includegraphics[width = 0.31\textwidth]{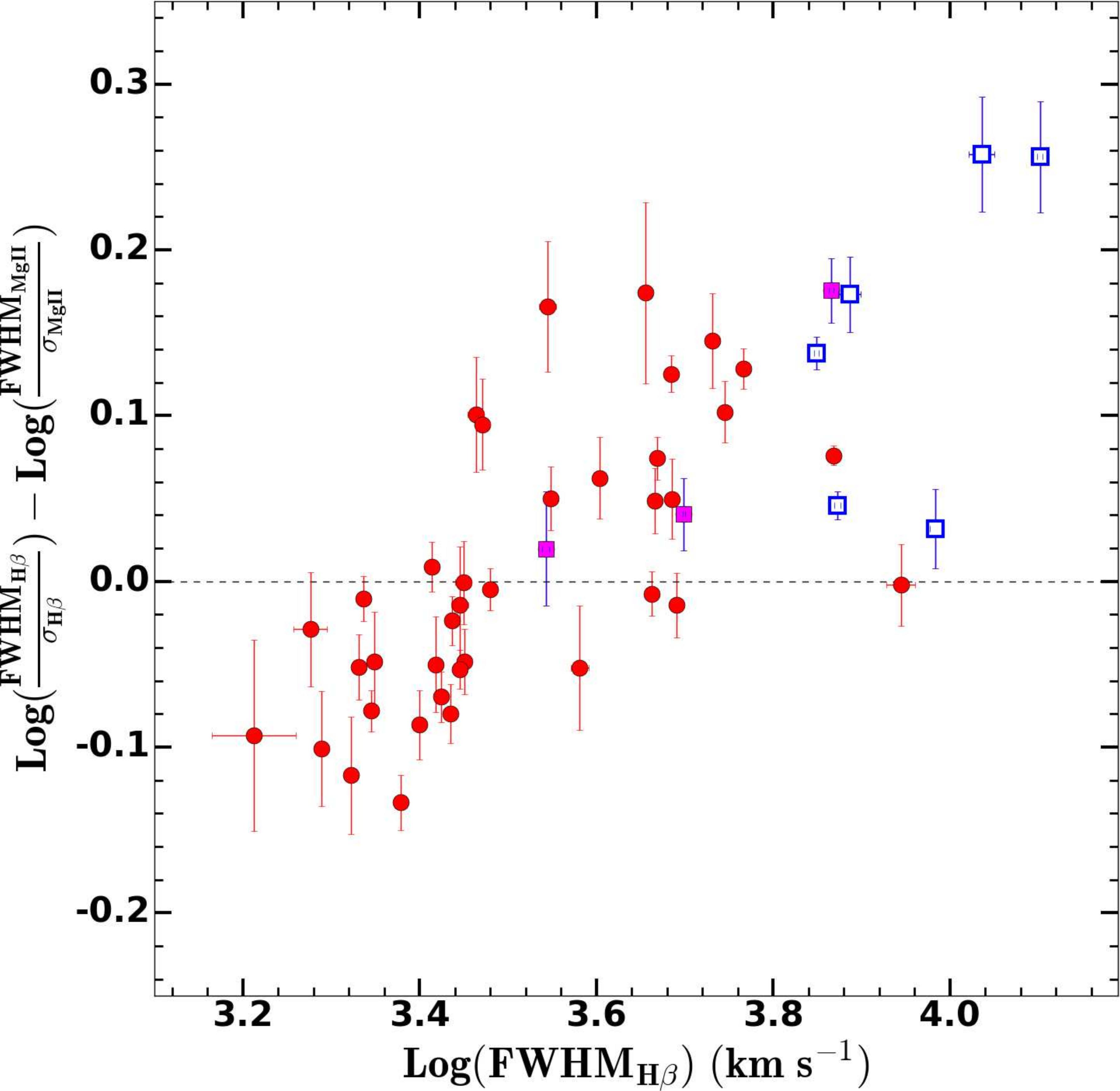}
	\caption{Comparing the FWHM and line dispersion ($\sigma$) of \ion{Mg}{2} (left) and \Hb\ (center).  The objects with strong absorption line features in the \MgII\ line are denoted as magenta squares, showing that these objects are not deviating from the distribution of other AGNs. The six AGNs with strong discrepancy of line profiles between \Hb\ and \MgII\ are shown as open blue squares, while the best fit is represented by red solid lines. Investigating the line profiles as a function of FWHM of the \ion{Mg}{2} and \Hb\ (right). 
	\label{fig:mgcom}}	
	\endcenter
\end{figure*}

\section{Calibration} \label{section:scaling}

In this section we perform correlation analysis, respectively for line widths (\S~4.1), luminosities (\S~4.2), and \mbh\ estimates (\S~4.3),
and present the best-fit results. For the regression, we use the FITEXY method as implemented by \citet{Park+12b} using MPFIT \citep{markwardt09}, which finds the best-fit parameters (intercept, slope and intrinsic scatter $\sigma_{\mathrm{int}}$) by minimizing a reduced $\chi^2$, after accounting for the measurement errors (see the detailed descriptions in \citealp{Park+12b}). From the Keck sample, we exclude five objects with strong internal extinction, namely, S16, SS15, S21, W09, and W10 as we described in Section 3.2, since the measured luminosities and line velocities are biased. 

\subsection{Line width comparison}\label{section:wcompare}

\begin{figure}
\figurenum{6}
   \includegraphics[width=0.23\textwidth]{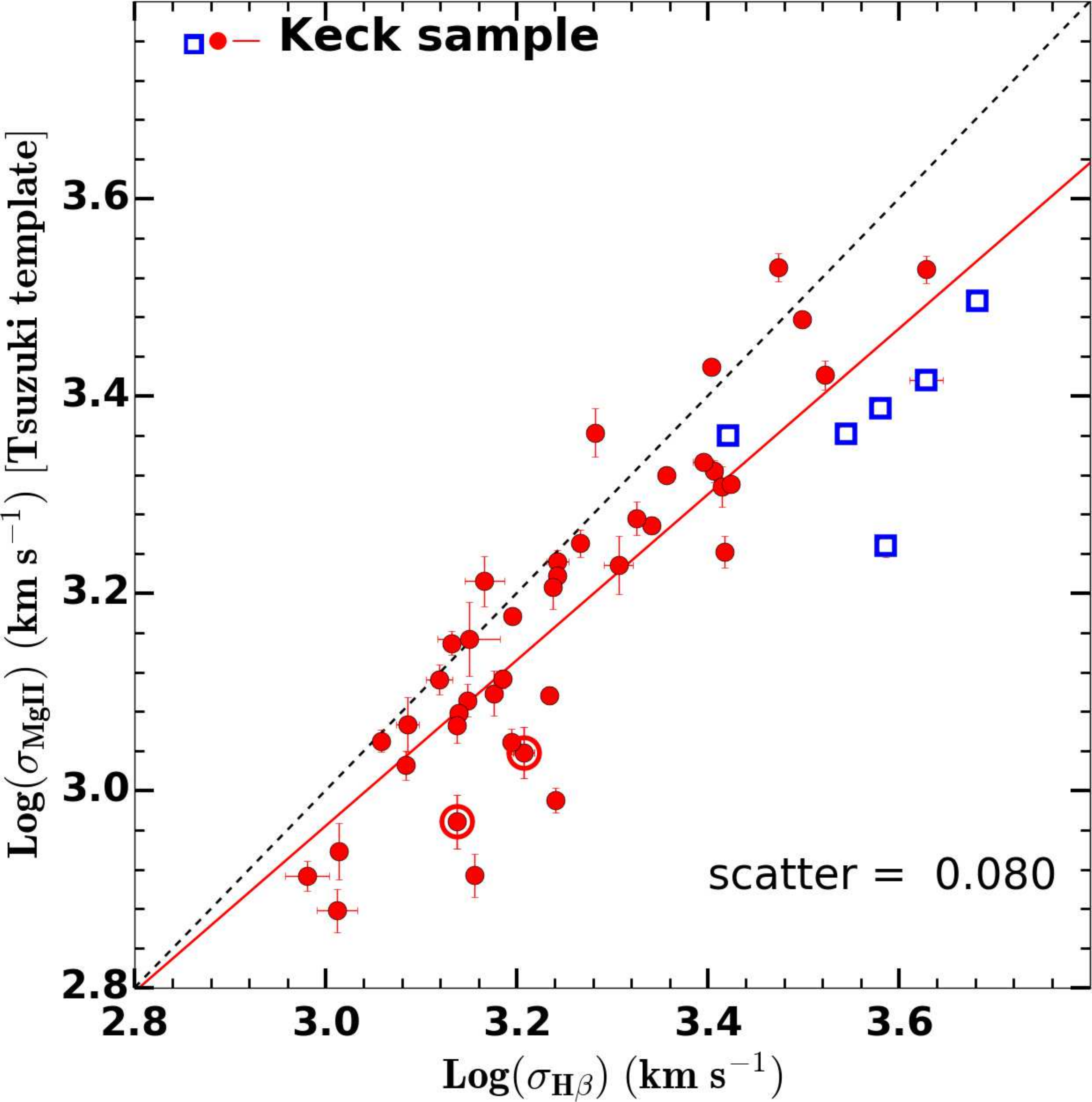} 
   \includegraphics[width=0.23\textwidth]{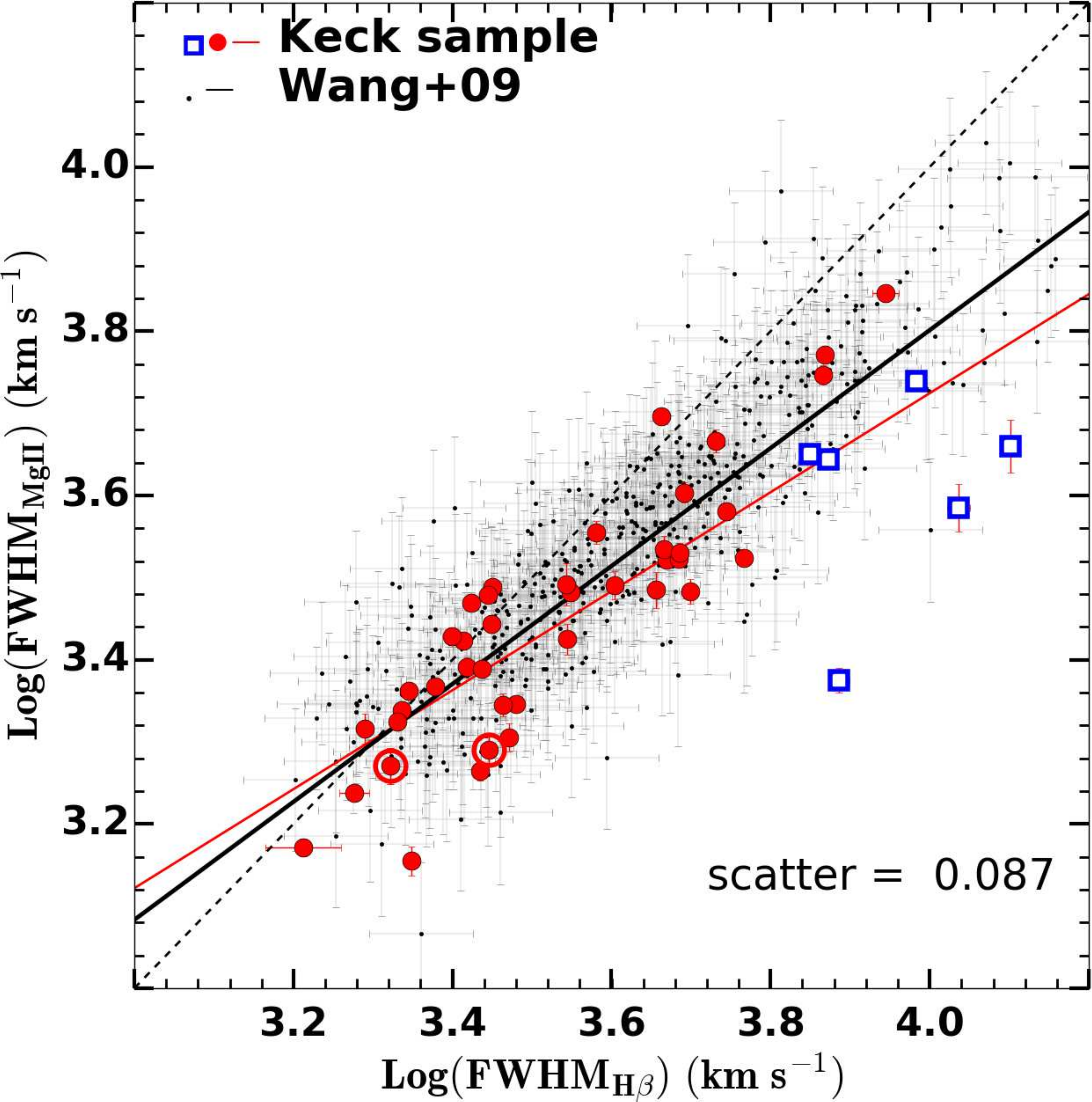}\\
   \includegraphics[width=0.23\textwidth]{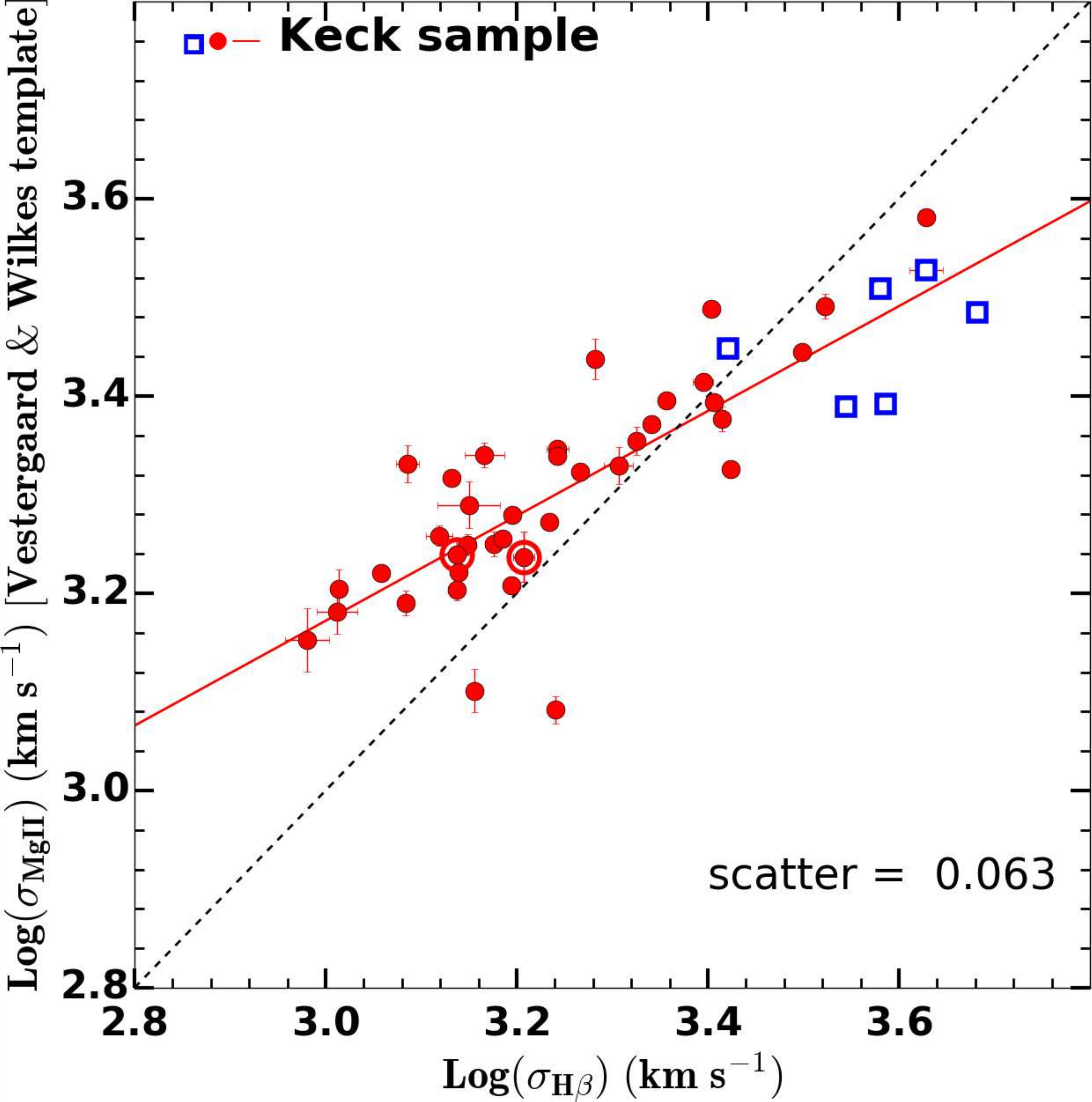} 
   \includegraphics[width=0.23\textwidth]{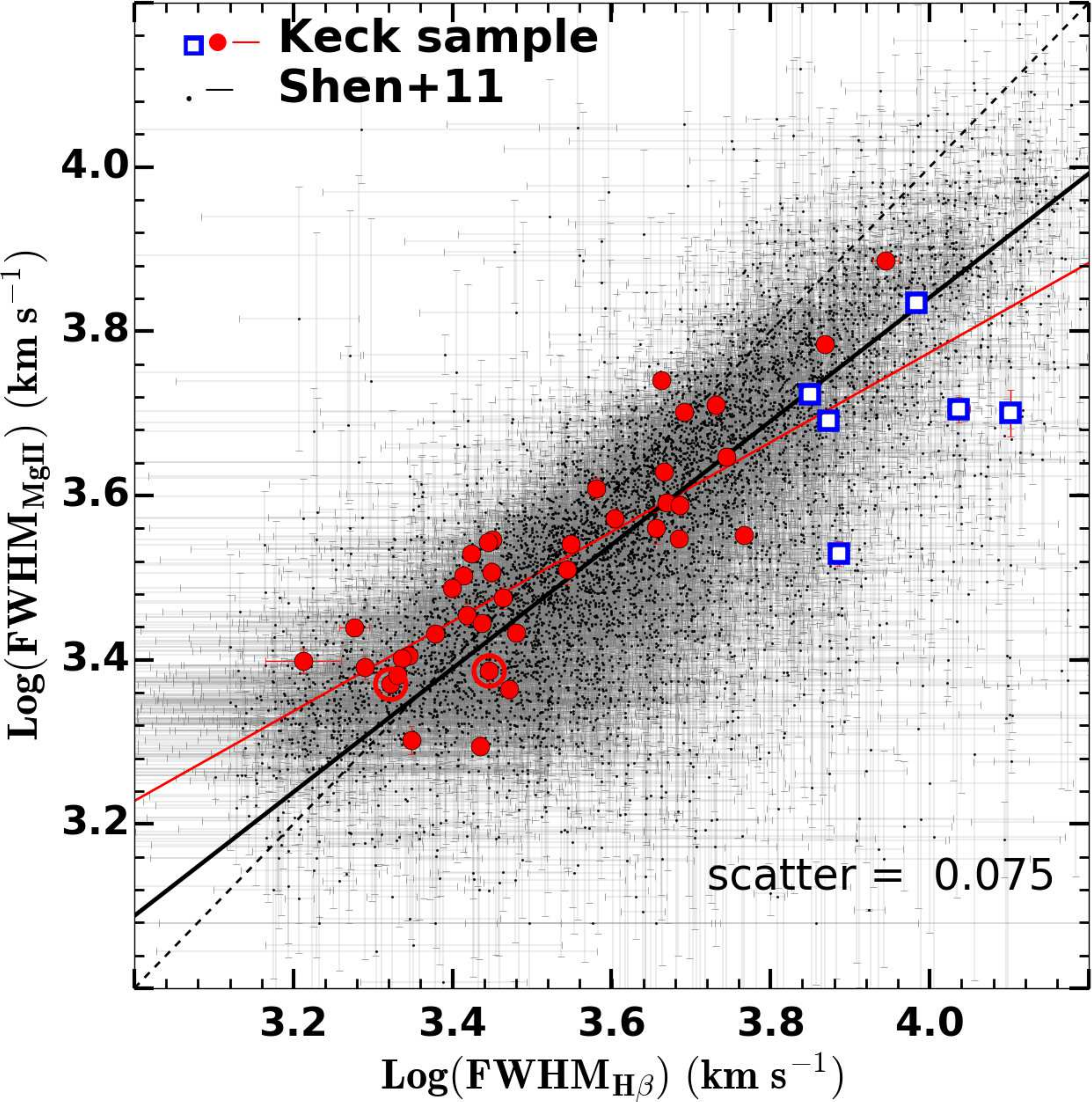}
   \centering
  \caption{Comparison of the line widths of \Hb\ and \MgII\ measured based on Tsuzuki \FeII\ template (top panels)
  and  Vestergaard \&\ Wilkes template (bottom panels). For FWHMs, Keck sample (red) is plotted along with 
  SDSS AGNs (grey), respectively from \cite{Wang+09} (top) and \cite{Shen+11} (bottom). Note that AGNs with heavy extinction, namely, S16, SS15, S21, W09, and W10 were excluded in our line regression. The solid red line represents the best-fit slope. The six AGNs with very different line profiles between \Hb\ and \MgII\ are shown with open blue squares. The AGNs with a hint of internal extinction in the \MgII\ region (SS5 and SS12) are marked with larger red open circles. The rms scatter of the best fit is shown as text in the plot. 
  \label{fig:tempcomp}}
\end{figure}

We compared the width measurements of  \MgII\ with those of \Hb\ in Figure~\ref{fig:tempcomp}.
As the line profile of \ion{Mg}{2} is often different from that of \Hb\ as shown in Figure~\ref{fig:model1}, we assumed an intrinsic scatter between the widths of the two lines in the fitting process. 
 While we measured the uncertainty of the line width measurements based on the Monte Carlo simulations as discussed in Section 3.2,
we also considered the systematic uncertainty and tested the fitting results assuming an average error of 5$\%$, 10$\%$, and 20$\%$, respectively, as the uncertainty of the width measurements of the \MgII\ line
We found that regardless of the adopted errors, the best-fit slopes are consistent among each other, indicating that the fitting results are independent of the width measurement errors, unless the uncertainty is significantly large ($>>$ 20\%). 
For the final fitting results, we used the errors measured from the Monte Carlo simulations.

First, we compared \Hb\ line width and \MgII\ line width, that was measured from the spectral decomposition based on the \FeII\ template of \cite{Tsuzuki06} (top panels in Figure 6). We obtained the best-fit result as
\begin{equation}
	\log (\sigma_{\rm Mg \sevenrm II}) \propto 0.84 \pm0.07 \times \log (\sigma_{\rm H\beta}),
\end{equation}
with $\sigma_{\mathrm{int}}$ = 0.08 $\pm$ 0.01, indicating a sub-linear relationship that \MgII\ is somewhat narrower than \Hb, particularly for AGNs with a broader line. 
In the case of FWHM, we obtained a shallower slope,
\begin{equation}
	\log (\rm FWHM_{\rm Mg \sevenrm II})  \propto 0.60 \pm0.07 \times \log (\rm FWHM_{H \beta}),
\end{equation}
with $\sigma_{\mathrm{int}}$ = 0.09 $\pm$ 0.01. 
To investigate this correlation further, we adopted a sample of 495 SDSS AGNs with $S/N \ge 20$ for \MgII\ and \Hb\
from \citet{Wang+09}, who used the same \FeII\ template of \cite{Tsuzuki06} in modeling the \MgII\ line profile, and measured FWHM of \MgII\ and \Hb. For the combined sample of the Keck and SDSS AGNs, we obtained the best-fit slope of 0.72 $\pm$ 0.03 ($\sigma_{\mathrm{int}}$ = 0.04 $\pm$ 0.01), which is higher than that we obtained using the Keck sample only. Since \citet{Wang+09} modeled the \ion{Mg}{2} line profile after subtracting a narrow component of \MgII, it is possible that some of the FWHM measurements are systematically overestimated. 

To investigate the systematic effect due to the choice of the \FeII\ template, we also used \MgII\ line width measurements based on 
the \FeII\ template of \citet{VW01} (bottom panels in Figure 6). As described in Section~\ref{section:meas:MgII}, the subtraction using the template from \citet{VW01} introduces systematic uncertainties due to the lack of \FeII\ features at the location of the \MgII\ line. 
We found a more significant systematic difference of the line dispersions between \MgII\ and \Hb\ with the best-fit slope of 0.53 $\pm$ 0.05 ($\sigma_{\mathrm{int}}$ = 0.06 $\pm$ 0.01).
These results support the hypothesis that the \FeII\ template of \citet{VW01} introduces additional systematic uncertainties on the line width measurements, particularly for the line dispersion. 
In the case of FWHM between \MgII\ and \Hb, we obtained the best-fit slope 0.55 $\pm$ 0.06 ($\sigma_{\mathrm{int}}$ = 0.07 $\pm$ 0.01), which is similar to the slope 0.60$\pm$ 0.07, that we obtained using the Keck sample based on the \citet{Tsuzuki06} template, indicating that the choice of \FeII\ template makes less significant difference in comparing the FWHM of \MgII\ and \Hb. 
We also adopted the FWHM measurements from \citet{Shen+11}, who used the \FeII\ template of \citet{VW01} for the \MgII\ line fitting process for a large sample of type 1 AGNs, by selecting 6017 AGNs at $0.4 \le z \le 0.8$ with $S/N \ge 10$ in the continuum (4750$-$4950 \AA).
Using the combined sample of the Keck and SDSS AGNs, we obtained the best-fit slope 0.75 $\pm$ 0.01 ($\sigma_{\mathrm{int}}$ = 0.08 $\pm$ 0.01), which is again close to the slope that we obtained using the measurements of \MgII\ FWHM from \citet{Wang+09} based on the template of \citet{Tsuzuki06}. 
These results suggest that the choice of \FeII\ temple does not strongly affect the FWHM comparison, while it strongly changes the correlation between the line dispersions of \MgII\ and \Hb\ as expected from the fact that the two templates make a significant difference in the wing of the \MgII\ line profile. 

\begin{figure*}
\figurenum{7}
    \centering
    \includegraphics[width=0.43\textwidth]{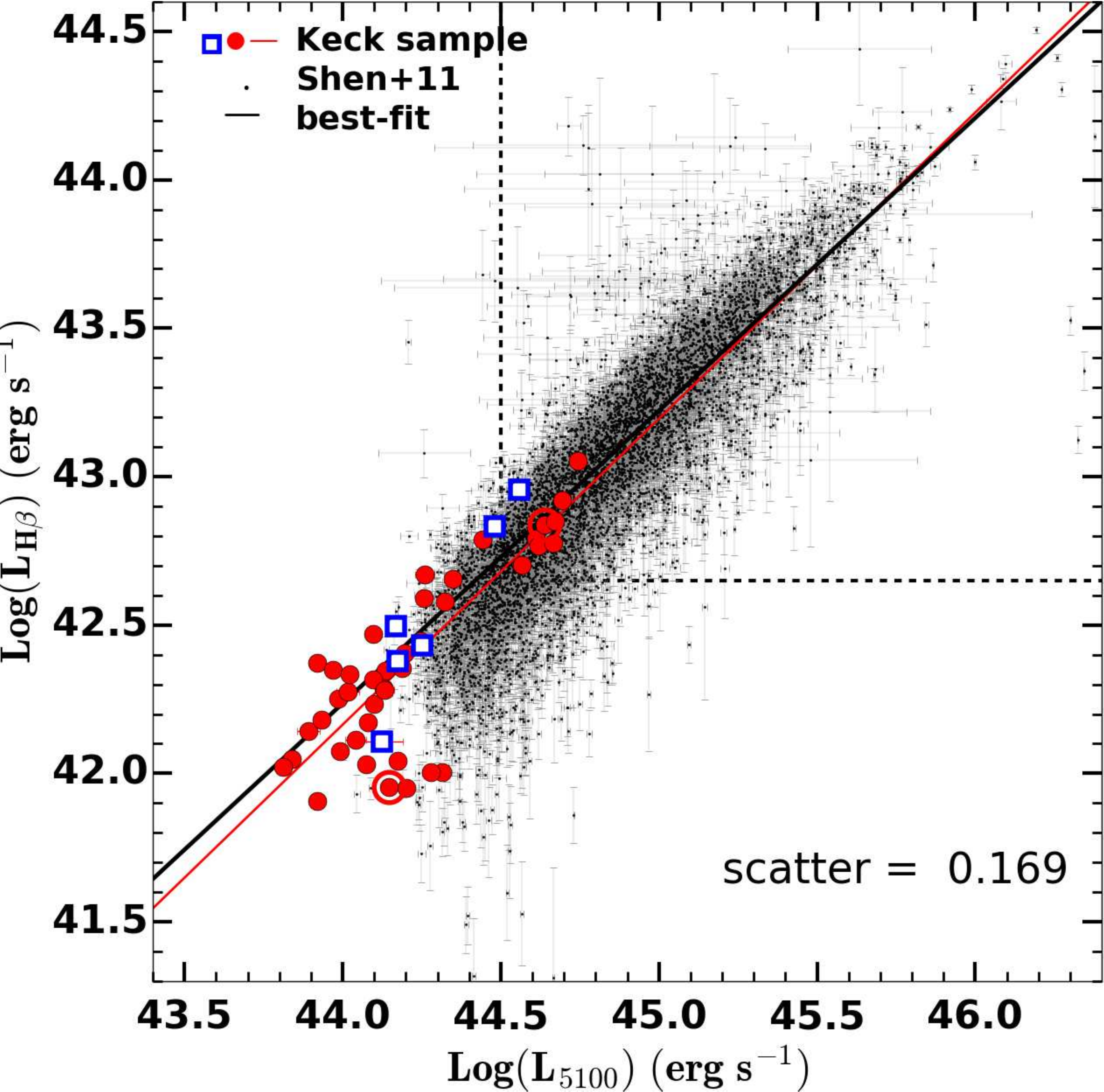} 
    \includegraphics[width=0.43\textwidth]{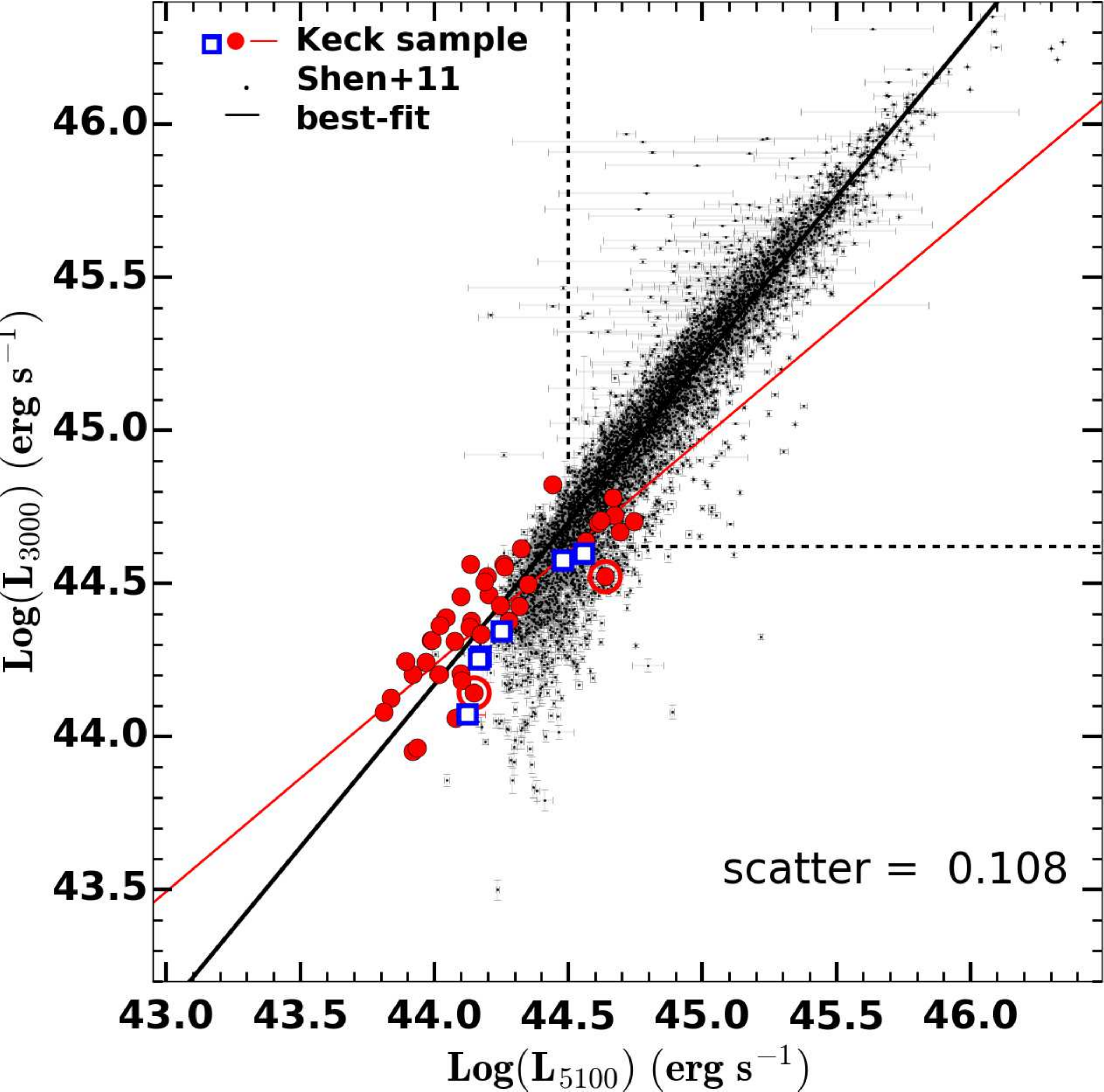}\\
    \includegraphics[width=0.43\textwidth]{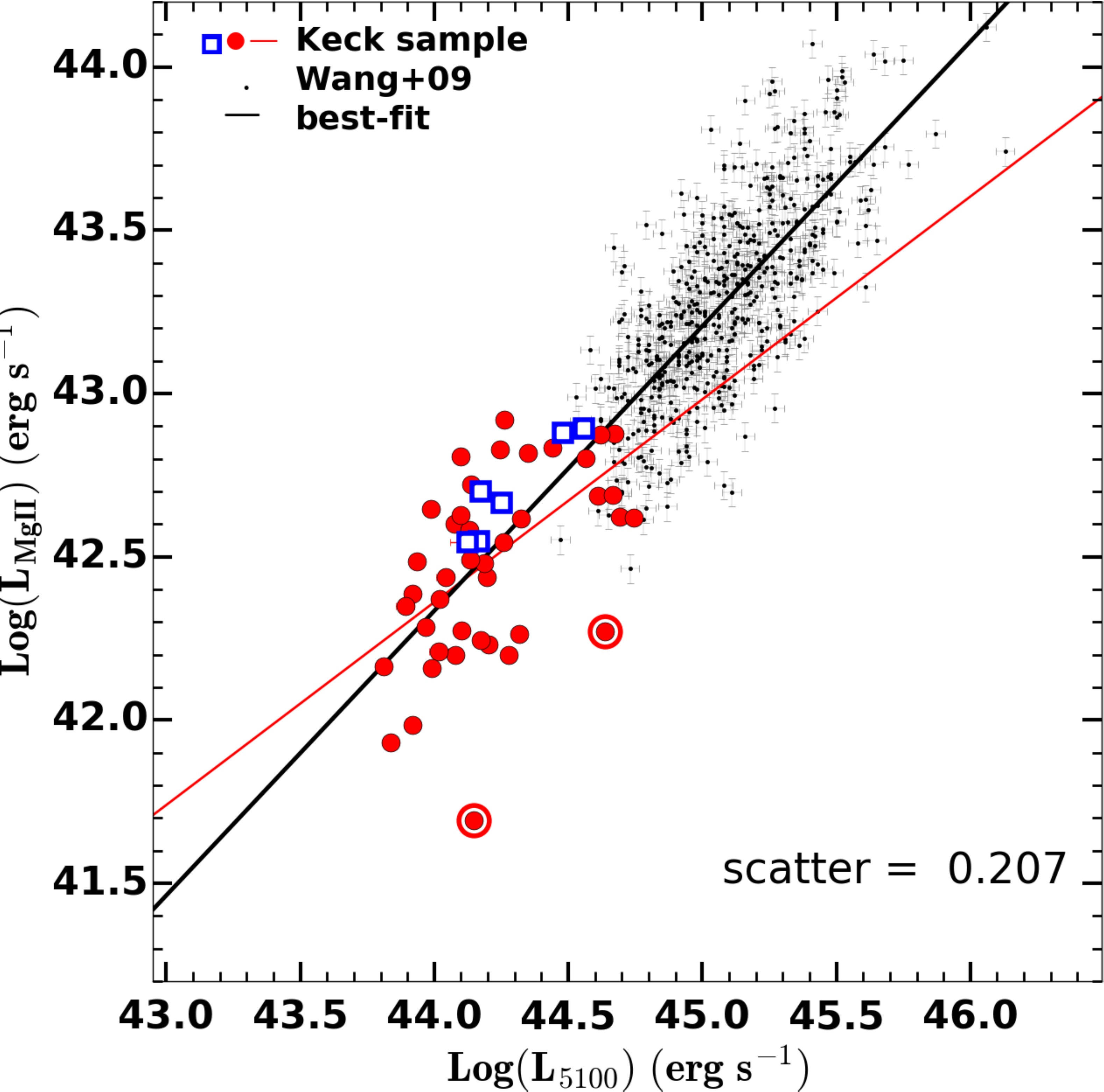} 
    \includegraphics[width=0.43\textwidth]{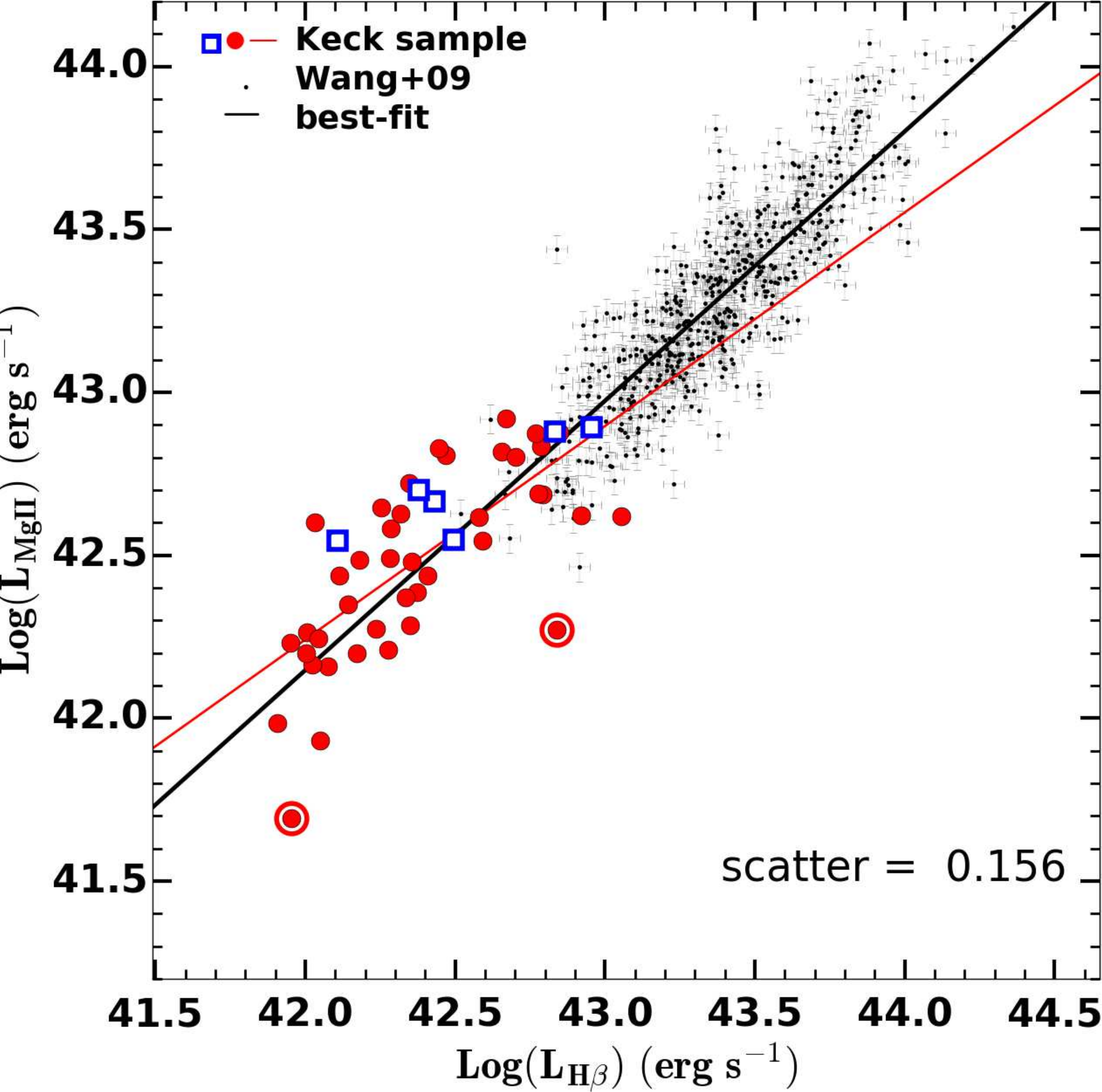}
    \caption{Comparison of various luminosities. Top panels show L$_{5100}$ vs. L$_{H\beta}$ (left) and L$_{5100}$ vs. L$_{3000}$ (right) with the Keck (red circles) and SDSS AGNs from \cite{Shen+11} (grey points). Bottom panels present L$_{5100}$ vs. L$_{MgII}$ (left) and L$_{H\beta}$ vs. L$_{MgII}$ (right) with the Keck AGNs (red) and SDSS AGNs from \cite{Wang+09} (gray cross). Red and black solid lines represent the best-fit for the Keck sample, and the combined sample of Keck and SDSS AGNs, respectively. The six AGNs with very different line profiles between \MgII\ and \Hb\ (blue squares) do not significantly affect the best-fit results. Two AGNs, SS5 and SS12 with a hint of internal extinction in the \MgII\ region are marked with larger red open circles. 
} 
    \label{fig:lcompare}     
\end{figure*}

\subsection{Luminosity comparison}\label{section:lcompare}

We compare various continuum and line luminosities with the best calibrated continuum luminosity at 5100\AA\ ($L_{5100}$), in order to use the UV luminosities as a proxy for BLR size.  Note that including or excluding the six AGNs with very different line profiles between \MgII\ and \Hb\ does not significantly change the result. Thus, we only present the best-fit result including these AGNs. 
First, we compare $\mathrm{L_{5100}}$ with $\mathrm{L_{H\beta}}$ for the Keck sample, obtaining the best-fit slope 1.03 $\pm$ 0.08 with $\sigma_{\mathrm{int}}$ = 0.18 $\pm$ 0.02. This result is consistent with but slightly shallower than the slope 1.13 $\pm$ 0.01 reported by \citet{Greene05}. 
In order to increase the dynamic range, we adopted the measurements of 6017 SDSS AGNs  from \citet{Shen+11}, which were used for the line width comparison in \S~4.1. For this comparison, we made an arbitrary luminosity cut at $\log(\mathrm{L_{H\beta}}) = 42.7$ and $\log(\mathrm{L_{5100}}) = 44.5$ for further selecting 4584 luminous  AGNs, in order to avoid uncertain $\mathrm{L_{5100}}$ measurements of lower luminosity AGNs since the potential contribution from stellar continuum can be significant. In fact, a systematic offset at low luminosity is clearly present in Figure~\ref{fig:lcompare} (grey points). By combining our Keck sample and the higher luminosity SDSS AGNs, we obtained the best-fit result, 
\begin{eqnarray}
	 \log(L_{H\beta}) \propto {0.99 \pm0.01} \times \log(L_{5100}),
\end{eqnarray}
with $\sigma_{\mathrm{int}}$ = 0.15 $\pm$ 0.01, again consistent with the linear relationship. 

Second, by comparing $\mathrm{L_{3000}}$ with $\mathrm{L_{5100}}$, we obtained the best-fit slope 0.74 $\pm$ 0.06 ($\sigma_{\mathrm{int}}$ = 0.12 $\pm$ 0.01) for the Keck sample. To increase the dynamic range, we also adopted 4800 luminous AGNs from \citet{Shen+11}, using the luminosity cut at $\log(\mathrm{L_{3000}}) = 42.62$ and $\log(\mathrm{L_{5100}}) = 45.5$, to avoid the systematic uncertainty due to stellar contribution to $\mathrm{L_{5100}}$. In this case we obtained the best-fit result, 
\begin{equation}
	\log(L_{3000}) \propto {1.06 \pm0.01} \times \log (L_{5100}),
\end{equation} 
with $\sigma_{\mathrm{int}}$ = 0.10 $\pm$ 0.01.

Third, by comparing the $\rm L_{Mg \sevenrm II}$ with $\mathrm{L_{5100}}$, we obtained the best-fit slope 0.62 $\pm$ 0.13 ($\sigma_{\mathrm{int}}$ = 0.24$\pm$ 0.03) based on the Keck sample (bottom left in Figure~\ref{fig:lcompare}). If we combined the Keck sample with the  SDSS AGNs from \citet{Wang+09}, who used the same \FeII\ template for the \MgII\ line fitting,  we obtained the best-fit result,
\begin{equation}
	\log(\rm L_{\rm Mg \sevenrm II})\propto {0.87 \pm0.03}\times \log (L_{5100}),
\end{equation}
with $\sigma_{\mathrm{int}}$ = 0.20 $\pm$ 0.01.
The sub-linear correlation between \MgII\ and UV continuum luminosities represents the Baldwin effect \citep{Baldwin77}, which can be explained due to the increase of the thermal component in the UV continuum, which is represented by the Big Blue Bump (BBB), for more luminous AGNs \citep{Malkan&Sargent82, Zheng&Malkan93}. Thus, for given photoionizing flux and the emission line luminosity, the continuum luminosity close to BBB will be higher for more luminous AGNs.  The sub-linear relation between \MgII\ and UV continuum luminosity is also reported to be related with physical parameters, i.e., Eddington ratio \citep[see, for example,][]{Dong+09}. The linear relation between \Hb\ and optical continuum luminosity at 5100\AA\ indicates that the effect of the increasing thermal component is relatively weak at 5100\AA, which is well off the BBB.

Last, we also compare $\rm L_{Mg \sevenrm II}$ with $\mathrm{L_{H\beta}}$ and obtained the best-fit slope 0.66 $\pm$ 0.11 ($\sigma_{\mathrm{int}}$ = 0.20 $\pm$ 0.02) for the Keck sample. For the combined sample of Keck an SDSS AGNs from \citet{Wang+09}, we obtained the best-fit slope of 0.83 $\pm$ 0.02 ($\sigma_{\mathrm{int}}$ = 0.15 $\pm$ 0.01), which is close to the slope of $\rm L_{Mg \sevenrm II}$ vs. $\mathrm{L_{5100}}$. 

In summary, we obtained an almost linear relation between $\mathrm{L_{5100}}$ and $\mathrm{L_{3000}}$ while the relation is sub-linear between
$\mathrm{L_{5100}}$ and $\rm L_{Mg \sevenrm II}$. Also, we find that the slope between UV and optical luminosities varies depending on the sample. 
For \mbh\ calibrations, we will use the correlation results expressed with Equation 4, 5, 6 in the next section. However, since the correlation slope depends on the sample and the dynamic range, we will also calibrate UV mass estimators without using these correlations.

\begin{figure}
\figurenum{8}
	\includegraphics[width=0.48\textwidth]{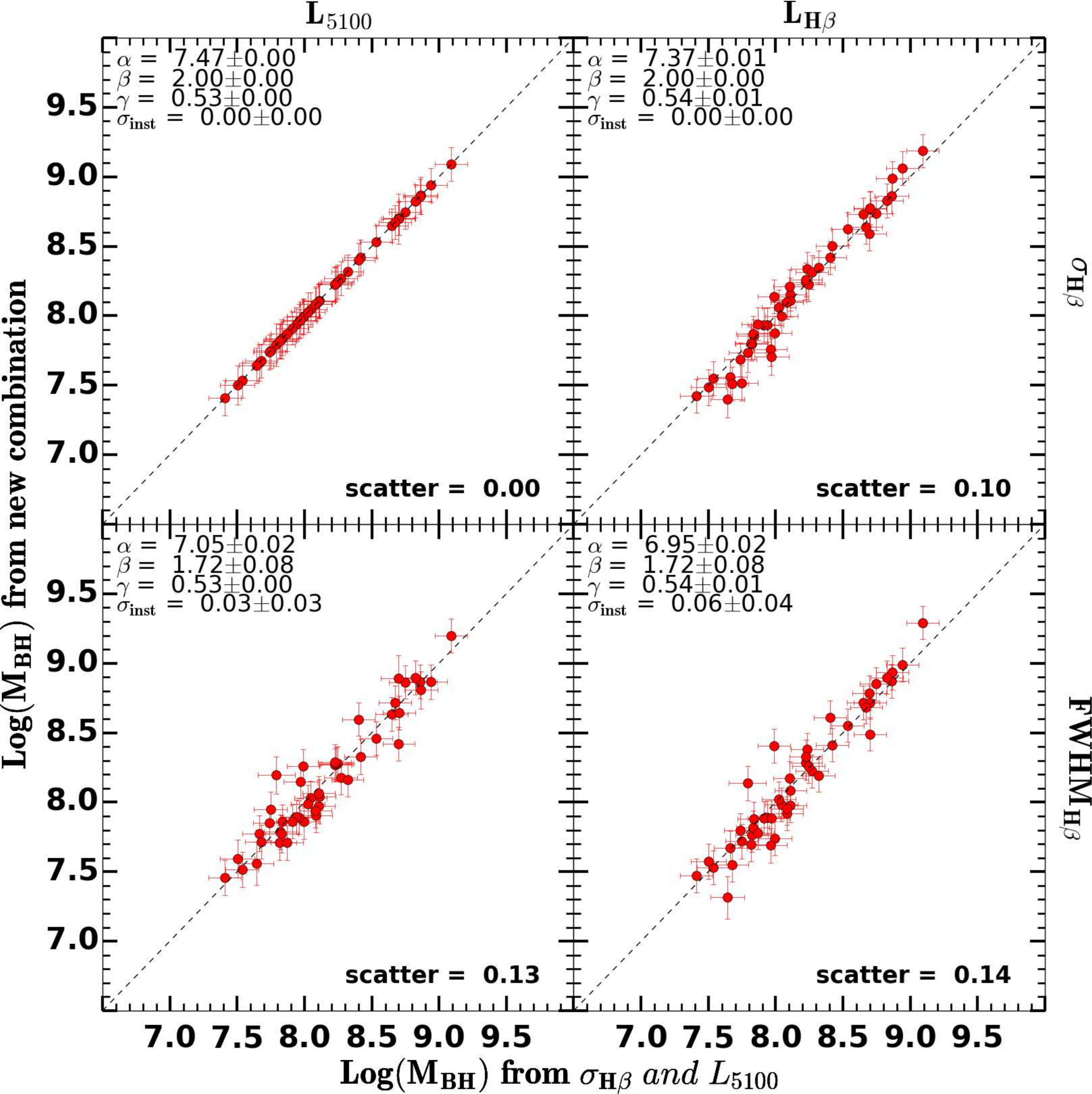}
	\centering
	\caption{Cross-calibration fitting between newly derived \mbh\ and fiducial mass with estimators from \Hb\ line.
	X axis data points represent our target's fiducial mass, while y axis data points are \mbh\ from
	$\alpha + \beta \log V_{1000} + \gamma \log L$. $V_{1000}$ means velocity estimator using $1000~ km~ s^{-1}$ unit,
	L is luminosity estimator having $10^{44}~erg~s^{-1}$ unit for continuum or $10^{42}~erg~s^{-1}$ unit for emission line.
	$\beta$ and $\gamma$ in each panel depend on a kind of estimators which are shown in upper left part of figure, and $\alpha$ is estimated by $\chi^{2}$ minimization fitting. 
	\label{fig:mbh}}
\end{figure}

\subsection{Calibrating \mbh \ estimators} \label{section:mbh}

To calibrate \mbh\ estimators we determine the coefficients in Eq. 1 for each pair of velocity and luminosity measurements based on \Hb, \MgII, UV and optical continuum, by comparing the UV-based mass with the reference \mbh. 
As the reference mass, we adopt the \mbh\ estimated based on \Hb\ line dispersion and $\mathrm{L_{5100}}$, by combining the virial theorem (i.e., $\beta$=2) and the \Hb\ size-luminosity relation from \cite{Bentz+13} (i.e., $\gamma$=0.533). For $\alpha$, we adopt the virial factor f=4.47 from \cite{Woo15} based on the calibration of AGN \mbh-stellar velocity dispersion relation, which corresponds to $\alpha$=7.47 \citep[see Appendix in][]{Woo15}.

\subsubsection{\Hb-based mass estimators}

In Figure 8, we first calibrate optical mass estimators based on \Hb. 
For \Hb\ line dispersion, we fixed $\beta$ as 2 (top panels), while for \Hb\ FWHM we used $\beta$=2/1.16=1.72 since log FWHM$_{H\beta}$ $\propto$ 1.16 log $\sigma_{H\beta}$ (see Figure 5). Also, when we used L$_{H\beta}$, we adopted the correlation result between L$_{H\beta}$ and L$_{5100}$ from Eq. 4, which corresponds to $\gamma$=0.533/0.99=0.54. 
The rms scatter between two mass estimates is $\sim$0.1-0.14 dex, indicating that the choice of velocity measure (either FWHM or line dispersion) or the choice of luminosity (i.e., either continuum luminosity at 5100\AA\ or line luminosity of \Hb) adds small additional systematic uncertainties. However, this assessment only applies when the data quality, hence the measurement uncertainty is comparable to those of our Keck sample. 
In comparing our result based on the enlarged sample over a large dynamic range with our previous result based on the limited subsample (see Table 3 in \citealp{McGill+08}), we find that our new calibration is more reliable since the scatter is significantly reduced by $0.1 - 0.19$ dex.

\begin{figure*}
\figurenum{9}
	\includegraphics[width=0.335\textwidth]{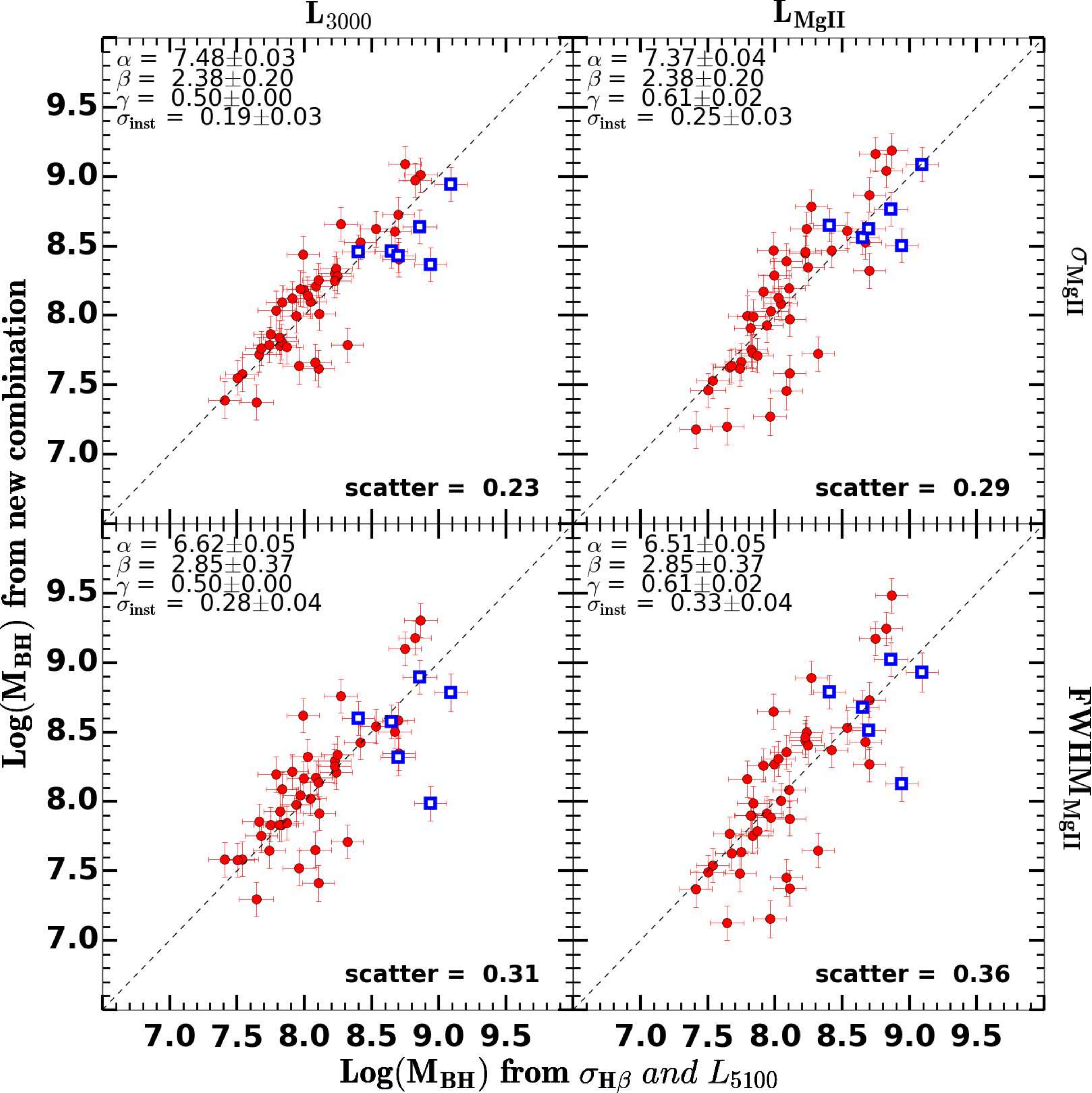}
	\includegraphics[width=0.32\textwidth]{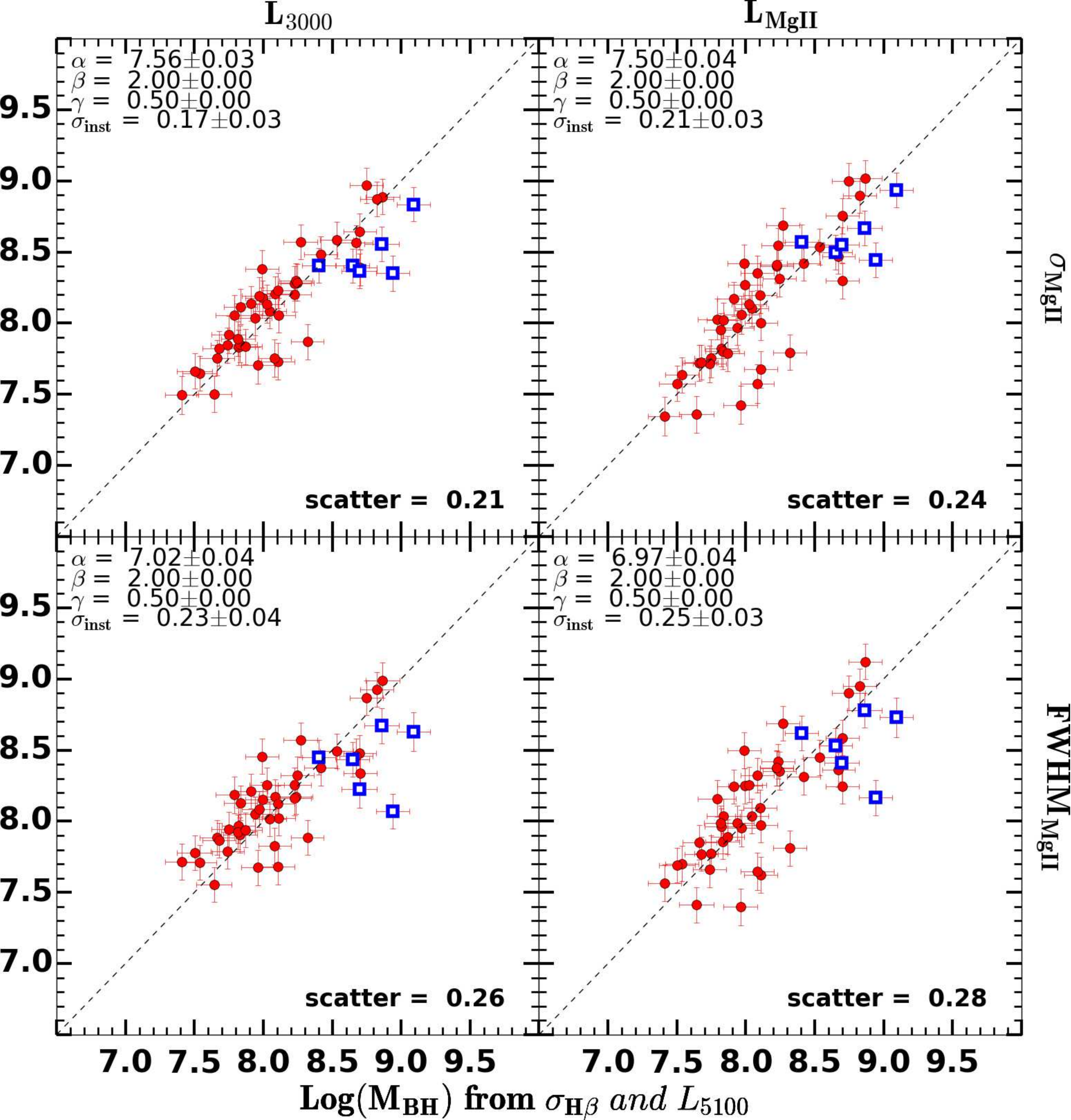}	
		\includegraphics[width=0.32\textwidth]{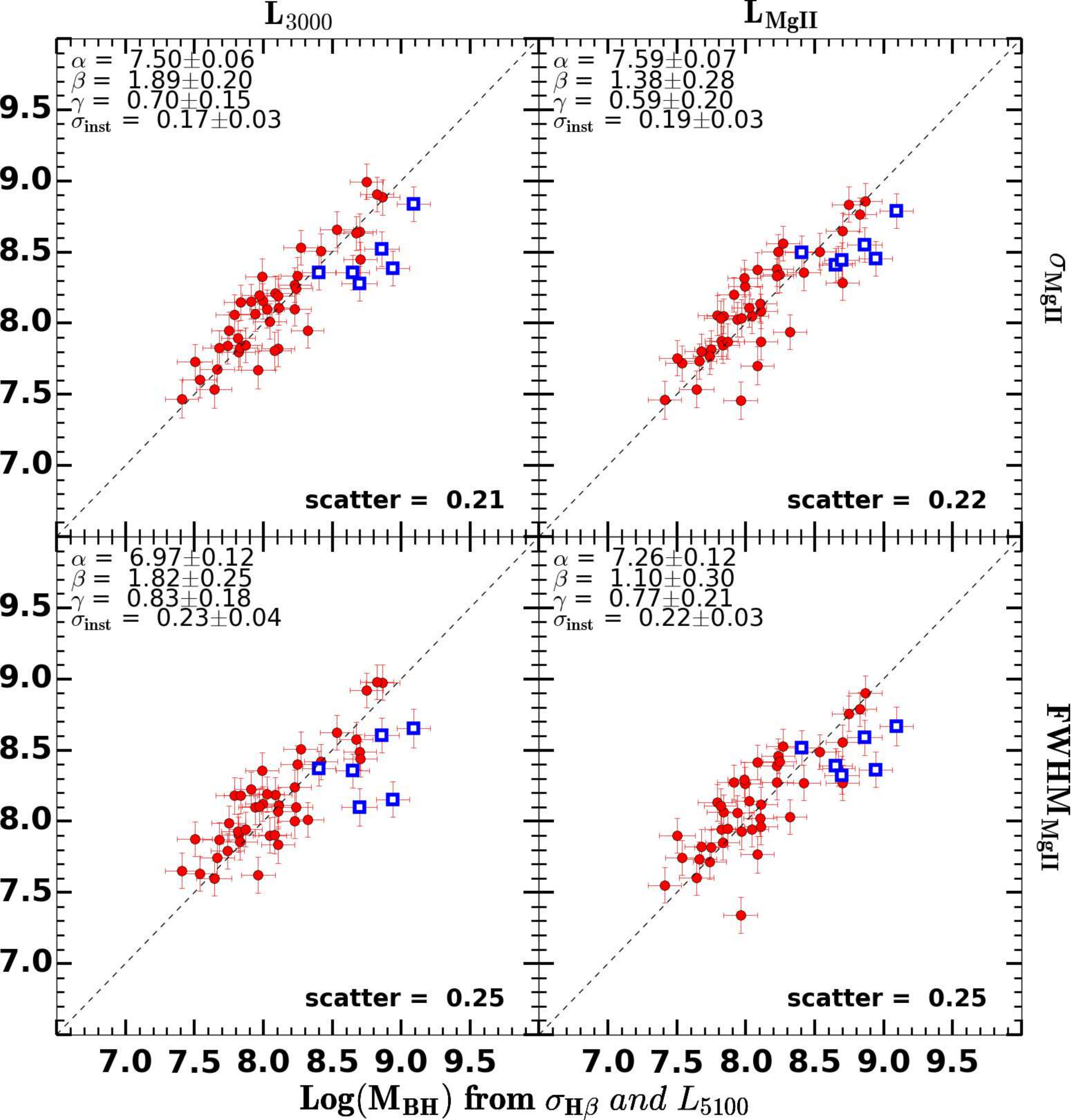}	
	\centering
	\caption{Similar as Figure 8 but with estimators from \ion{Mg}{2} line. Left panels: $\beta$ and $\gamma$ are obtained from the results of line width and luminosity comparison in Section 4.1 and 4.2.
	Middle panels: We fixed $\beta$=2 and $\gamma$=0.5, following a virial relation and the expected size-luminosity relation. Right panels: $\beta$ and $\gamma$ are from the best-fit results. Open blue 		squares show the six AGNs with very different line profiles between \Hb\ and \MgII.
	\label{fig:mbh}}
\end{figure*}

\subsubsection{\MgII-based mass estimators}

We calibrate UV mass estimators by comparing \mbh\ estimated based on \ion{Mg}{2} line with the 
reference \mbh\ based on \Hb. First, we adopt the $\beta$ value from the direct comparison of \Hb\ line dispersion with \MgII\ line dispersion as well as \MgII\ FWHM, respectively.  In other words, for \MgII\ line dispersion, $\beta$ = 2/0.84=2.38 since log $\sigma_{\rm MgII}$ $\propto$ 0.84 log $\sigma_{H\beta}$, while $\beta$ = 2/0.70 = 2.85 for \MgII\ FWHM since log FWHM$_{\rm MgII}$ $\propto$ 0.70 log $\sigma_{H\beta}$ (see \S~4.1).
For luminosity, we also use the results from Section 4.2. Since $\mathrm{L_{3000}}$ $\propto$ 1.06 $\mathrm{L_{5100}}$, we adopted $\gamma$=0.53/1.06 = 0.50 for $\mathrm{L_{3000}}$. In the case of L$_{Mg \sevenrm II}$, we used
$\gamma$= 0.53/0.87 = 0.61 since L$_{Mg \sevenrm II}$ $\propto$ 0.87 $\mathrm{L_{5100}}$.
Using these fixed $\beta$ and $\gamma$ values, we performed the $\chi^2$ minimization with the FITEXY method \citep{Park+12b} to determine $\alpha$ (see Figure 9). In general the scatter is larger than 0.2 dex and 
the consistency with \Hb-based mass is better for \MgII\ line dispersion than FWHM. Also, continuum luminosity at 3000 \AA\ provides a better calibration than the line luminosity of \MgII. 

Since the six AGNs with strong discrepancy of line profiles between \Hb\ and \MgII\ are more scattered
from the best-fit relation as explained (open blue squares in Figure 9), we investigate how the calibration improves if we exclude these 6 AGNs. 
By removing the six AGNs, we obtained slightly better calibration with smaller scatter as presented in Table 3. Note that after removing these 6 AGNs, the $\beta$ becomes close to 2 as expected from a virial relation. 

Second, instead of using the correlation analysis between UV and optical luminosities and line widths, we simply fix $\beta$=2 and $\gamma$=0.5, following a virial relation and the expected size-luminosity relation, regardless of the choice of velocity measure (either \MgII\ line dispersion or FWHM) and luminosity measure
(either L$_{3000}$ or L$_{\rm MgII}$) (middle panels in Figure 9). In this case, we obtained somewhat smaller scatter. For fixed $\beta$ and $\gamma$, the result does not change significantly with/without removing the six AGNs with very different line profiles between \Hb\ and \MgII.

Third, we also tried to calibrate the UV mass estimators by fixing $\beta$=2 or by fixing $\gamma$=0.5, respectively. For these cases, the scatter is similar to the case with fixed $\beta$ and $\gamma$ while the coefficient $\alpha$ varies by 0.1-0.2 dex. Finally, we let all coefficients, $\alpha$, $\beta$, and $\gamma$ freely vary, and obtain the best-fit results (right panel in Figure 9). Again, we do not find a significant improvement in scatter. 

Based on these results we find that the pair of \MgII\ line dispersion and L$_{3000}$ provides 
the best calibration with  a $\sim$0.2 dex scatter than any other pair of velocity and luminosity measures. 
Among the various choice of $\beta$ and $\gamma$, we find no significant difference or improvement, indicating that a simple approach assuming the virial relation (i.e., $\beta$=2)
and the expected size-luminosity relation (i.e., $\gamma$=0.5) is comparable to the calibration based on UV-optical comparisons of luminosities, and velocities, respectively, or to the calibration using the $\alpha$, $\beta$, and $\gamma$ coefficients as free parameters. 

Compared to our previous results based on a subsample of the current data \citep{McGill+08}, 
we obtained improved calibrations with smaller intrinsic scatters. 
The intrinsic scatter between \Hb-based and \MgII-based masses is around 0.17-0.28 dex while the rms scatter is around 0.2-0.3 dex. For the best calibration (i.e., based on \MgII\ line dispersion and L$_{3000}$), the intrinsic scatter between \Hb-based mass and \MgII-based mass is $\sim$0.17 dex, indicating that even with the measurements based on high quality spectra, the single-epoch mass determined from UV measurements suffers from additional uncertainties by more than 0.17 dex, if we assume the \Hb-based mass represents the true \mbh. 

As a consistency check, we also performed the same calibration for \MgII-based masses, using the fiducial mass determined from FWHM of \Hb\ and L$_{5100}$ \citep[for the issues on the FWHM vs. $\sigma$, see][]{Peterson+04, Collin+06, Denney+09, Park12}. As presented in Table 4, we obtained slightly worse calibrations with a larger scatter. Again,
the line dispersion of \MgII\ ($\sigma_{\rm MgII}$ ) and L$_{3000}$ provides the best calibration among all paris of velocity and luminosity measures.

\begin{table*}
\begin{center}
\tablewidth{1\textwidth}
\fontsize{7}{5}\selectfont
\caption{\mbh\ estimators based on \ion{Mg}{2}, using the fiducial mass from $\sigma_{H\beta}$ and $L_{51000}$}
\begin{tabular}{ccccccccccccc}

\tableline\tableline
Case & N & $\alpha$  & $\beta$  & $\gamma$  & $\sigma_{int}$  & rms   &  & $\alpha$  & $\beta$  & $\gamma$  & $\sigma_{inst}$  & rms \\
(1)&(2)&(3)&(4)&(5)& (6)&(7)& & (8)&(9)&(10)&(11) & (12)\\
\tableline
\tableline

~ \\
  & &  &  & {\bf L$_{3000}$ \& $\sigma_{\rm MgII}$ }&  &  & & &   & {\bf L$_{3000}$ \& FWHM$_{MgII}$ }\\
\cline{1-7} \cline{9-13} \\
1) $\beta$ \& $\gamma$ from scaling & 47 & 7.48 $\pm$ 0.03 & 2.38 $\pm$ 0.20 & 0.50 $\pm$ 0.00 & 0.19 $\pm$ 0.03 & 0.23
& & 6.62 $\pm$ 0.05 & 2.85 $\pm$ 0.37 & 0.50 $\pm$ 0.00 & 0.28 $\pm$ 0.04 & 0.31 \\
& 41 & 7.52 $\pm$ 0.03 & 1.98 $\pm$ 0.12 & 0.50 $\pm$ 0.00 & 0.13 $\pm$ 0.04 & 0.18
& & 6.83 $\pm$ 0.03 & 2.32 $\pm$ 0.23 & 0.50 $\pm$ 0.00 & 0.20 $\pm$ 0.03 & 0.23 \\
\cline{1-7} \cline{9-13} \\
{\bf 2) $\beta$=2 \& $\gamma$=0.5} & 47 & {\bf 7.56 $\pm$ 0.03} & {\bf 2.00} & {\bf 0.50} & {\bf 0.17 $\pm$ 0.03} & {\bf 0.21}
& & {\bf 7.02 $\pm$ 0.04} & {\bf 2.00} & {\bf 0.50} & {\bf 0.23 $\pm$ 0.04} & {\bf 0.26}  \\

\cline{1-7} \cline{9-13} \\
3) $\beta$ = 2 & 47 & 7.48 $\pm$ 0.06 & 2.00 & 0.69 $\pm$ 0.14 & 0.17 $\pm$ 0.03 & 0.21
& & 6.89 $\pm$ 0.07 & 2.00 & 0.83 $\pm$ 0.18 & 0.22 $\pm$ 0.04 & 0.25 \\

\cline{1-7} \cline{9-13} \\
4) $\gamma$=0.5 & 47 & 7.57 $\pm$ 0.05 & 1.92 $\pm$ 0.19 & 0.50 & 0.17 $\pm$ 0.03 & 0.21
& & 7.10 $\pm$ 0.13 & 1.83 $\pm$ 0.25 & 0.50 & 0.23 $\pm$ 0.04 & 0.26 \\

\cline{1-7} \cline{9-13} \\
Free $\beta$ \& $\gamma$ & 47 & 7.50 $\pm$ 0.06 & 1.89 $\pm$ 0.20 & 0.70 $\pm$ 0.15 & 0.17 $\pm$ 0.03 & 0.21
& & 6.97 $\pm$ 0.12 & 1.82 $\pm$ 0.25 & 0.83 $\pm$ 0.18 & 0.23 $\pm$ 0.04 & 0.25 \\

\tableline 
\tableline \\
~\\

  &  & &  &  {\bf L$_{\rm MgII}$ \& $\sigma_{MgII}$ } &  &  &  & &  &  {\bf L$_{\rm MgII}$ \& FWHM$_{MgII}$ } \\
\cline{1-7} \cline{9-13} \\
1)$\beta$ \& $\gamma$ from scaling & 47 & 7.37 $\pm$ 0.04 & 2.38 $\pm$ 0.20 & 0.61 $\pm$ 0.02 & 0.25 $\pm$ 0.03 & 0.29
& & 6.51 $\pm$ 0.05 & 2.85 $\pm$ 0.37 & 0.61 $\pm$ 0.02 & 0.33 $\pm$ 0.04 & 0.36 \\
 & 41 & 7.43 $\pm$ 0.04 & 1.98 $\pm$ 0.12 & 0.61 $\pm$ 0.02 & 0.21 $\pm$ 0.04 & 0.25
& & 6.74 $\pm$ 0.04 & 2.32 $\pm$ 0.23 & 0.61 $\pm$ 0.02 & 0.26 $\pm$ 0.04 & 0.29 \\
\cline{1-7} \cline{9-13} \\
{\bf 2) $\beta$=2 \& $\gamma$=0.5} & 47 & {\bf 7.50 $\pm$ 0.04} & {\bf 2.00} & {\bf 0.50} & {\bf 0.21 $\pm$ 0.03} & {\bf 0.24}
& & {\bf 6.97 $\pm$ 0.04} & {\bf 2.00} & {\bf 0.50} & {\bf 0.25 $\pm$ 0.03} & {\bf 0.28} \\

\cline{1-7} \cline{9-13} \\
3) $\beta$ = 2 & 47 & 7.57 $\pm$ 0.07 & 2.00 & 0.36 $\pm$ 0.13 & 0.21 $\pm$ 0.03 & 0.24
& & 6.97 $\pm$ 0.09 & 2.00 & 0.50 $\pm$ 0.17 & 0.25 $\pm$ 0.03 & 0.28 \\

\cline{1-7} \cline{9-13} \\
4) $\gamma$=0.5 & 47 & 7.61 $\pm$ 0.05 & 1.48 $\pm$ 0.19 & 0.50 & 0.19 $\pm$ 0.03 & 0.22
& & 7.27 $\pm$ 0.13 & 1.37 $\pm$ 0.24 & 0.50 & 0.23 $\pm$ 0.03 & 0.26 \\

\cline{1-7} \cline{9-13} \\
5) Free $\beta$ \& $\gamma$ & 47 & 7.59 $\pm$ 0.07 & 1.38 $\pm$ 0.28 & 0.59 $\pm$ 0.20 & 0.19 $\pm$ 0.03 & 0.22
& & 7.26 $\pm$ 0.12 & 1.10 $\pm$ 0.30 & 0.77 $\pm$ 0.21 & 0.22 $\pm$ 0.03 & 0.25 \\

\tableline
\tableline
\end{tabular}
\tablecomments{Col. (1): Method of calibration. Col. (2): Number of data use in the calibration.  Col. (3) \& (8): $\alpha$ values. Col. (4) \& (9): 
$\beta$ values. Col. (5) \& (10): $\gamma$ values. Col. (6) \& (11): intrinsic scatter. Col. (7) \& (12): rms scatter. The recommended estimator is represented by bold fonts.
}

\end{center}
\end{table*}


\begin{table*}
\begin{center}
\tablewidth{1\textwidth}
\fontsize{7}{5}\selectfont
\caption{\mbh\ estimators based on \ion{Mg}{2}, using the fiducial mass from $FWHM_{H\beta}$ and $L_{51000}$}
\begin{tabular}{ccccccccccccc}

\tableline\tableline
Case & N & $\alpha$  & $\beta$  & $\gamma$  & $\sigma_{int}$  & rms   &  & $\alpha$  & $\beta$  & $\gamma$  & $\sigma_{inst}$  & rms \\
(1)&(2)&(3)&(4)&(5)& (6)&(7)& & (8)&(9)&(10)&(11) & (12)\\
\tableline
\tableline

~ \\
  & &  &  & {\bf L$_{3000}$ \& $\sigma_{\rm MgII}$ }&  &  & & &   & {\bf L$_{3000}$ \& FWHM$_{MgII}$ }\\

\cline{1-7} \cline{9-13} \\
1) $beta$ \& $\gamma$ from scaling & 47 & 7.36 $\pm$ 0.04 & 2.89 $\pm$ 0.16 & 0.50 $\pm$ 0.00 & 0.21 $\pm$ 0.03 & 0.25
& & 6.38 $\pm$ 0.05 & 3.33 $\pm$ 0.27 & 0.50 $\pm$ 0.00 & 0.30 $\pm$ 0.05 & 0.33 \\
& 41 & 7.31 $\pm$ 0.04 & 2.89 $\pm$ 0.16 & 0.50 $\pm$ 0.00 & 0.19 $\pm$ 0.03 & 0.23
& & 6.33 $\pm$ 0.04 & 3.33 $\pm$ 0.27 & 0.50 $\pm$ 0.00 & 0.25 $\pm$ 0.04 & 0.28 \\

\cline{1-7} \cline{9-13} \\
2) $beta$=2 \& $\gamma$=0.5 & 47 & 7.54 $\pm$ 0.03 & 2.00 & 0.50 & 0.21 $\pm$ 0.03 & 0.24
& & 7.01 $\pm$ 0.04 & 2.00 & 0.50 & 0.26 $\pm$ 0.04 & 0.29 \\

\cline{1-7} \cline{9-13} \\
3) $beta$ = 2 & 47 & 7.50 $\pm$ 0.08 & 2.00 & 0.61 $\pm$ 0.16 & 0.21 $\pm$ 0.03 & 0.24
& & 6.91 $\pm$ 0.09 & 2.00 & 0.74 $\pm$ 0.20 & 0.26 $\pm$ 0.04 & 0.29 \\

\cline{1-7} \cline{9-13} \\
4) $\gamma$=0.5 & 47 & 7.47 $\pm$ 0.04 & 2.38 $\pm$ 0.21 & 0.50 & 0.20 $\pm$ 0.03 & 0.23
& & 6.84 $\pm$ 0.13 & 2.35 $\pm$ 0.27 & 0.50 & 0.26 $\pm$ 0.04 & 0.29 \\

\cline{1-7} \cline{9-13} \\
5) Free $\beta$ \& $\gamma$ & 47 & 7.44 $\pm$ 0.06 & 2.36 $\pm$ 0.21 & 0.58 $\pm$ 0.17 & 0.20 $\pm$ 0.03 & 0.23
& & 6.75 $\pm$ 0.12 & 2.35 $\pm$ 0.25 & 0.74 $\pm$ 0.21 & 0.26 $\pm$ 0.05 & 0.28 \\

\tableline 
\tableline \\
~\\

  &  & &  &  {\bf L$_{\rm MgII}$ \& $\sigma_{MgII}$ } &  &  &  & &  &  {\bf L$_{\rm MgII}$ \& FWHM$_{MgII}$ } \\
\cline{1-7} \cline{9-13} \\
1) $\beta$ \& $\gamma$ from scaling & 47 & 7.25 $\pm$ 0.05 & 2.89 $\pm$ 0.16 & 0.61 $\pm$ 0.02 & 0.28 $\pm$ 0.03 & 0.31
& & 6.27 $\pm$ 0.06 & 3.33 $\pm$ 0.27 & 0.61 $\pm$ 0.02 & 0.35 $\pm$ 0.04 & 0.38 \\

\cline{1-7} \cline{9-13} \\
2) $\beta$=2 \& $\gamma$=0.5 & 47 & 7.49 $\pm$ 0.04 & 2.00 & 0.50 & 0.21 $\pm$ 0.03 & 0.25
& & 6.96 $\pm$ 0.04 & 2.00 & 0.50 & 0.26 $\pm$ 0.03 & 0.29 \\

\cline{1-7} \cline{9-13} \\
3) $\beta$ = 2 & 47 & 7.53 $\pm$ 0.06 & 2.00 & 0.42 $\pm$ 0.13 & 0.22 $\pm$ 0.03 & 0.25
& & 6.93 $\pm$ 0.08 & 2.00 & 0.57 $\pm$ 0.17 & 0.26 $\pm$ 0.03 & 0.29 \\

\cline{1-7} \cline{9-13} \\
4) $\gamma$=0.5 & 47 & 7.51 $\pm$ 0.05 & 1.93 $\pm$ 0.21 & 0.50 & 0.22 $\pm$ 0.03 & 0.25
& & 7.01 $\pm$ 0.12 & 1.89 $\pm$ 0.25 & 0.50 & 0.26 $\pm$ 0.03 & 0.29 \\

\cline{1-7} \cline{9-13} \\
5) Free $\beta$ \& $\gamma$ & 47 & 7.53 $\pm$ 0.06 & 2.02 $\pm$ 0.29 & 0.42 $\pm$ 0.19 & 0.22 $\pm$ 0.03 & 0.25
& & 7.01 $\pm$ 0.12 & 1.75 $\pm$ 0.34 & 0.64 $\pm$ 0.23 & 0.26 $\pm$ 0.03 & 0.28 \\

\tableline
\tableline
\end{tabular}
\tablecomments{Col. (1): Method of calibration. Col. (2): Number of data use in the calibration.  Col. (3) \& (8): $\alpha$ values. Col. (4) \& (9): 
$\beta$ values. Col. (5) \& (10): $\gamma$ values. Col. (6) \& (11): intrinsic scatter. Col. (7) \& (12): rms scatter. 
}

\end{center}
\end{table*}

\section{Discussion}

\begin{figure*}
\figurenum{10}
\center
	\includegraphics[width=0.29\textwidth]{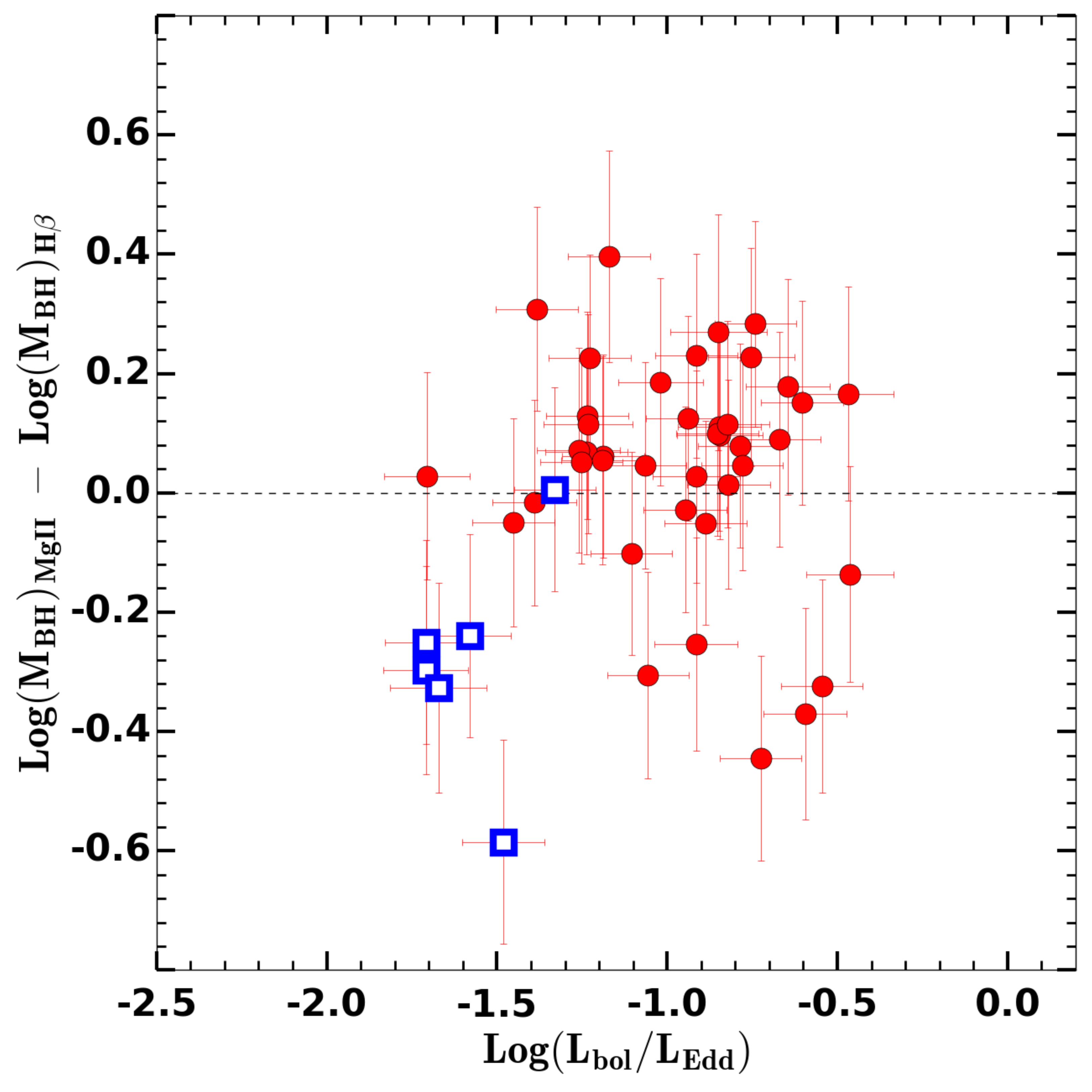}
	\includegraphics[width=0.29\textwidth]{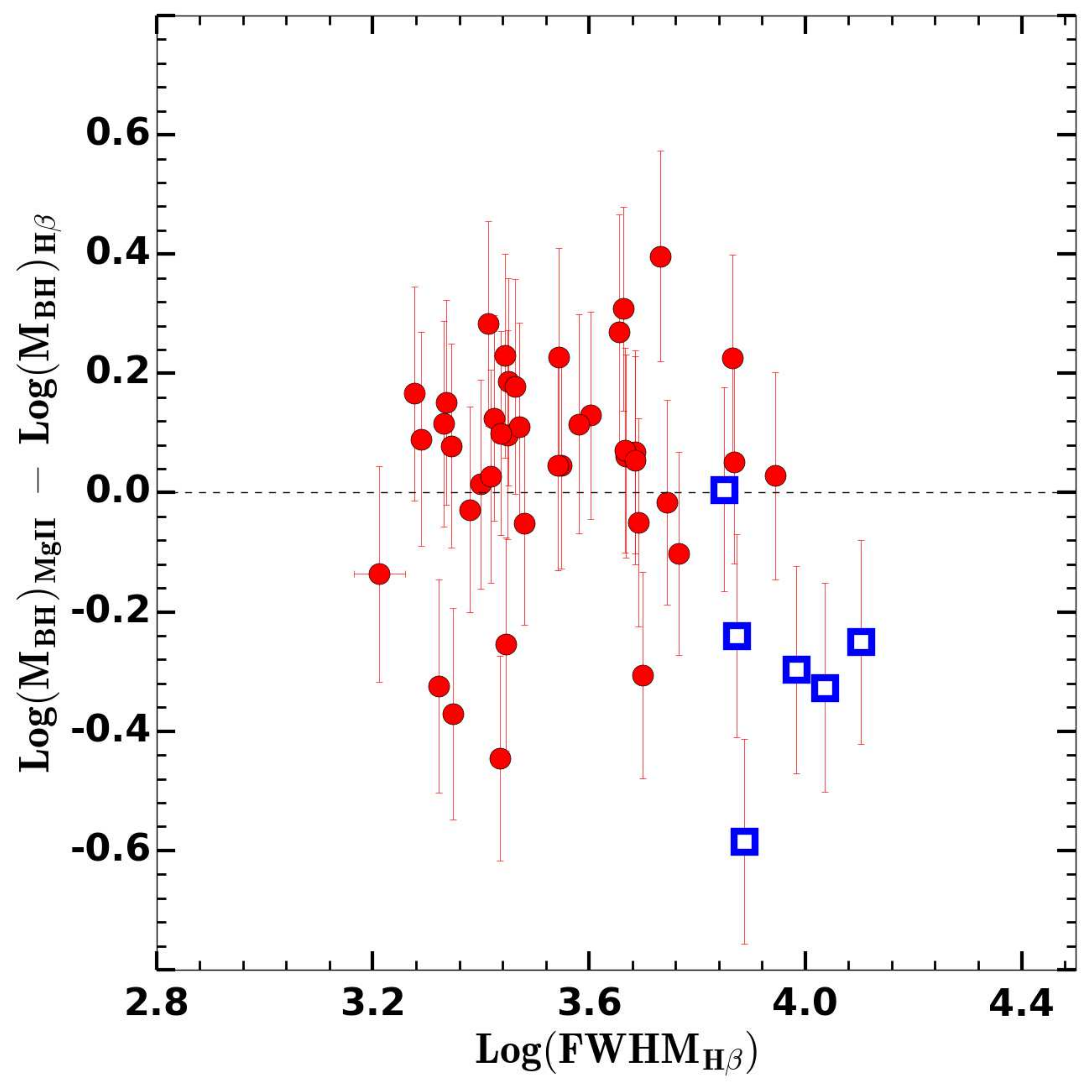}
	\includegraphics[width=0.283\textwidth]{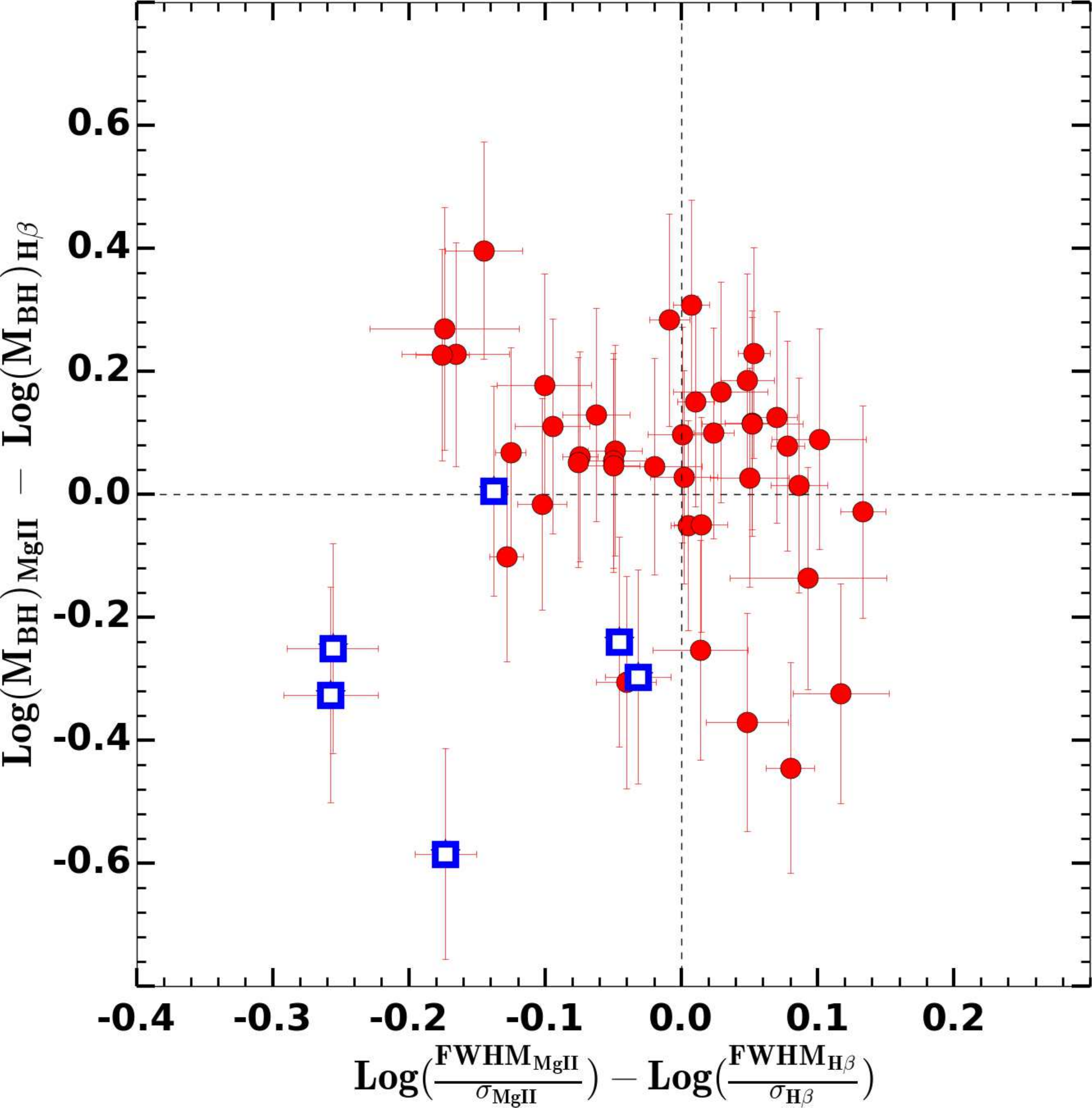} \\
	\includegraphics[width=0.29\textwidth]{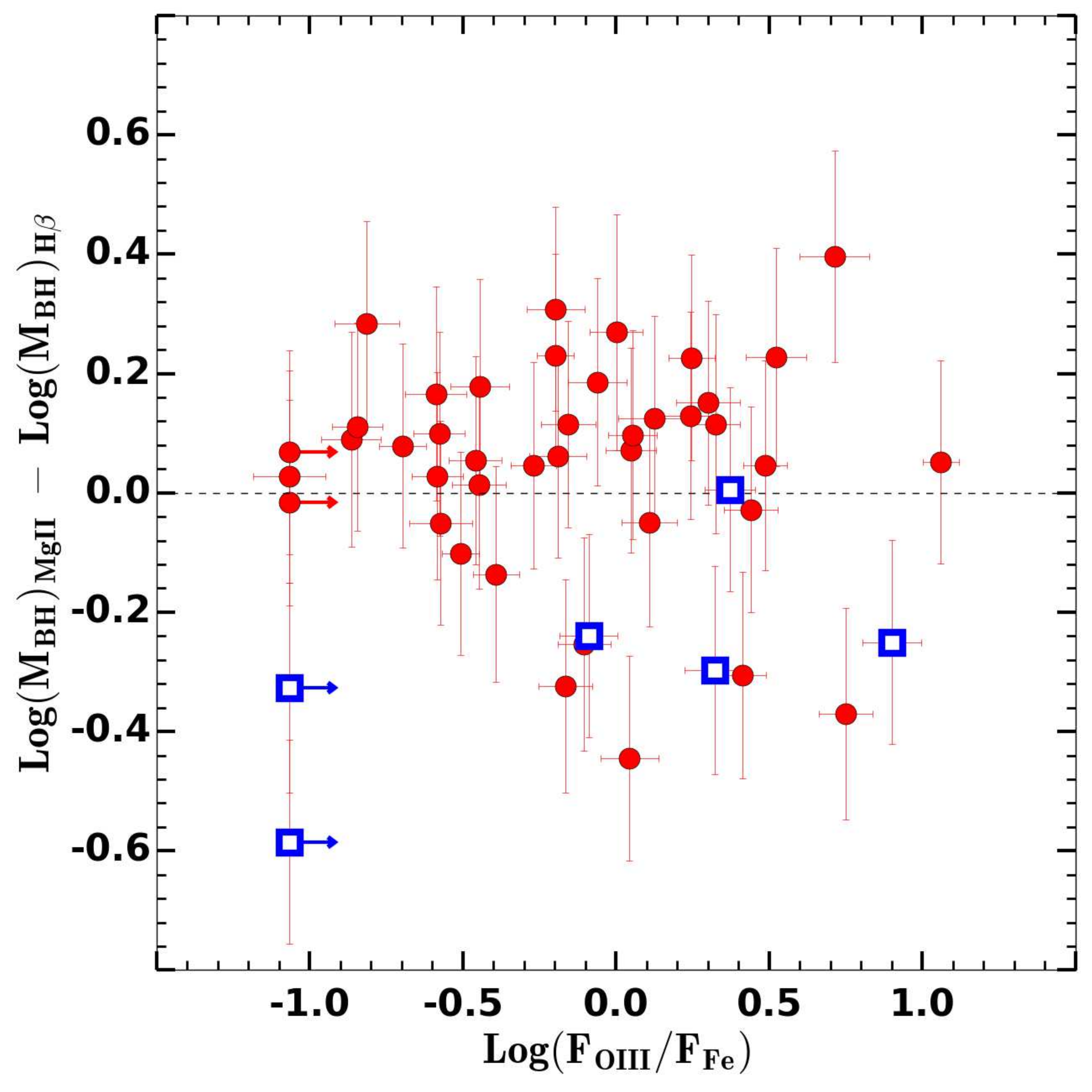}
	\includegraphics[width=0.29\textwidth]{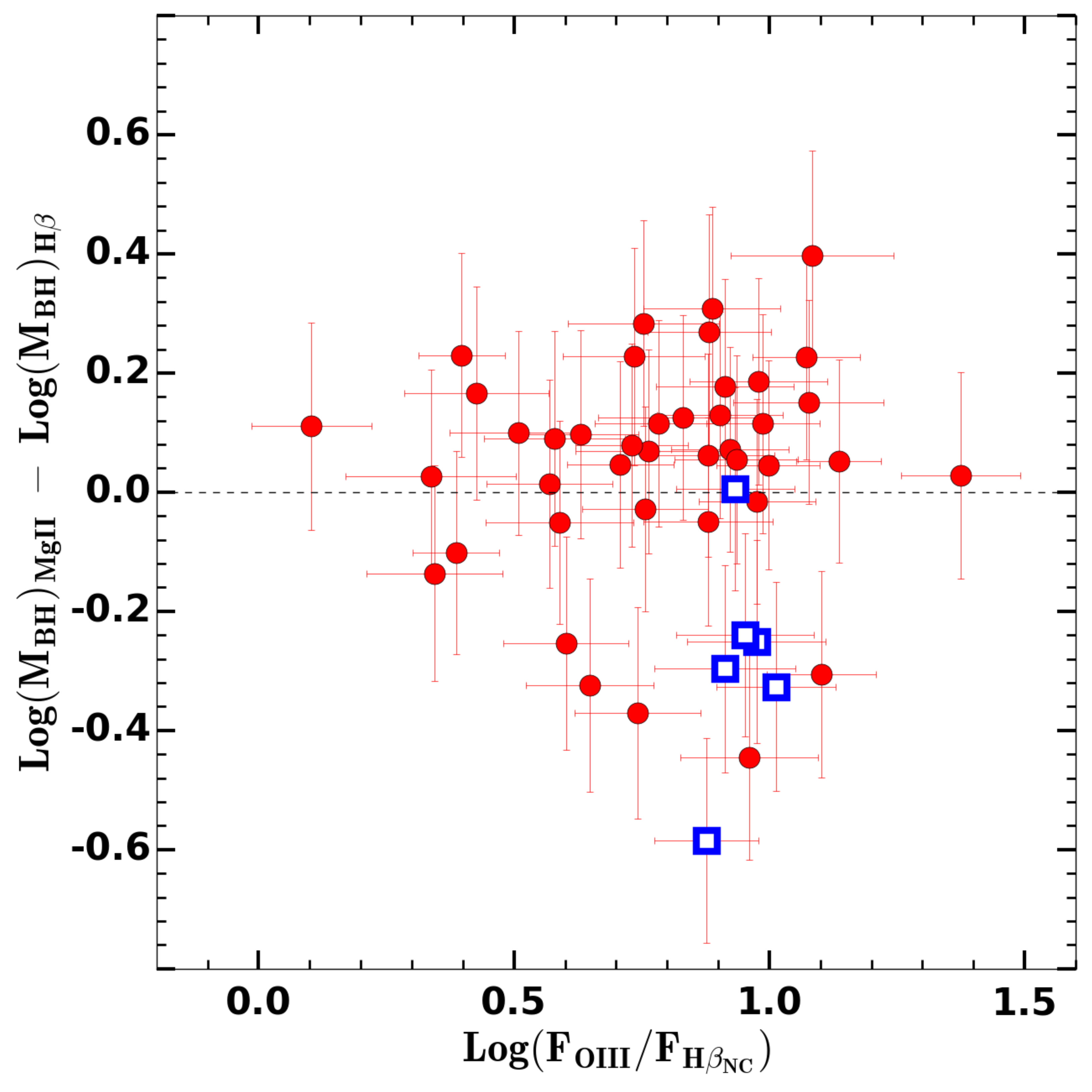}
	\includegraphics[width=0.29\textwidth]{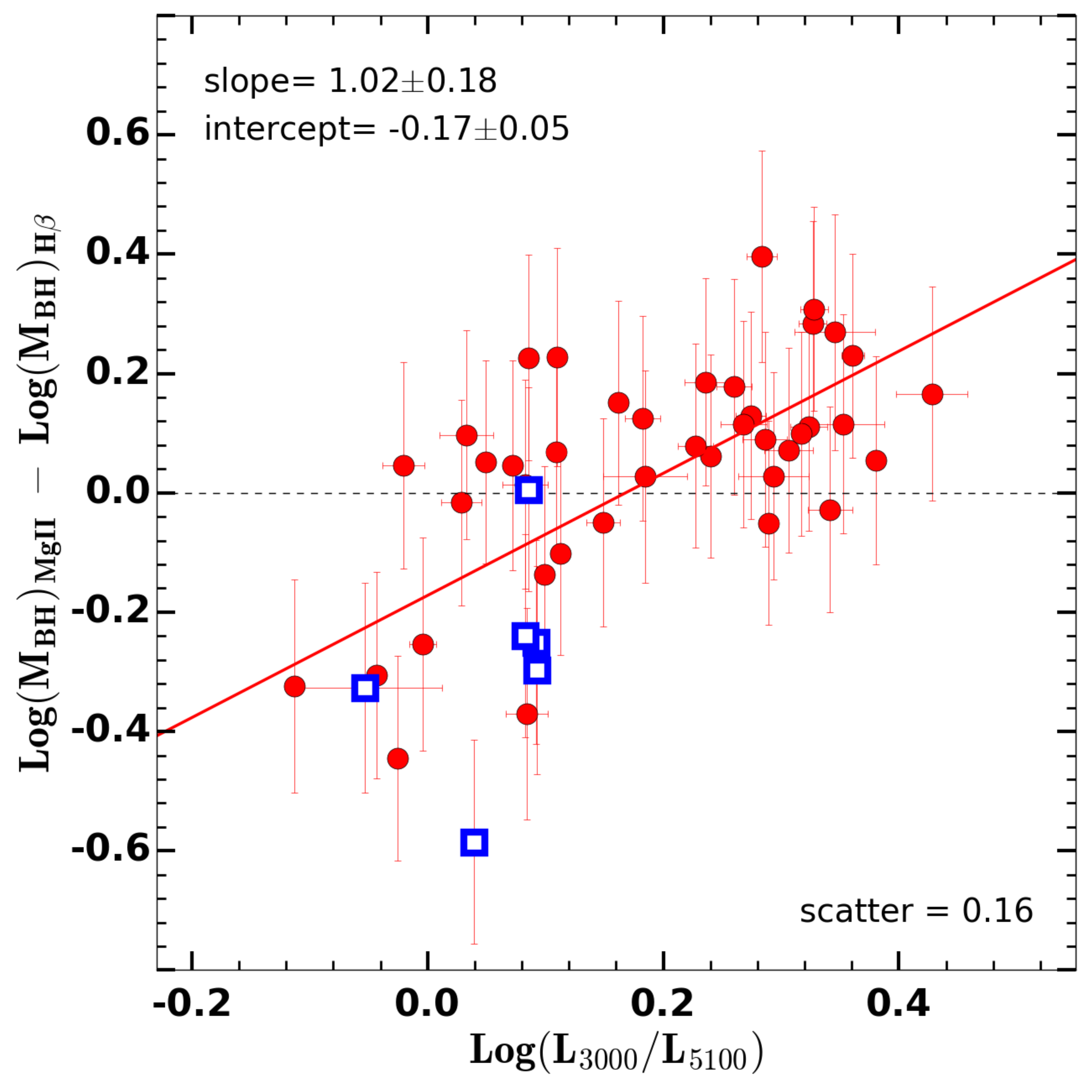}
	\caption{Testing systematic trends with Eddington ratio (top left), \Hb\ FWHM (top middle), the difference of line profiles between \Hb\ and \MgII\ (top right),
	F$_{\rm OIII}$/F$_{\rm FeII}$ (bottom left), F$_{\rm OIII}$/F$_{\rm H\beta, narrow}$ (bottom middle), L$_{3000}$/L$_{5100}$ (bottom right). Open blue squares show the six AGNs, S23, S24, W02, W03, W12, W20, with very different line profiles between \MgII\ and \Hb.}
	\endcenter
\end{figure*}

\subsection{Systematic uncertainties in \MgII-based mass estimates}

Although the single-epoch estimators are powerful in determining \mbh\ of a large sample, the uncertainty of the mass estimates is much more significant than that of the reverberation masses, due to the uncertainty and scatter of the size-luminosity relation  \citep[e.g., 0.19 dex reported by][]{Bentz+13}. Moreover, there are additional sources of uncertainties in \MgII-based mass estimates since these mass estimates are based on further calibrations of the \MgII\ line velocity and UV luminosities against \Hb-based \mbh\ since there is no available \MgII-based size-luminosity relation. 

Here, we discuss several issues to consider in understanding the systematic uncertainties
of \MgII-based mass estimates. First, while the variability is the key to measure the BLR size,
it also causes a difficulty in calibrating mass estimators. Since for given AGNs with a fixed \mbh, 
luminosity and velocity varies over time, simultaneous observations of the rest-frame UV and optical are required to properly compare the widths of \Hb\ and \MgII\ or the UV and optical luminosities. The non-simultaneity often causes difficulties in comparing \CIV\ and \Hb\ line widths \citep[see][]{Denney12}, while it can be avoided by selecting AGNs at optimal redshifts so that both \MgII\ and \Hb\ lines can be simultaneously obtained in the observed spectral range
\citep[e.g.,][]{McGill+08, Wang+09, Shen+11}. 

Instead of the continuum luminosity at 3000\AA, the line luminosity of \MgII\ can be utilized to determine \mbh\ by assuming that \MgII\ line luminosity varies in response to UV continuum luminosity
\citep[for example, see a recent study by][]{Zhu+17}. As we have shown in Figure 7, the scatter increases by a factor of 2 when we compare L$_{5100}$ with L$_{\rm MgII}$ instead of L$_{3000}$. Due to this less tight correlation, the uncertainty of \mbh\ estimates based on L$_{\rm MgII}$ will be larger than those based on L$_{3000}$ (see the scatter in Figure 9). 

Second, Balmer continuum present in the UV spectral range may cause a systematic uncertainty in measuring UV continuum luminosity. Without a proper fitting and subtraction of Balmer continuum, L$_{3000}$ may be overestimated, resulting in a higher \mbh. However, the systematic effect of Balmer continuum is limited since the contribution of Balmer continuum at 3000\AA\ is relatively small. For example, \cite{Kovacevic+17} reported that L$_{3000}$ changes by $\sim$10\% on average, hence, \mbh\ is overestimated by an average $\sim$5\%. 

Third, various studies reported a non-linear relationship between \Hb\ and \MgII\ line widths \citep{Salviander07, Wang+09, Shen+11}. The FWHM of \Hb\ is larger than that of \MgII\ by more than 20\% for AGNs with a very broad \Hb\ line \citep{Marziani+13} while the line widths of \Hb\ and \MgII\ are more consistent in AGNs with a narrower \Hb\ line. This non-linear relationship may cast doubts on \MgII-based mass estimates since while $\beta$=2 is used for \Hb\ line in Eq. 1, $\beta$ is forced to be larger than 2 for \MgII, violating the virial assumption. It is not clear whether \Hb\ line width overestimates the true velocity of BLR gas when the FWHM of \Hb\ is larger than, for example, 4000 \kms\ \citep[see the discussion on Population B in][]{Marziani+13}, or \MgII\ line width underestimates the velocity of BLR gas. The fact that the line profile of \MgII\ is rather similar to each other, regardless of the width of the line (see Figure 4) may imply that there is a systematic issue in measuring the \Hb\ line width, particularly, FWHM, when the line is extremely broad. 
In practice, we performed the calibration with/without using the correlation results between \Hb\ and \MgII\ line widths. It seems better to assume a virial relation (i.e., $\beta$=2) since the \MgII-\Hb\ width correlation results depend on the sample and dynamic range, suffering from systematic effects. 

Fourth, it is not clear whether the narrow component of \MgII\ should be separately fitted and subtracted to properly measure the width of the broad component of \MgII. Subtracting the narrow component originated from the narrow-lie region is a typical process in fitting and measuring the width of the broad component of \Hb, by assuming the narrow \Hb\ profile is identical to other narrow lines, i.e., \OIII\ $\lambda$5007. In the case of \MgII, however, it is practically difficult to constrain the profile of the narrow component. Thus, most previous studies do not attempt to subtract the narrow component. In contrast, \cite{Wang+09} used two components, respectively for the broad and narrow components in \MgII, and measured the FWHM of the broad component to determine \mbh. However, FWHM measurements suffer from significant uncertainties since the peak of the line profile will be strongly dependent of the amount of flux assigned to the narrow component. 

In our study, we see no clear sign of the presence of a narrow component in \MgII\ even when we see a strong and clear narrow component in \Hb. Note that this can be partly due to the lower spectral resolution in the \MgII\ area (i.e., $\sim$145 \kms), although a narrow line with a typical velocity dispersion of a few hundred \kms\ \citep[i.e., in the case of the \ion{O}{3};][]{Woo+16} can be resolved in our Keck spectra.
To test the potential effect of the narrow component in \MgII, we used the \OII\ line profile to constrain the narrow component of \MgII, assuming the profiles of narrow lines in the NLR (i.e., \OII\ and narrow \MgII) are similar. As \citet{Malkan+17} reported that the typical value of \MgII-to-\OII\ ratio in Seyfert 2 galaxies is $\sim$0.1, we take the \OII\ line profile, after multiplying by 0.1, as a narrow component of \MgII\ and subtract it from the \MgII\ line to calculate the line width of the broad component of \MgII. However, we find that this practice makes no difference in the line width measurements since \OII\ is much weaker than \MgII, hence the narrow component of \MgII\ is negligible in most objects. Thus, we did not subtract the potential narrow component and used the total line profile to measure the FWHM and line dispersion. Note that the measurements of line dispersion are not significantly affected by the subtraction or inclusion of the narrow component. 

Fifth, we investigated whether the systematic difference between \MgII- and \Hb-based masses shows any dependency on other AGN parameters, i.e., Eddington ratio, FWHM of \Hb, the systematic difference of the line profiles between \MgII\ and \Hb\ in Figure 10 (top panels). We also check whether the systematic difference of \mbh\ is due to the eigenvector 1 by calculating the flux ratio between \OIII\ and \FeII, which is integrated in the spectral range and the spectral slope 4434-4684\AA\ \citep[e.g., see][]{Woo15}, and the flux ratio between \OIII\ and the narrow component of \Hb\ (bottom panel in Figure 10). For the first five parameters, we find no significant trend, suggesting that \MgII-based masses are not significantly biased due to the Eddington ratio, line width, and the \FeII\ strength. 

In contrast, we expect to see a broad trend between the UV to optical mass ratio and the UV to optical luminosity ratio (i.e., L$_{3000}$/L$_{5100}$) since single-epoch \mbh\ correlates with continuum luminosity as far as the size-luminosity relation (i.e., \mbh\ $\propto$ L$^{0.5}$) is used for determining \mbh. For given L$_{5100}$ and \Hb-based mass, for example, if the spectral slope becomes steeper (i.e., higher L$_{3000}$/L$_{5100}$ ratio), then L$_{3000}$, and consequently, \MgII-based mass will be systematically higher. We see this trend in Figure 10. To correct for this systematic trend, we obtain the best-fit slope 1.02$\pm$0.18 and the intercept - 0.17$\pm0.05$ (last panel in Figure 10), and add the following color correction term to Eq. 1: 
\begin{equation}
\Delta C  = -1.02 \times \log (L_{3000}/L_{5100}) + 0.17. 
\end{equation}
Since L$_{5100}$ will not be available for \mbh\ determination for high-z AGNs, we can instead use the spectral slope $\alpha_{\lambda}$, with which we model
the local UV/optical AGN continuum as a power law,  f$_{\lambda}$ $\propto$ $\lambda^{\alpha_{\lambda}}$. 
Using L$_{3000}$/L$_{5100}$ =  3000 f$_{3000}$/ 5100 f$_{5100}$ = (3000/5100)$^{1+\alpha_{\lambda}}$, we derive the correction term as a function of $\alpha_{\lambda}$:
\begin{equation}
\Delta C = 0.24 (1+\alpha_{\lambda}) + 0.17.
\end{equation}
Note that the mean $\alpha_{\lambda}$ of the Keck sample is -1.73$\pm$0.60 (i.e., $\alpha_{\nu}$=-0.27$\pm$0.60), which is slightly bluer than the average
spectral slope $\alpha_{\nu}$=-0.44 of the SDSS quasars \citep{VandenBerk+01}.
Once applied, this correction term will reduce the systematic uncertainty of \MgII-based masses due to the large range of the spectral slope between the UV and optical wavelength range. However, the correction is relatively small. For example, if the spectral slope changes from $\alpha_{\lambda}$=-1.5 to $\alpha_{\lambda}$ =-2, the correction on \mbh\ is $\sim$0.1 dex. Thus, if the spectral slope is difficult to determine due to the low S/N, strong \FeII\ blends, the limited spectral range, or internal dust extinction, this correction can be ignored.

\begin{figure}
\figurenum{11}
\center
	\includegraphics[width=0.42\textwidth]{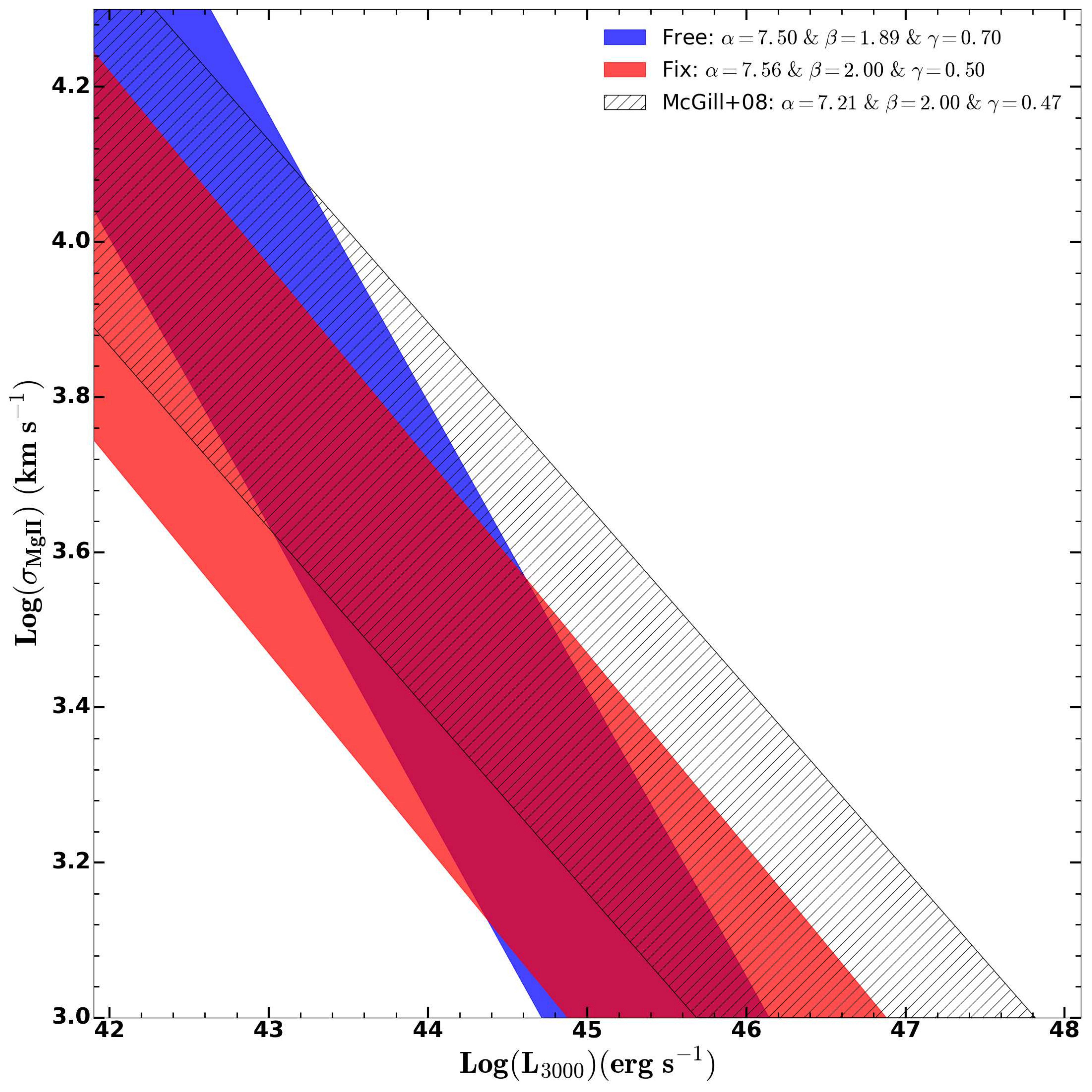}
	\includegraphics[width=0.42\textwidth]{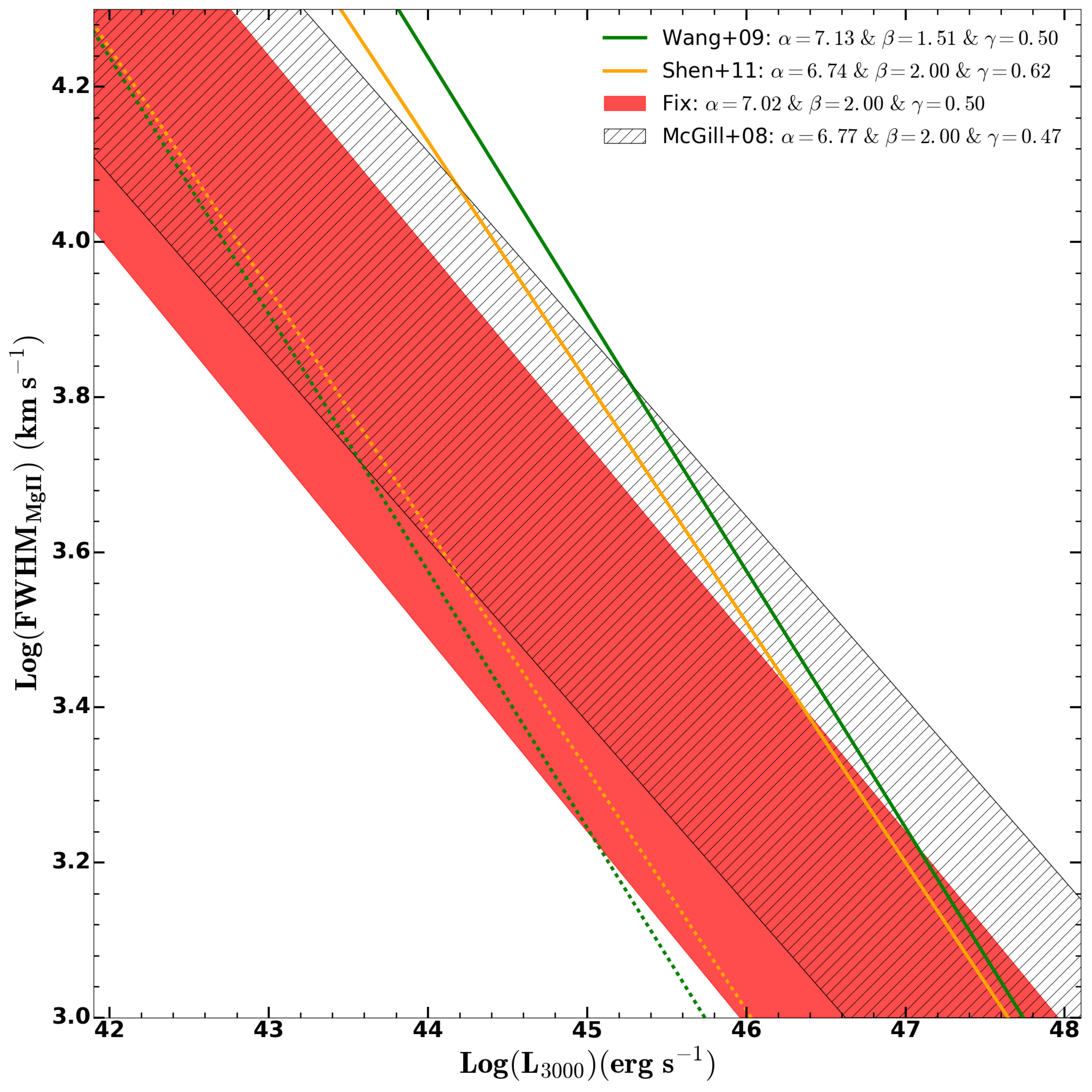}
	\centering
	\caption{Comparison of \mbh\ estimates for given paris of velocity and luminosity measures depending on the choice of UV mass estimator. Top: \mbh\ based on \MgII\ line dispersion from our estimators with fixed $\beta$=2 \& $\gamma$=0.5 (red area) or free fit results (blue area), compared to that of \cite{McGill+08} (hatched area). For each mass estimator, the area is defined by two equi-mass lines at \mbh\= 10$^{8}$ and \mbh\= 10$^{9}$, showing the systematic difference among various mass estimators. Bottom: \mbh\ based on \MgII\ FWHM compared to those of \cite{Wang+09} (green line) and \cite{Shen+11} (orange line).}	
\endcenter
\end{figure}

\subsection{Comparison of various \MgII-based mass estimators}

There have been various calibrations of \mbh\ estimators based on \MgII\ in the literature, and here we investigate how \mbh\ changes depending on the choice of the estimators. In Figure 11, we present the calculated \mbh\ for given pairs of \MgII\ line width and L$_{3000}$
based on several UV mass estimators. From our calibrations, we choose Case 2 as the best calibration, and Case 5 as an extreme calibration. 
In the case of \mbh\ based on \MgII\ line dispersion, \citet{McGill+08} reported the mass calibrators, and we compare our estimators with theirs in the top panels.  Case 2 with the fixed $\beta$ and $\gamma$ provides a similar \mbh\ compared to \citet{McGill+08}, with a systematic offset by 0.2-0.3 dex. The difference of the normalization is mainly due to the change of the width measurements. Since for given objects, \MgII\ line dispersion becomes smaller due to the \citet{Tsuzuki06} template, that we used in our analysis, while \citet{McGill+08} used the \FeII\ template of \citet{VW01}. Thus, $\alpha$ becomes larger for UV-based \mbh\ to be consistent with given \Hb-based mass.

In the case of FWHM, we compare our estimator with those of \citet{McGill+08, Wang+09, Shen+11}. Compared to Case 2 (red in Figure 11), other estimators derive somewhat lower \mbh, and the equi-mass line is steeper because of lower $\beta$ than 2 or higher $\gamma$ than 0.5. Note that depending on the choice of the estimators, \mbh\ will be systematically larger or smaller for AGNs with very broad lines (or lower luminosity). These results indicate that the inferred shape of the mass function of high-z AGNs will be sensitive to the choice of UV \mbh\ estimator. Note that for estimating \mbh\ using a large survey data, FWHM of \MgII, rather than line dispersion, is often used since the spectral quality in survey data is not enough to measure the line dispersion of broad lines. Thus, a careful interpretation is required to understand the mass distribution and mass function of high-z AGNs.

\section{Conclusions}\label{section:sum}

In this paper, we present a new calibration of \mbh\ estimators, using a sample of 52 AGNs at $z \sim 0.36$  and $z \sim 0.52$ over the \mbh\ range, 7.4 $<$ log \mbh\ $<$ 9, for which high quality Keck spectra are available to properly measure line widths and UV and optical luminosities. In addition, we utilize the measurements of SDSS AGNs from literature to increase the dynamic range. The main results are summarized as follows:

\smallskip
(1) There are a number of AGNs, for which \Hb\ is much broader than \MgII, particularly for AGNs with a large \Hb\ FWHM \citep[see also][]{Wang+09, Marziani+13}. Consequently, we obtain a sub-linear relationship between \MgII\ and \Hb\ both for FWHM as well as line dispersion.
 
\smallskip
(2) By comparing optical continuum luminosity at 5100\AA\ with UV continuum luminosity and
\Hb\ line luminosity, we find a correlation close to a linear relationship, while the relation with \MgII\ line luminosity
is somewhat sub-linear, reflecting the Baldwin effect in the UV.

\smallskip
(3) We perform a cross-calibration of \mbh\ estimators using various combinations of velocity and
luminosity indicators measured from the rest-frame UV and optical spectra, using the the mass based on \Hb\ line dispersion 
and L5100 as a reference mass.  \mbh\ from the new calibrations using \Hb\ line and optical luminosities are consistent with each other with an intrinsic scatter less than 0.1 dex and a rms scatter of $\sim$0.1 dex. 

\smallskip
(4) In the case of UV mass estimators based on \MgII\ line, the comparison with the reference \Hb-based masses shows an intrinsic scatter of 0.17-0.28 dex and a rms scatter of $\sim$0.2-0.3 dex, suggesting that there is an additional uncertainty larger than $\sim$0.2 dex, depending on a choice of line width (i.e., line dispersion or FWHM) and luminosity measures (i.e., L$_{3000}$ or L$_{\rm MgII}$). Over all, we find that the pair of \MgII\ line dispersion and L$_{3000}$ provides the best calibration with an additional 0.17$\pm$0.03 dex uncertainty.
In the case of \Hb\ single-epoch mass estimates, the uncertainties are mainly introduced by three sources. First, the uncertainty of the virial factor, which is 0.12-0.15 dex based on the comparison of the \mbh-$\sigma_*$ relation between the reverberation-mapped AGNs and quiescent galaxies \citep{Woo+10, Woo15}, or 0.4 dex based on the dynamical modeling of the five reverberation-mapped AGNs with velocity-resolved measurements \citep{Pancoast+14}. Second, the scatter in the \Hb\ size-luminosity relation is 0.13-0.19 dex, depending on the choice of more reliable measurements \citep{Bentz+13}. Third, the random variability of the line width and luminosity introduces $\sim$0.1 dex scatter \citep{Park12}. Compared to the total uncertainty of \Hb-based mass estimates, which can be 0.3-0.4 dex, the additional 0.17-0.28 dex uncertainty from the calibration of UV mass estimators is somewhat smaller. However, the overall uncertainty of \MgII-based mass is larger than that of \Hb\ masses. 

\smallskip
In this paper, we calibrated UV and optical \mbh\ estimators based on single-epoch measurements. 
While future direct measurements of the \MgII\ time lag for a sizable sample of AGNs will enable to reduce the systematic uncertainties in the single-epoch mass estimates, the updated and calibrated \MgII\ mass estimators in this paper will be useful for revisiting \mbh\ related issues for high-z AGNs. 


\acknowledgements
This work has been supported by the Basic Science Research Program through 
the National Research Foundation of Korea government (2016R1A2B3011457 and No.2017R1A5A1070354). We thank the anonymous referee for useful suggestions.
\\



\begin{thebibliography}{}

\bibitem[Baldwin(1977)]{Baldwin77} Baldwin, J.~A.\ 1977, \apj, 214, 679 

\bibitem[Barth et al.(2011)]{Barth2011} Barth, A.~J., Pancoast, A., Thorman, S.~J., et al.\ 2011, \apjl, 743, L4 

\bibitem[Barth et al.(2015)]{Barth2015} Barth, A.~J., Bennert, 
V.~N., Canalizo, G., et al.\ 2015, \apjs, 217, 26 

\bibitem[Bennert et al.(2010)]{Bennert+10} Bennert, V.~N., Treu, T., Woo, J.-H., et al.\ 2010, \apj, 708, 1507 



\bibitem[Bentz et al.(2006)]{Bentz+06} Bentz, M.~C., Peterson, 
B.~M., Pogge, R.~W., Vestergaard, M., \& Onken, C.~A.\ 2006, \apj, 644, 133 

\bibitem[Bentz et al.(2009)]{Bentz+09} Bentz, M.~C., Peterson, 
B.~M., Netzer, H., Pogge, R.~W., \& Vestergaard, M.\ 2009, \apj, 697, 160 

\bibitem[Bentz et al.(2013)]{Bentz+13} Bentz, M. C., Denney, 
K. D., Grier, C. J., et al.\ 2013, \apj, 767, 149

\bibitem[Blandford \& McKee(1982)]{BM82} Blandford, R.~D., \& McKee, C.~F.\ 1982, \apj, 255, 419 

\bibitem[Boroson \& Green(1992)]{BG92} Boroson, T.~A., \& Green, R.~F.\ 1992, \apjs, 80, 109 

\bibitem[{{Bruzual} \& {Charlot}(2003)}]{BC03}
{Bruzual}, G., \& {Charlot}, S. 2003, \mnras, 344, 1000 

\bibitem[Cackett et al.(2015)]{Cackett+15} Cackett, E.~M., G{\"u}ltekin, K., Bentz, M.~C., et al.\ 2015, \apj, 810, 86 

\bibitem[Collin et al.(2006)]{Collin+06} Collin, S., Kawaguchi, T., Peterson, B.~M., \& Vestergaard, M.\ 2006, \aap, 456, 75 

\bibitem[Denney et al.(2009)]{Denney+09} Denney, K.~D., Peterson, B.~M., Dietrich, M., Vestergaard, M., \& Bentz, M.~C.\ 2009, \apj, 692, 246 

\bibitem[Denney(2012)]{Denney12} Denney, K. D.\ 2012, \apj, 759, 44

\bibitem[Dong et al.(2009)]{Dong+09} Dong, X.-B., Wang, T.-G., Wang, J.-G., et al.\ 2009, \apjl, 703, L1 


\bibitem[Fausnaugh(2017)]{Fausnaugh+17} Fausnaugh, M.~M.\ 2017, \pasp, 129, 024007 


\bibitem[Greene 
\& Ho(2005)]{Greene05} Greene, J.~E., \& Ho, L.~C.\ 2005, \apj, 630, 122 

\bibitem[Grier et al.(2013)]{Grier2013} Grier, C.~J., Martini, 
P., Watson, L.~C., et al.\ 2013, \apj, 773, 90 

\bibitem[Karouzos et al.(2015)]{Karouzos+15} Karouzos, M., Woo, J.-H., Matsuoka, K., et al.\ 2015, \apj, 815, 128 


\bibitem[Kaspi et al.(2000)]{Kaspi+00} Kaspi, S., Smith, P.~S., Netzer, H., et al.\ 2000, \apj, 533, 631 

\bibitem[Kaspi et al.(2005)]{Kaspi+05} Kaspi, S., Maoz, D., 
Netzer, H., et al.\ 2005, \apj, 629, 61 

\bibitem[Kormendy \& Ho(2013)]{Kormendy&Ho13} Kormendy, J., \& Ho, L.~C.\ 2013, ARA\&A, 51, 511

\bibitem[Kova{\v c}evi{\'c}-Doj{\v c}inovi{\'c} et al.(2017)]{Kovacevic+17} Kova{\v c}evi{\'c}-Doj{\v c}inovi{\'c}, J., Mar{\v c}eta-Mandi{\'c}, S., \& Popovi{\'c}, L.~{\v C}.\ 2017, arXiv:1707.08251 

\bibitem[Malkan et al.(2017)]{Malkan+17} Malkan, M.~A., Jensen, L.~D., Rodriguez, D.~R., Spinoglio, L., \& Rush, B.\ 2017, \apj, 846, 102 

\bibitem[Malkan \& Sargent(1982)]{Malkan&Sargent82} Malkan, M.~A., \& Sargent, W.~L.~W.\ 1982, \apj, 254, 22 

\bibitem[{Markwardt(2009)}]{markwardt09}
Markwardt, C.~B. 2009, in Astronomical Data
Analysis Software and Systems XVIII, ed. D. A. Bohlender, D. Durand, \&
P. Dowler (San Francisco: ASP), 251


\bibitem[Marziani et al.(2013)]{Marziani+13} Marziani, P., Sulentic, J.~W., Plauchu-Frayn, I., \& del Olmo, A.\ 2013, \aap, 555, A89 


\bibitem[McGill et al.(2008)]{McGill+08} McGill, K.~L., Woo, 
J.-H., Treu, T., \& Malkan, M.~A.\ 2008, \apj, 673, 703 

\bibitem[McLure 
\& Dunlop(2004)]{MD04} McLure, R.~J., \& Dunlop, J.~S.\ 2004, \mnras, 352, 1390 

\bibitem[McLure 
\& Jarvis(2002)]{MJ02} McLure, R.~J., \& Jarvis, M.~J.\ 2002, \mnras, 337, 109 

\bibitem[Metzroth et al.(2006)]{Metzroth+06} Metzroth, K.~G., Onken, C.~A., \& Peterson, B.~M.\ 2006, \apj, 647, 901 

\bibitem[Oke et al.(1995)]{Oke+95} Oke, J.~B., Cohen, J.~G., Carr, M., et al.\ 1995, \pasp, 107, 375 


\bibitem[Pancoast et al.(2014)]{Pancoast+14} Pancoast, A., Brewer, B.~J., Treu, T., et al.\ 2014, \mnras, 445, 3073 

\bibitem[Park et al.(2012a)]{Park12} Park, D., Woo, J.-H., 
Treu, T., et al.\ 2012a, \apj, 747, 30 

\bibitem[Park et al.(2012b)]{Park+12b} Park, D., Kelly, B. C., 
Woo, J.-H., \& Treu, T. \ 2012b, \apj, 203, 6

\bibitem[Park et al.(2015)]{Park15} Park, D., Woo, J.-H., Bennert, V. et al.\ 2015, \apj, 799, 164

\bibitem[Park et al.(2017)]{Park+17} Park, S., Woo, J.-H., Romero-Colmenero, E., et al.\ 2017, \apj, 847, 125 

\bibitem[Peterson(1993)]{Peterson93} Peterson, B.~M.\ 1993, \pasp, 
105, 247 

\bibitem[Peterson et al.(2004)]{Peterson+04} Peterson, B.~M., 
Ferrarese, L., Gilbert, K.~M., et al.\ 2004, \apj, 613, 682 

\bibitem[Reichert et al.(1994)]{Reichert+94} Reichert, G.~A., Rodriguez-Pascual, P.~M., Alloin, D., et al.\ 1994, \apj, 425, 582 


\bibitem[Salviander et al.(2007)]{Salviander07} Salviander, S., 
Shields, G.~A., Gebhardt, K., \& Bonning, E.~W.\ 2007, \apj, 662, 131 

\bibitem[Schlegel et al.(1998)]{schlegel98} Schlegel, D.~J., 
Finkbeiner, D.~P., \& Davis, M.\ 1998, \apj, 500, 525 

\bibitem[Shen et al.(2011)]{Shen+11} Shen, Y., Richards, G.~T., 
Strauss, M.~A., et al.\ 2011, \apjs, 194, 45 

\bibitem[Shen et al.(2016)]{Shen+16} Shen, Y., Horne, K., Grier, C.~J., et al.\ 2016, \apj, 818, 30 


\bibitem[Treu et al.(2004)]{Treu04} Treu, T., Malkan, M.~A., 
\& Blandford, R.~D.\ 2004, \apjl, 615, L97 

\bibitem[Tsuzuki et al.(2006)]{Tsuzuki06} Tsuzuki, Y., Kawara, 
K., Yoshii, Y., et al.\ 2006, \apj, 650, 57

\bibitem[Vanden Berk et al.(2001)]{Vanden01} Vanden Berk, D.~E., 
Richards, G.~T., Bauer, A., et al.\ 2001, \aj, 122, 549 


\bibitem[Vestergaard 
\& Peterson(2006)]{VP06} Vestergaard, M., \& Peterson, B.~M.\ 2006, \apj, 641, 689 

\bibitem[Vestergaard 
\& Wilkes(2001)]{VW01} Vestergaard, M., \& Wilkes, B.~J.\ 2001, \apjs, 134, 1 

\bibitem[Wandel et al.(1999)]{Wandel1999} Wandel, A., Peterson, B.~M., \& Malkan, M.~A.\ 1999, \apj, 526, 579 

\bibitem[Vanden Berk et al.(2001)]{VandenBerk+01} Vanden Berk, D.~E., Richards, G.~T., Bauer, A., et al.\ 2001, \aj, 122, 549 



\bibitem[Wang et al.(2009)]{Wang+09} Wang, J.-G., Dong, X.-B., 
Wang, T.-G., et al.\ 2009, \apj, 707, 1334 

\bibitem[Woo \& Urry(2002)]{Woo&Urry02} Woo, J.-H., \& Urry, C.~M.\ 2002, \apj, 579, 530 

\bibitem[Woo et al.(2006)]{Woo06} Woo, J.-H., Treu, T., 
Malkan, M.~A., \& Blandford, R.~D.\ 2006, \apj, 645, 900 

\bibitem[Woo(2008)]{Woo+08} Woo, J.-H.\ 2008, \aj, 135, 1849 

\bibitem[Woo et al.(2010)]{Woo+10} Woo, J.-H., Treu, T., Barth, A.~J., et al.\ 2010, \apj, 716, 269 
\bibitem[Woo et al.(2013)]{Woo+13} Woo, J.-H., Schulze, A., Park, D., et al.\ 2013, \apj, 772, 49 


\bibitem[Woo et al.(2015)]{Woo15} Woo, J.-H., Yoon, Y., Park, S. et al.\ 2015, \apj, 801, 38

\bibitem[Woo et al.(2016)]{Woo+16} Woo, J.-H., Bae, H.-J., Son, D., \& Karouzos, M.\ 2016, \apj, 817, 108 

\bibitem[Zheng \& Malkan(1993)]{Zheng&Malkan93} Zheng, W., \& Malkan, M.~A.\ 1993, \apj, 415, 517 

\bibitem[Zhu et al.(2017)]{Zhu+17} Zhu, D., Sun, M., \& Wang, T.\ 2017, \apj, 843, 30 


\end{thebibliography}
\end{document}